\documentclass[12pt]{article}

\usepackage{amssymb}
\usepackage{./localARAstroBib}
\usepackage{./psfig}
\usepackage{ifthen}
\usepackage{lscape}
\usepackage[margin=1in]{geometry}

\newboolean{HideNotes}
\setboolean{HideNotes}{true}

\newboolean{HideFigures}
\setboolean{HideFigures}{false}

\newcommand{\be}{\begin{eqnarray}}
\newcommand{\ee}{\end{eqnarray}}

\newcommand{\ts}{\mbox{${t_{\rm s}}$}}

\newcommand{\tdyn}{{\mbox{$t_{\rm dyn}$}}}

\newcommand{\trh}{{\mbox{$t_{\rm rh}$}}}

\newcommand{\tcc}{{\mbox{$t_{\rm cc}$}}}

\newcommand{\tdis}{\mbox{${t_{\rm dis}}$}}
\newcommand{\rvir}{\mbox{${r_{\rm vir}}$}}
\newcommand{\mvir}{\mbox{${M_{\rm vir}}$}}

\newcommand{\rJ}{{r_{\rm J}}}
\newcommand{\rj}{{r_{\rm J}}}
\newcommand{\rcore}{\mbox{$r_{\rm c}$}}

\newcommand{\rhm}{\mbox{$r_{\rm hm}$}}
\newcommand{\reff}{\mbox{$r_{\rm eff}$}}

\newcommand{\mmean}{\mbox{${\langle m \rangle}$}}
\newcommand{\rhocore}{\mbox{${\rho_{\rm c}}$}}
\newcommand{\rhohm}{\mbox{${\rho_{\rm hm}}$}}

\newcommand{\rhonot}{\mbox{${\rho_0}$}}

\newcommand{\sigoned}{\mbox{${\sigma_{\rm 1D}}$}}
\newcommand{\sigmaoned}{\mbox{${\sigma_{\rm 1D}}$}}

\newcommand{\Mto}{\mbox{${m_{\rm to}}$}}
\newcommand{\mtot}{\mbox{${M}$}}

\newcommand{\tenc}{\mbox{${t_{\rm enc}}$}}

\newcommand{\kms}{\mbox{${\rm km~s}^{-1}$}}
\newcommand{\msun}{\mbox{${\rm M}_\odot$}}
\newcommand{\Msun}{\mbox{${\rm M}_\odot$}}

\newcommand{\Rsun}{\mbox{${\rm R}_\odot$}}
\newcommand{\Zsun}{\mbox{${\rm Z}_\odot$}}
\newcommand{\trlx}{\mbox{$t_{\rm rh}$}}
\newcommand{\trl}{\mbox{$t_{\rm rl}$}}

\def\apgt{\ {\raise-.5ex\hbox{$\buildrel>\over\sim$}}\ }
\def\aplt{\ {\raise-.5ex\hbox{$\buildrel<\over\sim$}}\ }
\def\lt{\ {\raise-.5ex\hbox{$\buildrel>$}}\ }
\def\gt{\ {\raise-.5ex\hbox{$\buildrel<$}}\ }

\newcommand{\rtide}{\mbox{$r_{\rm t}$}}

\newcommand{\mphot}{\mbox{$M_{\rm phot}$}}
\newcommand{\mdyn}{\mbox{$M_{\rm dyn}$}}
\newcommand{\tcr}{\mbox{$t_{\rm dyn}$}}

\newcommand{\myrpkpc}{\mbox{${\rm Myr}^{-1}\,{\rm kpc}^{-2}$}}

\newcommand{\msmyrkpc}{\mbox{$\msun\,{\rm Myr}^{-1}\,{\rm kpc}^{-2}$}}

\newcommand{\dndl}{\mbox{${\rm d}N/{\rm d}L$}}
\newcommand{\lmax}{\mbox{$L_{\rm max}$}}

\newcommand{\mmax}{\mbox{$M_{\rm max}$}}

\newcommand{\sfrarea}{\mbox{$\Sigma_{\rm SFR}$}}
\newcommand{\rt}{\mbox{$r_{\rm t}$}}
\newcommand{\gammathreed}{\mbox{$\gamma_{\rm 3D}$}}
\newcommand{\ncl}{\mbox{$N_{\rm cl}$}}

\newcommand{\msunpc}{M_\odot\,{\rm pc}^{-3}}
\newcommand{\mg}{M_{\rm g}}
\newcommand{\ms}{M_*}

\newcommand{\rh}{{r_{\rm hm}}}
\newcommand{\rhogmc}{\rho_{\rm gas}}

\newcommand{\sigmagas}{\Sigma_{\rm gas}}

\newcommand{\tdisgmc}{t_{\rm dis}^{\rm GMC}}
\newcommand{\trem}{\mbox{$t_{\rm exp}$}}

\newcommand{\vrec}{\mbox{$v_{\rm rec}$}}
\newcommand{\vesc}{\mbox{$v_{\rm esc}$}}
\newcommand{\vrms}{\mbox{$\langle v^2 \rangle^{1/2}$}}
\newcommand{\vrmssq}{\mbox{$\langle v^2 \rangle$}}

\newcommand{\xie}{\xi_{\rm e}}

\begin{document}


\title{Young massive star clusters}

\author{
 Simon F. Portegies Zwart \\
       Leiden Observatory, Leiden University, \\
       P.O. Box 9513, NL-2300 RA  Leiden, The Netherlands\\
       email: spz@strw.leidenuniv.nl \\
\\
 Stephen L.W. McMillan\\
       Department of Physics, Drexel University, Philadelphia, PA
       19104, USA\\
       email: steve@physics.drexel.edu\\
\\
 Mark Gieles\\
       European Southern Observatory, Casilla 19001, Santiago 19, Chile\\
       email: mgieles@eso.org
}

\maketitle

\begin{abstract}

  Young massive clusters are dense aggregates of young stars that form
  the fundamental building blocks of galaxies.  Several examples exist
  in the Milky Way Galaxy and the Local Group, but they are
  particularly abundant in starburst and interacting galaxies.  The
  few young massive clusters that are close enough to resolve are of
  prime interest for studying the stellar mass function and the
  ecological interplay between stellar evolution and stellar dynamics.
  The distant unresolved clusters may be effectively used to study the
  star-cluster mass function, and they provide excellent constraints
  on the formation mechanisms of young cluster populations.  Young
  massive clusters are expected to be the nurseries for many unusual
  objects, including a wide range of exotic stars and binaries.  So
  far only a few such objects have been found in young massive
  clusters, although their older cousins, the globular clusters, are
  unusually rich in stellar exotica.

  In this review we focus on star clusters younger than
  $\sim100$\,Myr, more than a few current crossing times old, and more
  massive than $\sim10^4$\,\Msun, irrespective of cluster size or
  environment.  We describe the global properties of the currently
  known young massive star clusters in the Local Group and beyond, and
  discuss the state of the art in observations and dynamical modeling
  of these systems.  In order to make this review readable by
  observers, theorists, and computational astrophysicists, we also
  review the cross-disciplinary terminology.

\end{abstract}


\newpage
  
\section{Introduction}
\label{Sect:Introduction}

Stars form in clustered environments \citep{2003ARA&A..41...57L}.  In
the Milky Way Galaxy, evidence for this statement comes from the
global clustering of spectral O-type stars
\citep{2007MNRAS.380.1271P}, of which $\sim70$\% reside in young
clusters or associations \citep{1987ApJS...64..545G}, and $\sim50$\%
of the remaining field population are directly identified as runaways
\citep{2005A&A...437..247D}. \cite{2005A&A...437..247D} found that
only $\sim4$\% of O-type stars can be considered as having formed
outside a clustered environment; further analysis has shown that a few
of even those 4\% may actually be runaway stars
\citep{2008A&A...490.1071G, 2008A&A...489..105S}, further
strengthening the case for clustered formation.  Additional evidence
that clusters are the primary mode of star formation comes from the
observed formation rate of stars in embedded clusters \citep[$\sim
3\times10^3\,\msmyrkpc$;][]{2003ARA&A..41...57L} which is comparable
to the formation rate of field stars
\citep[$\sim3$--$7\times10^3\,\msmyrkpc$;][]{1979ApJS...41..513M}.
Finally, some 96\% of the stars in the nearby Orion~B star-forming
region are clustered \citep{2000prpl.conf..151C}.

In nearby young starburst galaxies at least 20\%, and possibly all, of
the ultraviolet light appears to come from young star clusters
\citep{1995AJ....110.2665M}; this also seems to be the case for the
observed $H\alpha$ and B-band luminosities in interacting galaxies,
such as the Antennae \citep{2005ApJ...631L.133F} and NGC~3256
\citep{1999AJ....118..752Z}.  The fossil record of an early episode of
star formation is evidenced by the present-day population of old
globular clusters, although the first (population~III) stars seem not
to have formed in clusters \citep{2002Sci...295...93A}.

\subsection{Scope of this review}\label{Sect:Definitions}

In this review we focus on the {\em young massive clusters} (hereafter
YMCs) found in an increasing number of galaxies.  We adopt a
deliberately broad definition of this term, concentrating on
observations of star clusters younger than about 100\,Myr and more
massive than $10^4$\,\msun.  In addition, implicit throughout most of
our discussion is the assumption that the clusters under study are
actually {\em bound}.  This may seem obvious but, as we will discuss
in \S\ref{sec:observations}, some very young objects satisfying our
age and mass criteria may well be unbound, expanding freely into space
following the loss of the intracluster gas out of which they formed.
We find that imposing a third requirement, that the age of the cluster
exceed its current dynamical time (the orbit time of a typical
star---see \S\ref{Sect:TimeScales}) by a factor of a few, effectively
distinguishes between bound clusters and unbound associations.

Our adopted age limit is somewhat arbitrary, but represents roughly
the epoch at which a cluster can be said to have survived its birth
(phase~1, see \S\ref{Sec:Dynamics}) and the subsequent early phase of
vigorous stellar evolution (phase~2, see \S\ref{Sect:SCSurvival}), and
to be entering the long-term evolutionary phase during which its
lifetime is determined principally by stellar dynamical processes and
external influences (phase~3, see \S\ref{Sec:Dynamics}). This latter,
``stellar dynamical'' phase generally starts after about 100\,Myr and
is not the principal focus of this review.

The mass limit is such that lower-mass clusters are unlikely to
survive for more than 1 Gyr.  Based on the lifetimes presented in
\S\ref{Sect:SCSurvival} (in particular see Eq.\,\ref{eq:tdis3}), we
estimate that a cluster with an initial number of stars $N \simeq
10^5$ will survive for $\sim10$\,Gyr.  Since young clusters more
massive than $10^5$\,{\msun} are relatively rare, we relax our
criterion to include star clusters with masses as low as
$10^4$\,\msun.  We place no limits on cluster size, metallicity, or
galactic location, for the practical reason that this would further
reduce our already small sample of YMCs, even though it seems likely
that clusters such as the Arches and Quintuplet systems near the
Galactic center (see Tab.\,\ref{Tab:Galactic_Clusters}) are likely to
dissolve within a gigayear.

Thus, any young cluster massive enough to survive for a significant
fraction of a Hubble time---regardless of its current location---meets
our criteria for inclusion in this review.  Within the current context
we cannot make a strong connection between cluster properties and
environment, but from the discussion around
Fig.\,\ref{Fig:cluster_mass_functions} below it is evident that
environment has at least some influence on global cluster
characteristics (see \S\ref{Sect:CIMF}).

The masses and projected lifetimes of YMCs coincide with those of the
old globular clusters (hereafter GCs) that populate the bulges and
halos of many galaxies, including our own.  Indeed, YMCs are sometimes
referred to in the literature as ``young globular clusters.''  This
possible connection between YMCs and GCs offers the exciting prospect
of studying in the local universe physical processes that may have
occurred during the otherwise practically unobservable formative
stages of the GC population.\footnote{``Real'' young globular clusters
  at $z\sim5$ are expected to be $\sim$2 magnitudes brighter than the
  detection limit of the James~Webb Space Telescope.}  However,
although the characterization is highly suggestive, the extent to
which today's YMCs will someday come to resemble GCs remains unclear.

As we will see, the size distribution of YMCs does appear to be
consistent with them evolving into GCs \citep{2002IAUS..207..697M},
and we can reasonably expect that after (say) 10 Gyr they will be
roughly spherical in shape and will have surface brightness
distributions similar in character to those of the GCs, but other
properties are not so easy to assess.  Except for a few nearby cases,
such as Westerlund~1 and the Arches cluster (see
Tab.\,\ref{Tab:Galactic_Clusters} in \S\ref{sec:observations}),
observations of YMCs are limited to stellar masses $\apgt 1$\,\msun,
whereas the most massive stars observed in GCs are less than 1\,\msun.
Thus there is no guarantee that the stellar mass function in YMCs
below $\sim 1${\msun} resembles the initial mass functions of known
GCs, although we note that the stellar mass functions in nearby open
clusters are consistent with the distribution of low-mass stars in GCs
\citep{2003PASP..115..763C,2010arXiv1001.2965B}.  Similar
uncertainties apply to cluster binary populations.  The initial binary
fractions inferred for GCs are generally small, whereas YMCs appear to
be binary-rich (see \S\ref{Sect:binaries}).  But the known binaries in
massive clusters tend to be found among massive stars, while the
binary properties of low-mass stars in YMCs are unknown.  Again, any
comparison is complicated by the absence of any significant overlap in
the observed stellar mass spectra.

Our objective in this review is to summarize the current state of
knowledge of YMCs, to describe the key physical processes governing
their evolution and survival, and to assess the extent to which we
expect them to evolve into systems comparable to the old GCs observed
today.

In this review we will use the terms {\em young}, {\em dense}, and
{\em massive} in relation to star clusters. Although not precise,
these descriptions do have specific connotations.  As already
indicated, ``young'' means star clusters that are still in the early,
violent mass-loss phase during the first 100 Myr (see
\S\ref{Sect:SCSurvival}).  ``Dense'' indicates that in some clusters
the stars are packed together so closely that stellar collisions start
to play an important role (see \S\ref{Sect:Collisions}).  In terms of
the Safronov number,\footnote{The \citet{1969QB981.S26......} number
  $\Theta$ is defined as the square of the ratio of the escape
  velocity from the stellar surface to the rms (\vrms) velocity in the
  cluster:
  \begin{equation}
    \Theta = {1 \over 2} \left( {v_{\rm \star, esc}} \over { \vrms } \right)^2.
  \label{Eq:Safronov}\end{equation}}
young dense clusters have $\Theta \aplt 10^2$.  As a practical matter,
we consider a cluster to be dense if its half-mass relaxation time
(Eq.\,\ref{Eq:trtd}) is less than $\sim10^8$\,years
\citep{2007MNRAS.374...95P}. ``Massive'' indicates that we expect the
cluster to survive for $\sim10$ Gyr, into the ``old globular cluster''
regime.  Tab.\,\ref{Tab:Comparison} summarizes the main parameters of
the three different populations of star clusters.

\begin{table}[htbp!]
\begin{tabular}{lrrrrcrlrr}
\hline
cluster& age       &\Mto& $M$       &\rvir&$\rho_c$& Z     & location & \tdyn & \trlx \\
    &[Gyr]      &[\msun]&[\msun]     &[pc]&[\msun/pc$^3$]&[\Zsun]&          & [Myr] & [Myr] \\
\hline
OC &$\aplt ~~0.3$&$\aplt 4~~$&$\aplt 10^3$ & 1&$\aplt10^3$&$\sim 1$&disk  &$\sim 1$&$\aplt 100$ \\
GC &$\apgt 10~~~$ &$\sim 0.8$&$\apgt 10^5$& 10&$\apgt10^3$&$<1$&halo  &$\apgt 1$&$\apgt 1000$\\
YMC&$\aplt ~~0.1$&$\apgt 5~~$&$\apgt 10^4$ & 1&$\apgt10^3$&$\apgt 1$&galaxy&$\aplt 1$&$\aplt 100$  \\
\hline
\end{tabular}
\caption{Comparison of fundamental parameters for star cluster
  families relevant to this review: open cluster (OC), globular
  cluster (GC), and young massive cluster (YMC).  The numbers in the
  columns are intended to be indicative of the population and are
  rounded-off, and should be used with care, but they provide some
  flavor of the various cluster types.  The second column gives
  cluster age, followed by the turn-off mass (in \msun), the total
  cluster mass (in \msun), the virial radius {\rvir} (see
  \S\ref{Sect:TheRadius}), the core density, and the metallicity.  The
  last three columns give the location in the Galaxy where these
  clusters are found, and the dynamical and relaxation time scales,
  defined in \S\ref{Sect:TimeScales}.}
 \label{Tab:Comparison}
\end{table}

\begin{figure}
\begin{tabular}{c}
\psfig{figure=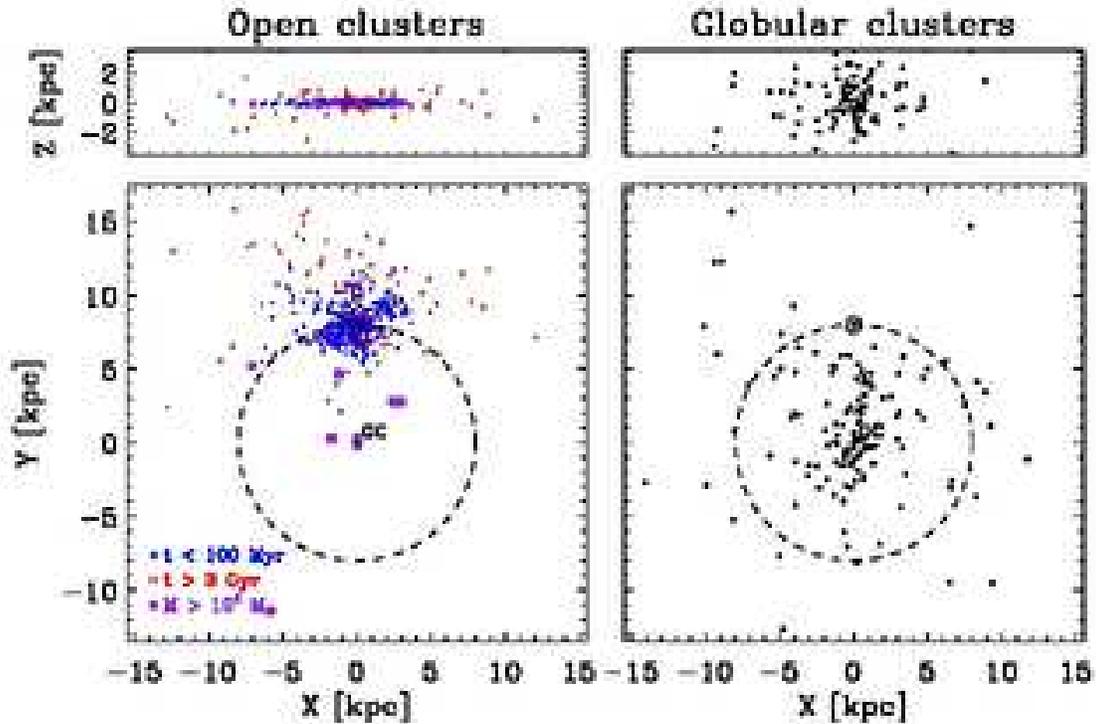,width=\columnwidth}
\end{tabular}
\caption{Left: Distribution of young ($<100$\,Myr, bullets) and old
  ($>3$\,Gyr, circles) open clusters in the Galactic plane, based on
  the catalog of \citet{2002A&A...389..871D}.  The old open clusters
  are found preferentially towards the Galactic anti-center and above
  the plane.  The young massive clusters (squares) seem to be
  concentrated in the same quadrant as the Sun, which is an
  observational selection effect.  Right: Distribution of old globular
  clusters, data from the \cite{1996AJ....112.1487H} catalog.}
   \label{Fig:distribution_open_globular_clusters}
\end{figure}

\begin{figure}[!t]
\begin{tabular}{c}
\hspace{0.15\columnwidth}
\psfig{figure=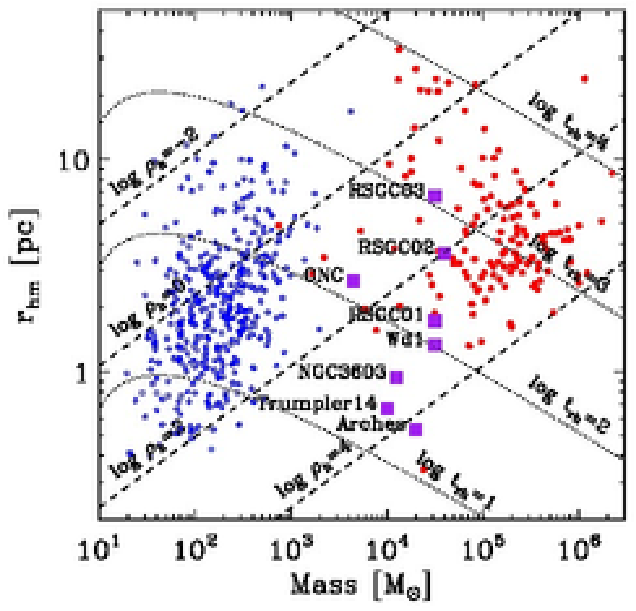,width=0.6\columnwidth}
\end{tabular}
\caption{Radius--mass diagram of Milky Way open clusters, young
  massive clusters and old globular clusters. Open cluster half-mass radii
  $\rhm$ (see \S\ref{Sect:TheRadius}) and masses are taken from
  \citet{2002A&A...389..871D} and \citet[][private
  communication]{2005A&A...441..117L}, respectively.  Data for the
  YMCs are discussed in more detail in
  \S\ref{sec:observations}. Globular cluster data are taken from the
  Harris catalog.  Dashed and dotted lines represent constant
  half-mass density $\rho_{\rm h}=3M/8\pi\rhm^3$ and half-mass
  relaxation time $\trlx$ (Eq.~\ref{Eq:trlx_YMC}), respectively.}
   \label{Fig:radius_mass_all_clusters}
\end{figure}

Figs.\,\ref{Fig:distribution_open_globular_clusters} and
\ref{Fig:radius_mass_all_clusters} compare the distributions of
massive open star clusters, YMCs, and GCs in the Milky Way Galaxy.
The spatial distribution of the YMCs
(Fig.\,\ref{Fig:distribution_open_globular_clusters}) clearly identify
them as a disk population comparable to the open clusters, but in the
mass-radius diagram (Fig.\,\ref{Fig:radius_mass_all_clusters}), YMCs
seem more closely related to GCs.

\subsection{Properties of Cluster Systems}\label{Sect:ClusterProperties}

The Milky Way Galaxy contains some 150 GCs, with mass estimates
ranging from $\sim 10^3$\,{\msun} (for AM4, a member of the Sgr dwarf
spheroidal galaxy) to $2.2\times 10^6$\,{\msun} (for NGC\,5139, Omega
Centauri)\footnote{These estimates are made assuming a constant
  mass-to-light ratio $M/L = 2$, with data from the
  \citep{1996AJ....112.1487H} catalog of Milky Way GCs {\tt
    http://www.physics.mcmaster.ca/Globular.html}. Another useful
  catalog for Milky Way globular cluster data is available online {\tt
    http://www.astro.caltech.edu/$\sim$george/glob/data.html}
  \citep{1993ASPC...50.....D}.}.  If we assume constant $M/L = 2$, the
current total mass in GCs in the \cite{1996AJ....112.1487H} catalog is
$\sim 3.5\times 10^7$\,\msun, or $\sim 0.07$\% of the baryonic mass of
the Galaxy and 0.005\% of the total mass (including dark matter).  The
luminosity function of GCs in the Galaxy peaks at
$M_V\approx-7.4\,$mag, corresponding to a typical mass of
$\sim2\times10^5\,\msun$, and has a (Gaussian) width of $\simeq
1.2\,$mag \citep{2001stcl.conf..223H}.  The total initial mass for GCs
is estimated to be $\sim4-8 \times
10^8$\,\msun\ \citep{2001ApJ...561..751F}.  Apparently, more than 90\%
of all globular star clusters have been disrupted during the last
$\sim 12$\,Gyr \citep{1990ApJ...351..121C}; the inferred total mass in
disrupted clusters is comparable to the total mass of the Galactic
stellar halo \citep[][see also
  \S\ref{Sect:SCSurvival}]{1992Natur.359..806H,2008ApJ...680..295B},
although we note that the spatial distributions of halo stars and GCs
differ \citep{2009MNRAS.397.1003F}.

The open cluster databases of \cite{2005A&A...440..403K} and
\cite{2008A&A...477..165P}, with 81 clusters, are probably complete to
a distance of $\sim 600$\,pc and have a mean cluster mass of $\sim
500$\,\msun.  With an open cluster birth rate of
$0.2$--$0.5\,\myrpkpc$
\citep{1991MNRAS.249...76B,2006A&A...445..545P}, the implied total
star formation rate in open clusters is $\sim 2\times10^2\,\msmyrkpc$.
For an average cluster age of $\sim250$\,Myr these estimates imply a
total of about 23,000--37,000 open star clusters currently in the
Galaxy.  However, the formation rate of embedded clusters (still
partly or completely enshrouded in the molecular cloud from which they
formed) is considerably higher---$2-4\,\myrpkpc$
\citep{2003ARA&A..41...57L}---and with a similar cluster mean mass the
total star formation rate in embedded clusters is $\sim
3\times10^3\,\msmyrkpc$, comparable to the formation rate of field
stars in the disk
\citep[$3-7\times10^3$\,\msmyrkpc;][]{1979ApJS...41..513M}.  Although
the uncertainties are large, this suggests that the majority of stars
form in embedded clusters, but only a relatively small fraction ($\sim
10$\%) of clusters survive the embedded phase.

\begin{figure}
\centering
\hspace{0.0\columnwidth}\psfig{figure=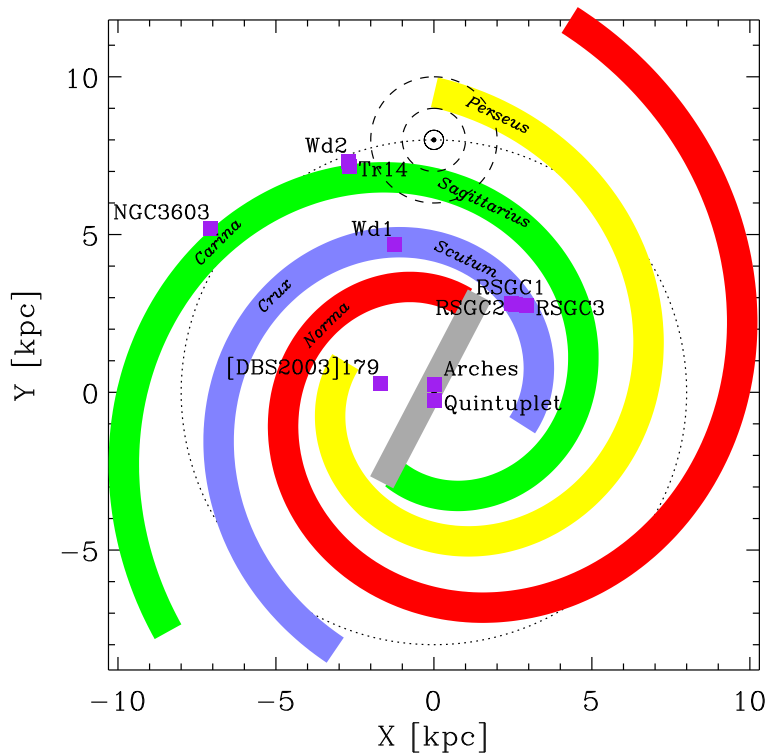,width=0.7\columnwidth}
\caption[]{ 
Top view of the Milky Way Galaxy, with the known spiral pattern
\citep{2008AJ....135.1301V} and young star clusters more massive than
$\gtrsim10^4\,\msun$ identified.  Basic cluster parameters are listed in
Tab.\,\ref{Tab:Galactic_Clusters}.  The location of the Sun is
indicated by $\odot$, and its orbit by the dotted circle.  Dashed
lines indicate circles of 1\,kpc and 2\,kpc around the sun. 
}\label{Fig:distribution_starburst_clusters}
\end{figure}

These estimates are sensitive to the underlying assumptions made about
the star-formation history of the Galaxy and the duration of the
embedded phase, as well as to observational selection effects.  For
example, the open cluster sample used by \citet{1991MNRAS.249...76B}
is based on a luminosity limited sample of 100 clusters from the
\citet{1982A&A...109..213L} catalog, which claims to be complete to a
distance of 2\,kpc, but the mass of open clusters in this catalog
between 600\,pc and 2\,kpc averages several thousand solar masses. The
much higher mean mass of open clusters at large distance indicates
that care has to be taken in using these catalogs, as there appear to
be selection effects with respect to distance.  Another problem arises
from confining the analysis to a distance of 600 pc around the Sun,
since the cluster sample does not include any nearby spiral arms,
where many young clusters form; \citet{2003ARA&A..41...57L} considered
a sample of clusters within 2 kpc of the sun, which therefore includes
many objects in the Perseus and Sagittarius arms
(Fig.~\ref{Fig:distribution_starburst_clusters}).  These differences
complicate direct comparison of cluster samples.

\subsection{Terminology}

The study of star clusters has suffered from conflicting terminology
used by theorists and observers.  In this section we attempt to
clarify some terms, with the goal of making this review more readable
by all.

\subsubsection{Cluster center}\label{Sect:TheCenter}

Determining the center of a star cluster sounds like a trivial
exercise, but in practice it is not easy.  The cluster center is not
well defined observationally, although theorists have reached some
consensus about its definition.

\cite{1963ZA.....57...47V} defined the center of a simulated ($N$-body)
cluster as a density-weighted average of stellar positions:
\begin{equation}
  \bar{x}_{d,j} = { \sum_i x_i \rho^{(j)}_i \over \sum_i \rho^{(j)}_i},
\label{Eq:DensityCenter}
\end{equation}
where $\rho^{(j)}_i$ is the density estimator of order $j$ around star
$i$, and $x_i$ is the (3-dimensional) position vector of star $i$.

In direct $N$-body simulations (see \S\ref{Sect:DirectNbody}),
alternatives to Eq.\,\ref{Eq:DensityCenter} are preferred due to the
computational expense of determining the local density $\rho^{(j)}_i$.
The center of mass, often used in simple estimates, is generally {\em
  not} a good measure of the cluster center, as distant stars tend to
dominate.  This has led to approximate, but more efficient,
estimators, such as the ``modified'' density center
\citep{2001MNRAS.321..199P}, which iteratively determines the weighted
mean of the positions of a specified Lagrangian fraction (typically
$\sim90$\%) of stars, relative to the modified density center
\citep{1994MNRAS.271..706H}.  In general, it agrees well with the
density center defined above.

Observationally, the cluster center is considerably harder to define,
both because of the lack of full 3-dimensional stellar positions, and
also because of observational selection effects, including the
influence of low-luminosity stars and remnants, crowding, and the
broad range in luminosities of individual stars.  Both number-averaged
and luminosity-averaged estimators are found in the literature.  In
principle, the 2-dimensional equivalent of Eq.\,\ref{Eq:DensityCenter}
could be used, but observers often prefer the point of maximal
symmetry of the observed projected stellar distribution.  An example
is the technique adopted by \citet{2006ApJS..166..249M} to determine
the center of GC 47~Tuc.

\subsubsection{Size scales}\label{Sect:TheRadius}

Massive star clusters tend to be approximately spherically symmetric
in space, or at least circular on the sky, so the radius of a cluster
is a meaningful measure of its size.  Theorists often talk in terms of
Lagrangian radii---distances from the center containing specific
fractions of the total cluster mass.  For observers, a similar
definition can be formulated in terms of isophotes containing given
fractions of the total luminosity.  The {\em half-mass radius} (\rhm;
the 50\% Lagrangian radius) is the distance from the cluster center
containing half of the total mass.  Observationally, the projected
half-light radius---the effective radius \reff---is often used, although
the total cluster light, which is obviously required to define the
Lagrangian radii, can be hard to determine and is not the same as
$\rhm$ in projection when the mass-to-light ratio varies with the
distance to the cluster centre.

Arguably more useful cluster scales are the virial radius \rvir, the
core radius {\rcore} (see Eq.\,\ref{Eq:theoryrcore1}), and the tidal
radius \rtide.  We now define them in turn.

The {\em virial radius} is defined as
\begin{equation}
  \rvir \equiv \frac{G \mtot^2}{2|U|}.
\end{equation}\label{Eq:rvir}
Here $\mtot$ is the total cluster mass, $U$ is the total potential
energy and $G$ is the gravitational constant.  This is clearly a
theoretical definition, as neither the total mass nor the potential
energy are actually observed.  The potential energy may be obtained
directly from the stellar masses and positions in an $N$-body simulation
(see \S\ref{Sec:Simulations}), or from a potential--density pair by
$U=2\pi\int\rho(r)\phi(r)\,r^2\,dr$.

\begin{figure}
\begin{tabular}{c}
 \psfig{figure=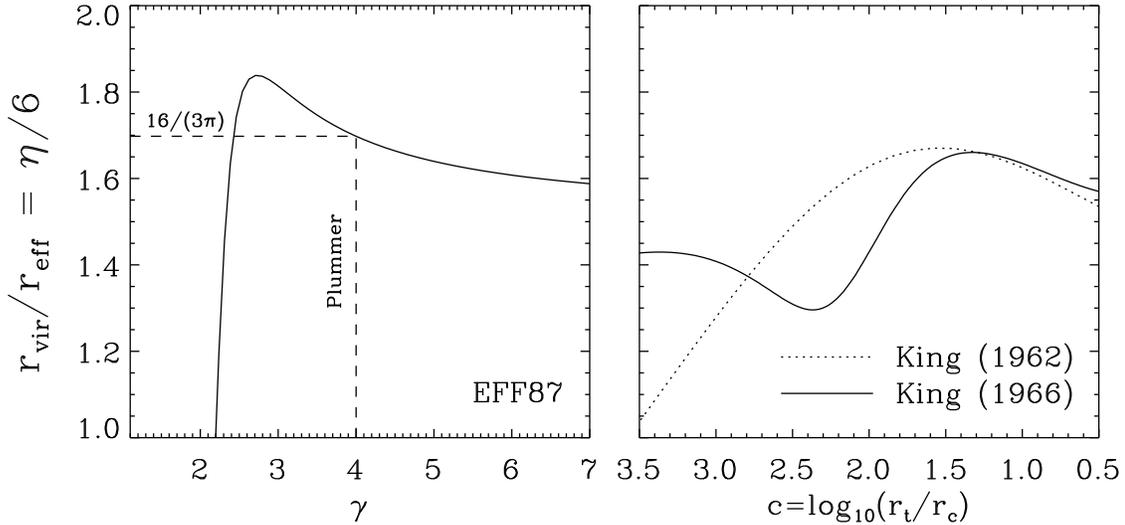,width=\columnwidth} 
\end{tabular}
\caption[]{The ratio $\rvir/\reff$ and the parameter $\eta$ used to
  convert an observed 1-D velocity dispersion and half-light radius
  into a dynamical mass (Eq.~\ref{Eq:Mvirial}) for Eq.\,\ref{Eq:EFF87}
  (left) and \cite{1962AJ.....67..471K} and
  \citet{1966AJ.....71...64K} models (right). The dashed line in the
  left panel indicates the analytical result for a
  \citet{1911MNRAS..71..460P} model ($\gamma=4$ in
  Eq.~\ref{Eq:EFF87}).}
\label{Fig:Observations_eta}
\end{figure}

From an observational point of view, the parameter $\eta \equiv 6
\rvir/\reff$ is generally introduced to determine the dynamical mass
of star clusters. In virial equilibrium ($U=-2T$, where $T$ is the
total kinetic energy of the cluster stars), $T/\mtot = \frac{1}{2}
\vrmssq = \frac{3}{2} \sigoned^2$ for an isotropic system.  The
line-of-sight velocity dispersion $\sigoned$ can be directly measured,
yielding the cluster mass
\begin{equation}
  \mvir = \eta \left( \frac{\sigoned^2\reff}{G} 
               \right).
\label{Eq:Mvirial}
\end{equation}
In Fig.\,\ref{Fig:Observations_eta} we present the dependence of
$\eta$ on the parameters of some typical density profiles: the
concentration parameter $c \equiv \log(\rt/\rcore)$ of a
\citet{1966AJ.....71...64K} model or the parameter $\gamma$ in an
\citet[][hereafter EFF87]{1987ApJ...323...54E} surface brightness
profile,
\begin{equation}
  \Sigma(r) = \Sigma_0\,\left(1+\frac{r^2}{a^2}\right)^{-\gamma/2},
\label{Eq:EFF87}
\end{equation}
where $a$ is a scale parameter, the 3-dimensional density profile has
a logarithmic slope of $-\gammathreed=-(\gamma+1)$ for $r\gg a$, and
{\rcore} and {\rtide} are, respectively, the core radius and tidal
radius of the cluster (to be defined below).

A \cite{1911MNRAS..71..460P} density profile has $\gamma=4$,
$\rvir/\reff=16/3\pi$, and therefore $\eta\simeq10$.  The value
$\eta=9.75$, corresponding to $\rvir = 1.625 \reff$, is a reasonable
and widely used choice for clusters with relatively shallow density
profiles---$\gamma\gtrsim 4$ or $c\lesssim 1.8$.  For $\gamma\le2$ the
EFF87 profile has infinite mass, and the ratio $\rvir/\reff$ drops
sharply for $\gamma\lesssim2.5$.  The choice for $\eta=9.75$ should be
made cautiously, since many young clusters tend to have relatively
shallow density profiles with $2\lesssim\gamma\lesssim3$, for which
$\eta \aplt 9$ (see Fig.\,\ref{Fig:Observations_eta} and also
\S\ref{sec:observations}). In addition, mass segregation can have a
severe effect on $\eta$, resulting in a variation of more than a
factor of $\sim$3 \citep{2005sf2a.conf..605F}.

Observers generally define the cluster {\em core radius}, \rcore, as
the distance from the cluster center at which the surface brightness
drops by a factor of two from the central value.
Unfortunately, theorists use at least two distinct definitions of
{\rcore}, depending on context.  When the central density $\rhonot$
and velocity dispersion $\vrmssq_0$ are easily and stably defined, as
is often the case for analytic, Fokker--Planck, and Monte Carlo models
(see \S\ref{Sec:Simulations} and the Appendix \ref{Sec:AppendixA} for
more details), the definition
\begin{equation}
  \rcore = \sqrt{\frac{3\vrmssq_0}{4\pi G\rhonot}}
\label{Eq:theoryrcore1}
\end{equation}
\citep{1966AJ.....71...64K} is often adopted.  For typical cluster
models this corresponds roughly to the radius at which the
three-dimensional stellar density drops by a factor of 3, and the
surface density by $\sim2$.

In $N$-body simulations, however, both $\rhonot$ and $\langle
v^2\rangle_0$ are difficult to determine, as they are subject to
substantial stochastic fluctuations.  As a result, a density-weighted
core radius is used instead.  Specifically, for each star a local
density $\rho_i$ is defined using the star's $k$ nearest neighbors
\citep{1985ApJ...298...80C}, where $k=12$ is a common choice.  A
density center is then determined, either simply the location of the
star having the greatest neighbor density, or as a mean stellar
position, as in Eq.\,\ref{Eq:DensityCenter}, except that the density
estimator is $\rho_i^2$.  [The square is used rather than the first
  power, as originally suggested by \cite{1985ApJ...298...80C}, to
  stabilize the algorithm and make it less sensitive to
  outliers.]  The core radius then is the $\rho_i^2$-weighted rms
stellar distance from the density center:
\begin{equation}
  \rcore = \sqrt{\frac{\sum_i\,\rho_i^2 r_i^2}{\sum_i\,\rho_i^2}}.
\label{Eq:theoryrcore2}
\end{equation}
Despite their rather different definitions, in practice the two
``theoretical'' core radii (Eqs. \ref{Eq:theoryrcore1} and
\ref{Eq:theoryrcore2}) behave quite comparably in simulations.

For simple models, the values of {\rcore} and {\rvir} determine the
density profile, which is generally assumed to be spherically
symmetric.  This is the case for the empirical
\cite{1962AJ.....67..471K} profiles and dynamical
\cite{1966AJ.....71...64K} models, both of which fit the observed
surface brightness distribution of many Milky Way GCs.  The dynamical
King models are often parameterized by a quantity $W_0$ representing
the dimensionless depth of the cluster potential well.  Centrally
concentrated clusters have $W_0\apgt8~ (c\apgt1.8)$, whereas shallow
models have $W_0 \aplt 4~ (c\aplt0.8)$.  The empirical and dynamical
King profiles are in good agreement for $W_0 \aplt 7~ (c\aplt1.5)$.
 
Galactic GCs are well fit by King models, but the observed surface
brightness profiles of young clusters in, for example, the LMC are not
\citep[see e.g.][]{2003MNRAS.338...85M}.  They are much better
represented by EFF87 profiles (Eq.~\ref{Eq:EFF87}), which have cores
(different from the King-model cores) and power-law halos.  For a King
model with concentration $c \gtrsim 1$ ($W_0\gtrsim5$), the surface
brightness drops to approximately half of its central value at
$r=\rcore$, as defined by Eq.\,\ref{Eq:theoryrcore1}, so the observed
core radius is a good measure of the core radius of the underlying
three-dimensional stellar density distribution.  The ``King'' core
radius of the EFF87 surface brightness profile often adopted by
observers is
\begin{equation}
  \rcore = a \left( 2^{2/\gamma}-1 \right)^{1/2}.
\label{Eq:rcore}
\end{equation}
Here $a$ is the scale parameter in the EFF87 profile.  Thus, when an
EFF87 surface brightness profile is fit to an observed cluster,
Eq.\,\ref{Eq:rcore} can be used to determine with good confidence the
3-dimensional core radius $\rcore$ of Eq.\,\ref{Eq:theoryrcore1}.

\begin{figure}
\begin{tabular}{cc}
  \hspace{0.15\columnwidth}
  \psfig{figure=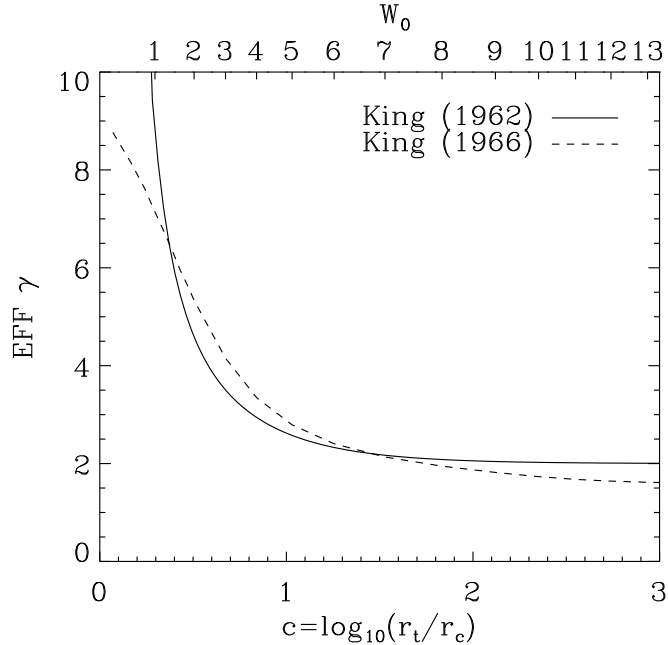,width=0.6\columnwidth}
\end{tabular}
\caption[]{EFF87 model fits (with power-law index $\gamma$) to King
  surface brightness profiles (characterized by King concentration
  parameter $c = \log_{10} r_t/\rcore$).  The fits of
  Eq.\,\ref{Eq:EFF87} to the King profiles/models are done within
  half the cluster tidal radius.}
\label{Fig:Comparison_EFF_King}
\end{figure}

In Fig.\,~\ref{Fig:Comparison_EFF_King} we compare the EFF87 profiles
with the empirical King profiles and (projected) King models, by
fitting Eq.\,\ref{Eq:EFF87} to each within the inner half-mass radius.
For $c\rightarrow\infty$ the \cite{1962AJ.....67..471K} surface
brightness profile tends to a power-law with index $-2$, which has
(logarithmically) infinite mass.  The \cite{1966AJ.....71...64K} model
in that case becomes an isothermal sphere ($\rho\propto r^{-2}$), also
with (linearly) infinite mass, corresponding to $\gamma=1$ in
Eq.\,\ref{Eq:EFF87}.

\begin{figure}
\begin{tabular}{cc}
  \hspace{0.15\columnwidth}
  \psfig{figure=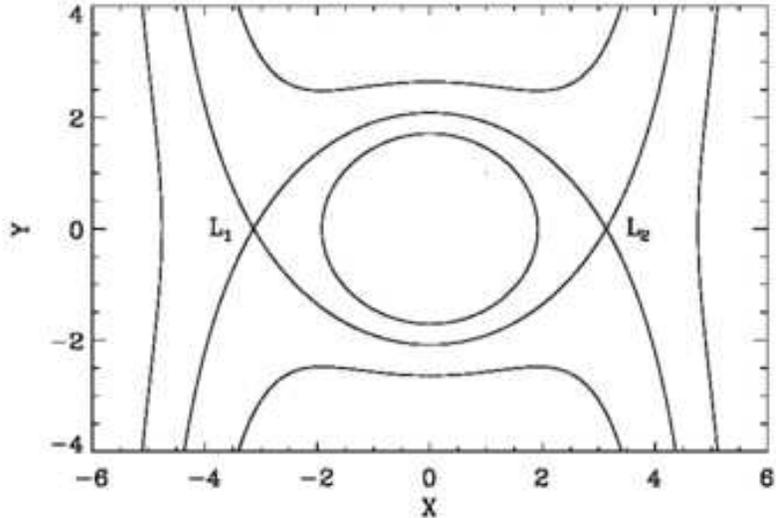,width=0.7\columnwidth}
\end{tabular}
\caption[]{Equipotential surfaces for a $W_0=3$ star cluster in an
  external point-mass tidal field.  The galactic center is to the
  left.  The Jacobi surface is delineated by the two crossing
  equipotentials, passing through the $L_1$ Lagrangian point to the
  left and the $L_2$ point to the right. \citep[Data from Fig.\,2 of
  ][]{2000MNRAS.318..753F}.  }
\label{Fig:Equipotentialsurface}
\end{figure}

The {\em tidal radius}, \rtide, is the distance from the center of a
star cluster where the gravitational acceleration due to the cluster
equals the tidal aceleration of the parent Galaxy
\citep{1957ApJ...125..451V}.  In Fig.\,\ref{Fig:Equipotentialsurface}
we show the equipotential curves of a cluster in the tidal potential
of its parent galaxy (the galactic center is to the left in the
figure).  The equipotential surface through the two Lagrangian points
($L_1$ and $L_2$ at $x\approx\pm3.1$ in the figure) is the {\em Jacobi
  surface}, defining the effective extent of the cluster in the
external field.  Stars within this surface may be regarded as cluster
members \citep{2006MNRAS.366..429R}.  Stars do not escape from the
cluster in random directions, but instead do so through the $L_1$ and
$L_2$ points \citep{2001ruag.conf..109H}.  As a practical matter, the
{\em Jacobi radius} $\rJ$, the distance from the center to the $L_1$
point, is a commonly used measure of the cluster's ``size.'' For
clusters on circular orbits $\rj$ is defined by the cluster mass, $M$,
the orbital angular frequency in the galaxy, $\omega$, and the
galactic potential, $\phi$, by \citep{1962AJ.....67..471K}
\begin{equation}
    \rj = \left(\frac{GM}{2\omega^2}\right)^{1/3}.
\end{equation}
Here, $\omega\equiv V_G/R_G$, where $R_G$ is galactocentric distance
and $V_G$ is the circular orbital speed around the galaxy center, and
we have assumed a flat rotation curve: $\phi(R_G)=V_G^2\ln(R_G)$.
Note that a factor $2/3$ is sometimes included to correct for the
elongation in the direction along the line connecting $L_1$ and $L_2$
\citep{1983AJ.....88..338I}.

For ``Roche lobe filling'' clusters that exactly fill their Jacobi
surfaces, the Jacobi radius is often identified with the truncation
radius of a \cite{1966AJ.....71...64K} model.  However, we emphasize
that there is no compelling reason why a star cluster should exactly
fill its Jacobi surface, nor is there necessarily any connection in
general between the King truncation radius and the properties (or even
the existence) of an external tidal field.  Observationally, the
truncation radius is not measured directly, but instead is inferred
from King model fits.

\subsubsection{Time scales}\label{Sect:TimeScales}

The two fundamental time scales of a self-gravitating system are the
dynamical time scale \tdyn, and the relaxation time scale \trl.

The dynamical time scale is the time required for a typical star to
cross the system; it is also the time scale on which the system
(re)establishes dynamical equilibrium.  A convenient formal definition
in terms of conserved quantities is
\begin{equation}
	\tdyn = \frac{G \mtot^{5/2}}{(-4E)^{3/2}},
\label{Eq:tdyn}
\end{equation}
where $E \equiv T+U$ is the total energy of the cluster.  In virial
equilibrium, $2T+U=0$ and this expression assumes the more familiar
form \citep{1987degc.book.....S}
\begin{eqnarray}
	\tdyn &=   & \left(\frac{G\mtot}{r^3_{\rm vir}}\right)^{-1/2}\\
	      &\sim& 2\times10^4\,{\rm yr}		
	              \left(\frac{\mtot}{10^6\,\msun}\right)^{-1/2}
	              \left(\frac{\rvir}{1\,{\rm pc}}\right)^{3/2},
\label{eq:dynamical}
\end{eqnarray}

The relaxation time, $\trl$, is typically much longer than
\tdyn. It the time scale on which two-body encounters transfer energy
between individual stars and cause the system to establish thermal
equilibrium.  The local relaxation time is
\citep{1987degc.book.....S}
\begin{equation}
  \trl = \frac{\langle v^2\rangle^{3/2}}{15.4 G^2m\rho\ln\Lambda},
\label{Eq:LocalRelaxationTime}
\end{equation}
where $m$ is the local mean stellar mass and $\rho$\, is the local
density.  The value of the parameter $\Lambda$ is $0.4N$ for the
theoretical case where all stars have the same mass and are
distributed homogeneously with an isotropic velocity distribution
\citep{1987degc.book.....S}.  \cite{1994MNRAS.268..257G} find
empirically $\Lambda\sim0.11N$ for systems where all stars have the
same mass.  For systems with a significant range of stellar masses,
the effective value of $\Lambda$ may be considerably smaller than this
value.

For a cluster in virial equilibrium we can replace all quantities by
their cluster-wide averages, writing $\langle v^2\rangle =
G\mtot/2\rvir$ and $\bar{\rho} \approx 3\mtot/8\pi r^3_{\rm vir}$,
where we ignore the distinction between virial and half-mass
quantities, so $\rhm\approx\rvir$.  We thus obtain the ``half-mass''
two-body relaxation time \citep{1987degc.book.....S}
\begin{eqnarray}
	\trlx &\simeq& \frac{0.065 \langle v^2\rangle^{3/2}}
                       {G^2 \mmean \bar{\rho}\ln\Lambda}
\label{Eq:Trlx}\\
              &=& 0.14 \frac{N^{1/2}\rvir^{3/2}}
                             {G^{1/2} \mmean^{1/2}\ln\Lambda}\\
              &\approx& \frac{N}{7\ln\Lambda}\,\tdyn\,,	
\label{Eq:trtd}
\end{eqnarray}
where $\mmean \equiv N/\mtot$ is the global mean stellar mass and $N$
is the total number of stars in the cluster.  If, for simplicity, we
adopt $\ln\Lambda=10$ as appropriate for the range in cluster masses
of interest in this review, Eq.~\ref{Eq:Trlx} becomes
\begin{equation}
	\trlx \sim 2\times10^8\,{\rm yr}~	
		\left(\frac{\mtot}{10^6\msun}\right)^{1/2}
		\left(\frac{\rvir}{1 {\rm pc}}\right)^{3/2}
		\left(\frac{\mmean}{\msun}\right)^{-1}\,.
\label{Eq:trlx_YMC}
\end{equation}

Finally, we note that in real stellar systems the one-parameter
simplicity of Eq.\,\ref{Eq:trtd} is broken by the introduction of a
third time scale independent of the dynamical properties of the
cluster---the stellar evolution time scale $t_S \sim 10$ Myr for YMCs,
but a complication here is that $t_S$ depends on time.  This simple
fact underlies almost all of the material presented in this review.


\section{Properties of young massive star clusters}
\label{sec:observations}

\subsection{General characteristics}
\label{Subsec:Populations}

Traditionally, astronomers have drawn a clear distinction between the
relatively young and low mass Milky Way open clusters associated with
the Galactic disk and the old and massive globular clusters that
reside mostly in the bulge and halo (see
Fig.~\ref{Fig:distribution_open_globular_clusters} and
Tab.\,\ref{Tab:Comparison}).  According to our definition in
\S\ref{Sect:Introduction}, the Milky Way hosts several clusters that
fill the gap between these populations, in terms of both mass and
density (Fig.~\ref{Fig:radius_mass_all_clusters}), indicating that the
formation of clusters with masses comparable to old GCs is not
restricted to the early universe. This becomes even more evident when
we look at the Magellanic clouds, which host massive clusters spanning
a broad range of ages \citep{2003AJ....126.1836H,2006MNRAS.366..295D},
as well as many YMCs that have received considerable attention since
the 1960s \citep{1961ApJ...133..413H}.

\begin{figure}
\begin{center}
\begin{tabular}{cc}
  \psfig{figure=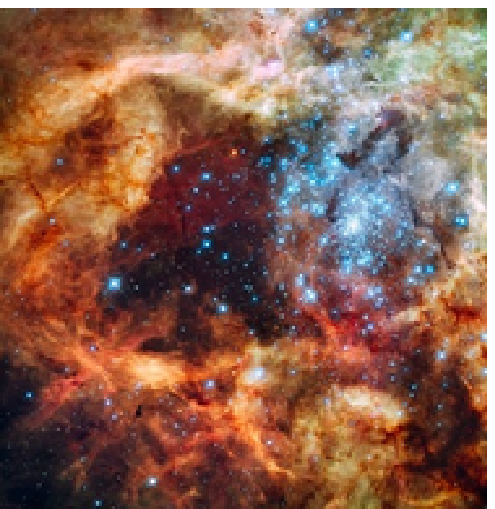,height=8cm}
  \psfig{figure=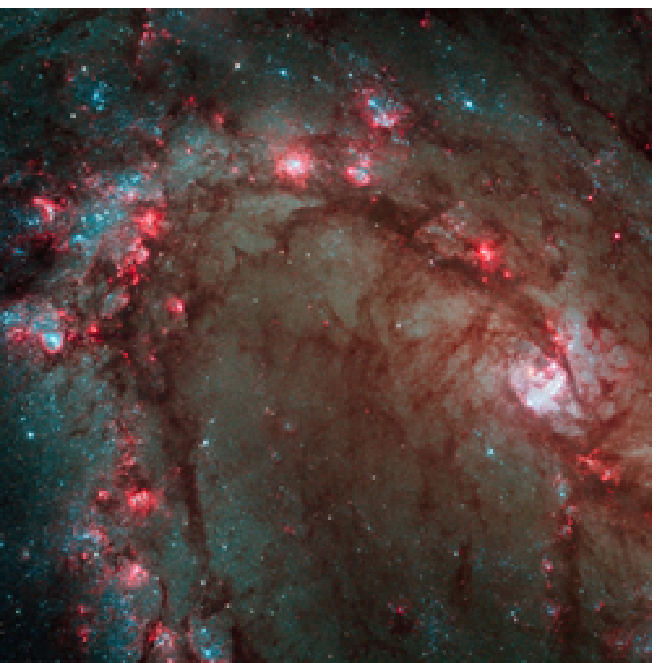,height=8cm}
\end{tabular}
\end{center}
\caption[]{Left: A region of $50\times50\,$pc$^2$ around the
  $\sim10^5\,\msun$ cluster R\,136 in the 30 Doradus region of the
  LMC, at a distance of $\sim50\,$kpc. Right: Many young star clusters
  forming in M83 at a distance of 3.6\,Mpc. Credit: NASA, ESA, the
  Wide Field Camera 3 Science Oversight Committee, and the Hubble
  Heritage Team (STScI/AURA) and F. Paresce
R. O'Connell 
  (R136) and R. O'Connell 
(M83).
}\label{Fig:Observations_R136_M83}
\end{figure}

The ages of the YMCs in the Magellanic Clouds are comparable to those
of many Milky Way open clusters (up to a few hundred megayears), but
the masses and core densities of these clusters exceed those of open
clusters in the Milky Way, in some cases by several orders of
magnitude \citep[e.g.][see
Table\,\ref{Tab:Comparison}]{1985ApJ...299..211E}.  A prominent
example is R\,136, whose core was once believed to be a single stellar
object at least 2000 times more massive and about $10^8$ times
brighter than the Sun. \citet{1985A&A...150L..18W} unambiguously
resolved it into a group of stars, and now we know it is a cluster of
$\sim 10^5$ young stars, rather than a single extraordinary object
\citep[][and left panel of
Fig.~\ref{Fig:Observations_R136_M83}]{1998ApJ...493..180M,2009ApJ...707.1347A,2010arXiv1002.0288C}. YMCs
such as R\,136 are responsible for the giant HII regions found in
other galaxies \citep[][and right panel in
Fig.~\ref{Fig:Observations_R136_M83}]{1988AJ.....95..720K} and appear
to be much more common phenomena than was previously thought.

YMCs more massive than R\,136 were already known several decades ago.
Ground-based observations revealed numerous ``bluish knots'' and
``super-star clusters'' in the starburst galaxies M82 and NGC~1569
\citep{1971A&A....12..474V, 1985AJ.....90.1163A}, and in the
relatively local ongoing galaxy mergers NGC~7252 and NGC~3597
\citep{1982ApJ...252..455S,1991A&A...245...31L}. HST has since
confirmed many extragalactic YMCs, starting with those in the
interacting galaxy NGC~1275 \citep{1992AJ....103..691H}. These and
later well studied examples have been cited as promising candidates
for the latest generation of ``young globular clusters''
\citep{2003dhst.symp..153W,2006pces.conf...35L}.

We review here the basic characteristics of known YMCs and attempt to
assess the similarities and differences between the young and old
populations (see also \S\ref{Sect:OtherStellarExotica} for a number of
well studied cases).  In the limited space available in this section
we highlight only the most relevant observed properties.  For more
in-depth reviews of YMC populations we refer to
\citet{2003dhst.symp..153W} and
\citet{2006pces.conf...35L,2009arXiv0911.0796L}. Some interesting
individual cases are discussed in depth in
\citet{2007ChJAA...7..155D}.

We have collected data on clusters for which estimates of the age,
(photometric) mass and half-light (or effective) radius exist in the
literature.  If more information on structural parameters is
available, we determine $\rvir$ using the relation presented in
Fig.~\ref{Fig:Observations_eta}. If only an estimate of $\reff$ is
available, we determine {\rvir} using the Plummer value of
$\rvir/\reff=16/(3\pi)\approx1.6$. We then calculate $\tdyn$ from
$\rvir$ and the mass using Eq.~\ref{eq:dynamical}. We assume here that
the observed clusters are in virial equilibrium, which results in an
overestimate for $\tdyn$ for unbound (expanding) clusters. The data
are presented in Tables~\ref{Tab:Galactic_Clusters},
\ref{Tab:Local_Group_Clusters} and \ref{Tab:Extra_Galactic_Clusters}.

As illustrated in Figure~2 of \citet{2009A&A...498L..37P}, young star
clusters appear to show two evolutionary sequences: (1) a compact
configuration starting at a radius of $\sim0.5\,$pc, which we refer to
as (dense) {\em clusters}, and (2) a sequence with radii $\sim5\,$pc,
which we call {\em associations}. (These two groups are referred to as
``starburst'' and ``leaky'' clusters by
\citealt{2009A&A...498L..37P}.) A more quantitative distinction may be
found in the ratio ${\rm age}/\tdyn$, since for bound objects this
ratio should be large and for unbound objects it is expected to remain
small. We note that the two groups identified by
\citet{2009A&A...498L..37P} can be divided in this way, with a
boundary at ${\rm age}/\tdyn\approx3$. This division is indicated in
the tables. More physical interpretations of these sequences are
presented in \S\ref{Sect:SCSurvival}.

\begin{table}[htbp!]
\resizebox{\textwidth}{!} {
\begin{tabular}{lrrrrrrrrrrrrrrr}
\hline
                (1)&                (2)&                (3)&                (4)&                (5)&                (6)&                (7)&                (8)&                (9)&               (10)&               (11)&              (12)\\
               Name&                Ref&                Age&       $\log\mphot$&        $\log\mdyn$&           $\rcore$&            $\reff$&           $\gamma$&       $\sigmaoned$&            $\rvir$&            $\tdyn$&       Age/$\tdyn$\\
                   &                   &              [Myr]&                   &                   &               [pc]&               [pc]&                   &           [$\kms$]&               [pc]&              [Myr]&                  \\
\hline
             Arches&1                  &               2.00&               4.30&                $-$&               0.20&               0.40&                $-$&                $-$&               0.68&               0.06&              33.86\\
            DSB2003&4                  &               3.50&               3.80&                $-$&                $-$&               1.20&                $-$&                $-$&               2.04&               0.55&               6.41\\
           NGC~3603&5                  &               2.00&               4.10&                $-$&               0.15&               0.70&               2.00&                $-$&               1.19&               0.17&              11.62\\
         Quintuplet&6                  &               4.00&               4.00&                $-$&               1.00&               2.00&                $-$&                $-$&               3.40&               0.93&               4.29\\
            RSGC~01&6                  &              12.00&               4.50&               4.70&                $-$&               1.50&                $-$&               3.70&               2.55&               0.34&              35.22\\
            RSGC~02&8                  &              17.00&               4.60&               4.80&                $-$&               2.70&                $-$&               3.40&               4.58&               0.73&              23.18\\
            RSGC~03&9                  &              18.00&               4.50&                $-$&                $-$&               5.00&                $-$&                $-$&               8.49&               2.07&               8.68\\
        Trumpler~14&10                 &               2.00&               4.00&                $-$&               0.14&               0.50&               2.00&                $-$&               0.85&               0.12&              17.15\\
               Wd~1&11                 &               3.50&               4.50&               4.80&               0.40&               1.00&               4.00&               5.80&               1.74&               0.19&              18.27\\
               Wd~2&4                  &               2.00&               4.00&                $-$&                $-$&               0.80&                $-$&                $-$&               1.36&               0.24&               8.48\\
               hPer&4                  &              12.80&               4.20&                $-$&                $-$&               2.10&                $-$&                $-$&               3.57&               0.80&              16.06\\
          $\chi$Per&4                  &              12.80&               4.10&                $-$&                $-$&               2.50&                $-$&                $-$&               4.24&               1.16&              11.02\\\hline
              CYgOB&4                  &               2.50&               4.40&                $-$&                $-$&               5.20&                $-$&                $-$&               8.83&               2.47&               1.01\\
            IC~1805&4                  &               2.00&               4.20&                $-$&                $-$&              12.50&                $-$&                $-$&              21.22&              11.58&               0.17\\
            I~Lac~1&4                  &              14.00&               3.40&                $-$&                $-$&              20.70&                $-$&                $-$&              35.14&              61.97&               0.23\\
     Lower~Cen-Crux&4                  &              11.50&               3.30&                $-$&                $-$&              15.00&                $-$&                $-$&              25.46&              42.89&               0.27\\
           NGC~2244&4                  &               2.00&               3.90&                $-$&                $-$&               5.60&                $-$&                $-$&               9.51&               4.90&               0.41\\
           NGC~6611&4                  &               3.00&               4.40&                $-$&                $-$&               5.90&                $-$&                $-$&              10.02&               2.98&               1.01\\
           NGC~7380&4                  &               2.00&               3.80&                $-$&                $-$&               6.50&                $-$&                $-$&              11.03&               6.88&               0.29\\
                ONC&13                 &               1.00&               3.65&                $-$&               0.20&               2.00&               2.00&                $-$&               3.40&               1.40&               0.72\\
             Ori~Ia&4                  &              11.40&               3.70&                $-$&                $-$&              16.60&                $-$&                $-$&              28.18&              31.50&               0.36\\
             Ori~Ib&4                  &               1.70&               3.60&                $-$&                $-$&               6.30&                $-$&                $-$&              10.70&               8.26&               0.21\\
             Ori~Ic&4                  &               4.60&               3.80&                $-$&                $-$&              12.50&                $-$&                $-$&              21.22&              18.35&               0.25\\
     Upper~Cen-Crux&4                  &              14.50&               3.60&                $-$&                $-$&              22.10&                $-$&                $-$&              37.52&              54.30&               0.27\\
              U~Sco&4                  &               5.50&               3.50&                $-$&                $-$&              14.20&                $-$&                $-$&              24.11&              31.38&               0.18\\
\hline
\end{tabular}}
\caption{
Properties of YMCs (top) and associations (bottom) in the Milky Way, with the distinction based on age/$\tdyn$. 
       1: \citet{1999ApJ...514..202F}; 
       2: \citet{2002ApJ...581..258F}; 
       3: \citet{2002A&A...394..459S}; 
       4: \citet{2009A&A...498L..37P}; 
       5: \citet{2008ApJ...675.1319H}; 
       6: \citet{2006ApJ...643.1166F}; 
       7: \citet{2008ApJ...676.1016D}; 
       8: \citet{2007ApJ...671..781D}; 
       9: \citet{2009A&A...498..109C}; 
      10: \citet{2007A&A...476..199A}; 
      11: \citet{2007A&A...466..151M}; 
      12: \citet{2008A&A...478..137B}; 
      13: \citet{1998ApJ...492..540H}. 
}
 \label{Tab:Galactic_Clusters}
\end{table}

\begin{table}[htbp!]
\resizebox{\textwidth}{!} {
\begin{tabular}{lrrrrrrrrrrrrr}
\hline
            (1)&            (2)&            (3)&            (4)&            (5)&            (6)&            (7)&            (8)&            (9)&           (10)&           (11)&           (12)&           (13)&          (14)\\
            Gal&           Name&            Ref&            Age&          $M_V$&   $\log\mphot$&    $\log\mdyn$&       $\rcore$&        $\reff$&       $\gamma$&   $\sigmaoned$&        $\rvir$&        $\tdyn$&   Age/$\tdyn$\\
               &               &               &          [Myr]&         [mag] &               &               &           [pc]&           [pc]&               &       [$\kms$]&           [pc]&          [Myr]&              \\
\hline
            LMC&           R136&1,2,3,4        &            3.0       &$-11.74$&           4.78&            $-$&           0.10&           1.70&           1.50&            $-$&           2.89&           0.30&          10.14\\
            LMC&       NGC~1818&2,3            &           25.1       &$ -9.62$&           4.42&            $-$&           2.07&           5.39&           3.30&            $-$&           9.76&           2.80&           8.96\\
            LMC&       NGC~1847&2,3            &           26.3       &$ -9.67$&           4.44&            $-$&           1.73&          32.58&           2.05&            $-$&          10.33&           2.98&           8.82\\
            LMC&       NGC~1850&2,3            &           31.6       &$-10.52$&           4.86&           5.22&           2.69&          11.25&           2.05&           3.00&           3.56&           0.37&          84.83\\
            LMC&       NGC~2004&2,3            &           20.0       &$ -9.60$&           4.36&            $-$&           1.41&           5.27&           2.90&            $-$&           9.81&           3.03&           6.59\\
            LMC&       NGC~2100&2,3            &           15.8       &$ -9.77$&           4.36&            $-$&           0.99&           4.41&           2.30&            $-$&           6.32&           1.57&          10.12\\
            LMC&       NGC~2136&2,3            &          100.0       &$ -8.60$&           4.30&            $-$&           1.59&           3.42&           3.50&            $-$&           6.10&           1.59&          62.85\\
            LMC&       NGC~2157&2,3            &           39.8       &$ -9.10$&           4.31&           4.90&           1.99&           5.39&           3.05&           2.80&           9.95&           3.27&          12.16\\
            LMC&       NGC~2164&2,3            &           50.1       &$ -8.65$&           4.18&           5.15&           1.48&           4.76&           2.95&           4.30&           8.84&           3.19&          15.72\\
            LMC&       NGC~2214&2,3            &           39.8       &$ -8.40$&           4.03&           5.28&           1.83&           8.13&           2.45&           3.90&          14.24&           7.74&           5.14\\
            LMC&       NGC~1711&2,3            &           50.1       &$ -8.82$&           4.24&            $-$&           1.91&           5.19&           2.70&            $-$&           9.70&           3.42&          14.66\\
            M31&          KW246&5              &           75.9       &$ -7.80$&           4.19&            $-$&            $-$&           3.20&            $-$&            $-$&           5.43&           1.52&          50.01\\
            M31&          B257D&6              &           79.4       &$ -8.84$&           4.45&            $-$&           3.16&          15.14&            $-$&            $-$&          25.70&          11.57&           6.86\\
            M31&           B318&6              &           70.8       &$ -8.76$&           4.38&            $-$&           0.19&           6.61&            $-$&            $-$&          11.22&           3.62&          19.57\\
            M31&           B327&6              &           50.1       &$ -8.95$&           4.38&            $-$&           0.20&           4.47&            $-$&            $-$&           7.59&           2.01&          24.91\\
            M31&           B448&6              &           79.4       &$ -9.20$&           4.58&            $-$&           0.20&          16.22&            $-$&            $-$&          27.54&          11.05&           7.19\\
            M31&           Vdb0&7              &           25.1       &$-10.00$&           4.85&            $-$&           1.40&           7.40&            $-$&            $-$&          12.56&           2.49&          10.07\\
            M31&     KW044/B325&5              &           58.9       &$ -9.20$&           4.59&            $-$&            $-$&          10.00&            $-$&            $-$&          16.98&           5.29&          11.14\\
            M31&          KW120&5              &           87.1       &$ -8.80$&           4.57&            $-$&            $-$&           2.60&            $-$&            $-$&           4.41&           0.72&         121.43\\
            M31&          KW208&5              &           56.2       &$ -7.70$&           4.01&            $-$&            $-$&           2.90&            $-$&            $-$&           4.92&           1.61&          34.93\\
            M31&          KW272&5              &           53.7       &$ -9.00$&           4.50&            $-$&            $-$&           9.00&            $-$&            $-$&          15.28&           5.01&          10.73\\
            M31&          B015D&6              &           70.8       &$ -9.71$&           4.76&            $-$&           0.24&          16.60&            $-$&            $-$&          28.18&           9.30&           7.61\\
            M31&           B040&6              &           79.4       &$ -9.00$&           4.50&            $-$&           0.55&          12.88&            $-$&            $-$&          21.87&           8.57&           9.27\\
            M31&           B043&6              &           79.4       &$ -8.81$&           4.43&            $-$&           0.72&           3.98&            $-$&            $-$&           6.76&           1.60&          49.77\\
            M31&           B066&6              &           70.8       &$ -8.43$&           4.25&            $-$&           0.38&           6.76&            $-$&            $-$&          11.48&           4.35&          16.29\\
       NGC~6822&      Hubble~IV&8,9            &           25.1       &$ -8.00$&           4.00&            $-$&           0.40&           2.00&            $-$&            $-$&           3.40&           0.93&          26.93\\
            SMC&        NGC~330&2,3            &           25.1       &$ -9.94$&           4.56&           5.64&           2.34&           6.11&           2.55&           6.00&          11.17&           2.92&           8.60\\\hline
            M31&          KW249&5              &            5.0       &$-10.50$&           4.30&            $-$&            $-$&          13.50&            $-$&            $-$&          22.92&          11.58&           0.43\\
            M31&          KW258&5              &            5.0       &$ -9.90$&           4.05&            $-$&            $-$&           3.40&            $-$&            $-$&           5.77&           1.95&           2.57\\
            M33&        NGC~595&10             &            4.0       &$-11.40$&           4.50&            $-$&            $-$&          26.90&            $-$&            $-$&          45.67&          25.87&           0.15\\
            M33&        NGC~604&10             &            3.5       &$-12.60$&           5.00&            $-$&            $-$&          28.40&            $-$&            $-$&          48.21&          15.78&           0.22\\
            SMC&        NGC~346&11             &            3.0       &$   $-$$&           5.60&            $-$&            $-$&           9.00&            $-$&            $-$&          15.28&           1.41&           2.13\\
\hline
\end{tabular}}
\caption{
Same as Table~\ref{Tab:Galactic_Clusters}, but now for the Local Group.
       1: \citet{1995ApJ...448..179H}; 
       2: \citet{2003MNRAS.338...85M}; 
       3: \citet{2005ApJS..161..304M}; 
       4: \citet{2009ApJ...707.1347A}; 
       5: \citet{2009ApJ...703.1872V}; 
       6: \citet{2009AJ....138.1667B}; 
       7: \citet{2009A&A...494..933P}; 
       8: \citet{2000PASP..112.1162W}; 
       9: \citet{2000AJ....120.3088C}; 
      10: \citet{2001ApJ...563..151M}; 
      11: \citet{2008AJ....135..173S}. 
}
  \label{Tab:Local_Group_Clusters}
\end{table}

\begin{table}[htbp!]
\resizebox{\textwidth}{!} {
\begin{tabular}{lrrrrrrrrrrrrr}
\hline
            (1)&            (2)&            (3)&            (4)&            (5)&            (6)&            (7)&            (8)&            (9)&           (10)&           (11)&           (12)&           (13)&          (14)\\
            Gal&           Name&            Ref&            Age&          $M_V$&   $\log\mphot$&    $\log\mdyn$&       $\rcore$&        $\reff$&       $\gamma$&   $\sigmaoned$&        $\rvir$&        $\tdyn$&   Age/$\tdyn$\\
               &               &               &          [Myr]&         [mag] &               &               &           [pc]&           [pc]&               &      [$\kms$] &           [pc]&          [Myr]&              \\
\hline
      ESO338-IG&             23&1              &           7.08&       $-15.50$&           6.70&           7.10&            $-$&           5.20&            $-$&          32.50&           4.26&           0.15&          39.22\\
            M51&          3cl-a&2              &          15.85&       $-11.10$&           5.04&            $-$&           1.60&           5.20&           2.00&            $-$&           7.13&           0.52&           6.10\\
            M51&          3cl-b&2              &           5.01&       $-12.25$&           5.91&            $-$&           0.86&           2.30&           2.60&            $-$&           8.83&           1.18&          46.07\\
            M51&             a1&2              &           5.01&       $-12.15$&           5.47&            $-$&           0.65&           4.20&           1.90&            $-$&           5.55&           0.24&          12.82\\
            M82&          MGG~9&3,4,5          &           9.55&       $-13.42$&           5.92&           6.36&            $-$&           2.60&            $-$&          15.90&           4.41&           0.15&          46.04\\
            M82&             A1&6,5            &           6.31&       $-14.85$&           5.82&           5.93&           1.30&           3.00&           3.00&          13.40&           5.77&           0.31&          28.36\\
            M82&              F&7              &          60.26&       $-14.50$&           6.70&           6.08&            $-$&           2.80&           3.00&          13.40&           3.90&           0.09&          99.10\\
       NGC~1140&              1&8              &           5.01&       $-14.80$&           6.04&           7.00&            $-$&           8.00&            $-$&          24.00&           5.86&           0.32&           9.40\\
       NGC~1487&              2&9              &           8.51&            $-$&           5.20&           5.30&           0.71&           1.20&            $-$&          11.10&           5.19&           0.40&          63.64\\
       NGC~1487&              1&9              &           8.32&            $-$&           5.18&           6.08&           0.97&           2.30&            $-$&          13.70&           2.72&           0.07&          23.40\\
       NGC~1487&              3&9              &           8.51&            $-$&           4.88&           5.78&           0.71&           2.10&            $-$&          14.30&           5.18&           0.08&          19.02\\
       NGC~1569&              A&10,11,12       &          12.02&       $-14.10$&           6.20&           5.52&            $-$&           2.30&            $-$&          15.70&           3.78&           0.15&          77.49\\
       NGC~1569&              C&13             &           3.02&            $-$&           5.16&            $-$&            $-$&           2.90&            $-$&            $-$&           4.50&           0.24&          15.13\\
       NGC~1569&              B&14             &          19.95&       $-12.85$&           5.74&           5.64&           0.70&           2.10&           2.50&           9.60&           8.83&           0.17&          49.47\\
       NGC~1569&             30&11,12          &          91.20&       $-11.15$&           5.55&            $-$&           0.75&           2.50&           2.50&            $-$&          13.58&           0.71&          33.37\\
       NGC~1705&              1&7              &          15.85&       $-13.80$&           5.90&           5.68&            $-$&           1.60&            $-$&          11.40&           6.11&           0.26&          96.15\\
       NGC~4038&          S2\_1&9              &           8.91&            $-$&           5.47&           5.95&           0.60&           3.70&            $-$&          11.50&          10.19&           0.90&          16.09\\
       NGC~4038&          W99-1&15             &           8.13&       $-14.00$&           5.86&           5.81&            $-$&           3.60&            $-$&           9.10&          13.58&           0.46&          26.11\\
       NGC~4038&         W99-16&15             &          10.00&       $-12.70$&           5.46&           6.51&            $-$&           6.00&            $-$&          15.80&           2.38&           0.08&           7.76\\
       NGC~4038&          W99-2&9              &           6.61&            $-$&           6.42&           6.48&           1.30&           8.00&            $-$&          14.10&           6.11&           0.27&          14.82\\
       NGC~4038&         W99-15&9              &           8.71&            $-$&           5.70&           6.00&            $-$&           1.40&            $-$&          20.20&           6.11&           0.32&          89.94\\
       NGC~4038&          S1\_1&9              &           7.94&            $-$&           5.85&           6.00&           0.58&           3.60&            $-$&          12.50&           1.53&           0.05&          25.77\\
       NGC~4038&          S1\_2&9              &           8.32&            $-$&           5.70&           5.90&           0.72&           3.60&            $-$&          11.50&           6.11&           0.17&          21.75\\
       NGC~4038&          S1\_5&9              &           8.51&            $-$&           5.48&           5.60&           1.35&           0.90&            $-$&          12.00&           6.28&           0.43&         135.26\\
       NGC~4038&        2000\_1&9              &           8.51&            $-$&           6.23&           6.38&           0.42&           3.60&            $-$&          20.00&           4.24&           0.21&          40.09\\
       NGC~4038&          S2\_2&9              &           8.91&            $-$&           5.60&           5.60&           0.40&           2.50&            $-$&           9.50&           5.09&           0.35&          33.64\\
       NGC~4038&          S2\_3&9              &           8.91&            $-$&           5.38&           5.40&           0.60&           3.00&            $-$&           7.00&           3.90&           0.30&          19.86\\
       NGC~4449&            N-1&13             &          10.96&            $-$&           6.57&            $-$&            $-$&          16.90&            $-$&            $-$&           2.04&           0.11&           5.92\\
       NGC~4449&            N-2&13             &           3.02&            $-$&           5.00&            $-$&            $-$&           5.80&            $-$&            $-$&           3.57&           0.36&           4.45\\
       NGC~5236&            805&16             &          12.59&       $-12.17$&           5.29&           5.62&           0.70&           2.80&           2.60&           8.10&           4.92&           0.43&          17.80\\
       NGC~5236&            502&16             &         100.00&       $-11.57$&           5.65&           5.71&           0.70&           7.60&           2.14&            $-$&          34.97&           9.10&          25.27\\
       NGC~5253&              I&13             &          11.48&            $-$&           5.38&            $-$&            $-$&           4.00&            $-$&            $-$&          44.65&          17.30&          13.11\\
       NGC~5253&             VI&13             &          10.96&            $-$&           4.93&            $-$&            $-$&           3.10&            $-$&            $-$&          33.27&          17.64&          11.41\\
       NGC~6946&           1447&16             &          11.22&       $-13.19$&           5.64&           6.25&           1.15&          10.00&           2.10&           8.80&          20.03&           2.83&          22.57\\\hline
       NGC~2403&            I-B&13             &           6.03&            $-$&           4.82&            $-$&            $-$&          26.30&            $-$&            $-$&          50.93&          15.81&           0.39\\
       NGC~2403&            I-C&13             &           6.03&            $-$&           4.42&            $-$&            $-$&          19.60&            $-$&            $-$&          28.01&           4.21&           0.38\\
       NGC~2403&            I-A&13             &           6.03&            $-$&           5.06&            $-$&            $-$&          20.60&            $-$&            $-$&          56.02&          12.47&           0.75\\
       NGC~2403&             II&13             &           4.47&            $-$&           5.35&            $-$&            $-$&          11.80&            $-$&            $-$&          25.97&           4.42&           2.35\\
       NGC~2403&             IV&13             &           4.47&            $-$&           5.07&            $-$&            $-$&          30.00&            $-$&            $-$&          36.84&          12.39&           0.42\\
       NGC~4214&             VI&13             &          10.96&            $-$&           4.93&            $-$&            $-$&          35.90&            $-$&            $-$&         142.43&          34.58&           0.29\\
       NGC~4214&              V&13             &          10.96&            $-$&           5.73&            $-$&            $-$&          83.90&            $-$&            $-$&          60.95&          24.31&           0.20\\
       NGC~4214&            VII&13             &          10.96&            $-$&           5.33&            $-$&            $-$&          40.40&            $-$&            $-$&          68.59&          18.31&           0.38\\
       NGC~4214&            I-A&13             &           3.47&            $-$&           5.44&            $-$&            $-$&          16.50&            $-$&            $-$&          28.69&           1.19&           1.55\\
       NGC~4214&            I-B&13             &           3.47&            $-$&           5.40&            $-$&            $-$&          33.00&            $-$&            $-$&           9.85&           1.46&           0.52\\
       NGC~4214&            I-D&13             &           8.91&            $-$&           5.30&            $-$&            $-$&          15.30&            $-$&            $-$&           6.79&           0.54&           1.57\\
       NGC~4214&           II-C&13             &           2.00&            $-$&           4.86&            $-$&            $-$&          21.70&            $-$&            $-$&          23.43&           7.38&           0.51\\
       NGC~5253&             IV&13             &           3.47&            $-$&           4.72&            $-$&            $-$&          13.80&            $-$&            $-$&           5.26&           0.62&           0.89\\
\hline
\end{tabular}}
\caption{
Same as Table~\ref{Tab:Galactic_Clusters}, but now for objects outside the Local Group.
       1: \citet{2007A&A...461..471O}; 
       2: \citet{2008MNRAS.389..223B}; 
       3: \citet{2003ApJ...596..240M}; 
       4: \citet{2006A&A...448..881B}; 
       5: \citet{2007ApJ...663..844M}; 
       6: \citet{2006MNRAS.370..513S}; 
       7: \citet{2001MNRAS.326.1027S}; 
       8: \citet{2007MNRAS.382.1877M}; 
       9: \citet{2008A&A...489.1091M}; 
      10: \citet{1996ApJ...466L..83H}; 
      11: \citet{2000AJ....120.2383H}; 
      12: \citet{2004MNRAS.347...17A}; 
      13: \citet{2001ApJ...563..151M}; 
      14: \citet{2008MNRAS.383..263L}; 
      15: \citet{2002A&A...383..137M}; 
      16: \citet{2004A&A...427..495L}. 
}
 \label{Tab:Extra_Galactic_Clusters}
\end{table}

\subsection{Mass segregation}
\label{Sect:Mass_Segregation}
Nearby clusters that can be resolved into individual stars are
excellent laboratories for studies of stellar populations, and their
derived properties can be used as input parameters to models of
cluster evolution.  Two well known examples of resolved YMCs are the
Arches and Quintuplet clusters near the Galactic center
\citep{2002ApJ...581..258F, 1999ApJ...514..202F}. The Arches has an
age of just 1--2 Myr and a central density of $\sim10^5\,\msun\,{\rm
  pc}^{-3}$; its stellar initial mass function (IMF) has been the
topic of much debate \citep{2002A&A...394..459S, 2005ApJ...628L.113S,
  2006ApJ...653L.113K}.  \citep[See also][who reconstruct the
cluster's initial conditions by iterated $N$-body
simulations.]{2009arXiv0911.3058H} We will refrain from commenting in
detail on this debate, and instead refer to the recent review by
\citet{2010arXiv1001.2965B} for an in-depth discussion of this
topic. The current consensus seems to be that ``evidence for IMF
variations is absent, although this is not evidence for their
absence.'' A \cite{1955ApJ...121..161S}\footnote{Here we use the term
  ``Salpeter'' as a possibly universal power-law mass function (with
  exponent -2.35) for stars more massive than $\sim1\,\msun$.  At
  present, given the range of detectable masses, it is not possible to
  distinguish more finely between the various mass functions commonly
  adopted by theorists.} distribution cannot be ruled out for the
stellar IMF of the Arches cluster \citep{2006ApJ...653L.113K}.

\citet{2002A&A...394..459S} find evidence for mass segregation within
the Arches cluster based on a steepening of the stellar mass function
(MF) with increasing distance from the cluster center.  However,
\cite{2009arXiv0903.2222E} suggest that the apparent mass segregation
could simply be an observational bias, and that a Salpeter MF in the
core is still consistent with their observations. The differential
extinction across the cluster and the total visual extinction of
$A_V\approx30$ magnitudes make this a challenging cluster to
study. See also \citet{2009A&A...495..147A} for a general discussion
on observational challenges in studies on mass segregation in the
dense central regions of clusters.

Currently the most massive young cluster known in our Galaxy is
Westerlund~1 \citep{2005A&A...434..949C}. Due to its relative
proximity ($\sim$4\,kpc), its lower extinction (although still a
considerable $A_V\approx10$ magnitudes), and slightly lower intrinsic
stellar density, it represents a somewhat easier target than the
Arches and Quintuplet for studies of its stellar content.  The MF and
structural parameters of Westerlund~1 and the somewhat less massive
NGC~3603 have been investigated by several space-based instruments and
from the ground using adaptive optics.  Evidence for an overabundance
of massive stars within the half-mass radius has been found in both
these clusters \citep{2008ApJ...675.1319H, 2008A&A...478..137B}.
Probably the best known example of a mass segregated cluster is the
(much closer) Orion Nebula Cluster (ONC) \citep{1998ApJ...492..540H,
  2006ApJ...644..355H}.  Radial variations of the stellar MF have also
been found for slightly older ($\sim10-25\,$Myr) star clusters in the
LMC \citep{2002MNRAS.331..245D, 2004A&A...416..137G}.

Clusters at much larger distances have also been used for studies of
mass segregation.  \citet{2005ApJ...621..278M} show that cluster F in
the starburst galaxy M82 appears to be smaller at red wavelengths than
at blue wavelengths, which they attribute to mass segregation. The
reasoning is that the most massive stars are red (super)giants and if
these are more centrally concentrated the cluster would appear smaller
in the red.  A note of caution is added by
\citet{2007MNRAS.379.1333B}, however, who point out that M82-F is a
complicated case because of differential extinction across the
cluster.  A negative correlation between radius and wavelength was
also found for NGC~1569-B \citep{2008MNRAS.383..263L}.
\citet{2008MNRAS.391..190G} modeled mass segregated clusters and
projected them in different filters, and showed that mass segregation
can only explain differences in measured $\rcore$ in different filters
at the 5\% level. This because the massive stars are much more
luminous than the average cluster member and dominate the light
profile at all wavelengths, so if a cluster is mass segregated it
should appear smaller at all wavelengths. It is thus not clear why
NGC~1569-B and M82-F should appear smaller at redder wavelengths.

If stars with masses greater than some mass $m$ are found to be more
centrally concentrated than the average stellar mass $\langle
m\rangle$, and if the cluster is much younger than the dynamical
friction time scale for stars of mass $m$ ($\sim\langle
m\rangle\trh/m$; see Eq.~\ref{eq:ts} in \S\ref{Sec:Dynamics}), then an
obvious conclusion is that the observed mass segregation is
primordial---that is, the cluster formed with stars more massive than
$M$ preferentially closer to the center. This is generally consistent
with simulations of cluster formation \citep{2001ApJ...556..837K,
  2006MNRAS.370..488B}.  However, the time scale argument only holds
for a spherically symmetric stellar system in virial equilibrium. If
clusters form through mergers of smaller sub-clumps, the degree of
mass segregation of the merger product is higher than would be
expected from these simple dynamical arguments
\citep{2007ApJ...655L..45M,
  2009ApJ...700L..99A,2009MNRAS.400..657M}. From studies of star
clusters that are still embedded in their natal molecular clouds, it
is evident that stars form in a clumpy hierarchical fashion
\citep{2003ARA&A..41...57L, 2005ApJ...632..397G}, and that some
merging has to occur during the early (embedded) evolution.  A YMC
with a clear lack of mass segregation would therefore be an very
interesting object to study.

The degree of mass segregation at a young age has important
consequences for the further evolution of a cluster, as we discuss in
more detail in \S\ref{Sec:Dynamics} and \S\ref{Sect:SCSurvival}.

\subsection{Structural Parameters}
\label{Sect:StructuralParameters}
Studies of the structural parameters of YMCs often involve the rich
cluster population of the LMC. Due to their vicinity, these clusters
can be spatially resolved and studied from the ground.  As mentioned
in \S\ref{Sect:TheRadius}, unlike the old GCs, whose surface
brightness profiles are best described by (tidally truncated) King
profiles, the surface brightness profiles of YMCs are best described
by power-law profiles with a core (Eq.~\ref{Eq:EFF87}; EFF87).  A
typical range for the power-law index is
$2.2\lesssim\gamma\lesssim3.2$ (EFF87;
\citealt{2003MNRAS.338...85M}). Similar slopes are found for
extragalactic clusters, with tentative evidence for an increase of
$\gamma$ with age \citep{2004A&A...416..537L}.

\begin{figure}
\begin{tabular}{cc}
  \psfig{figure=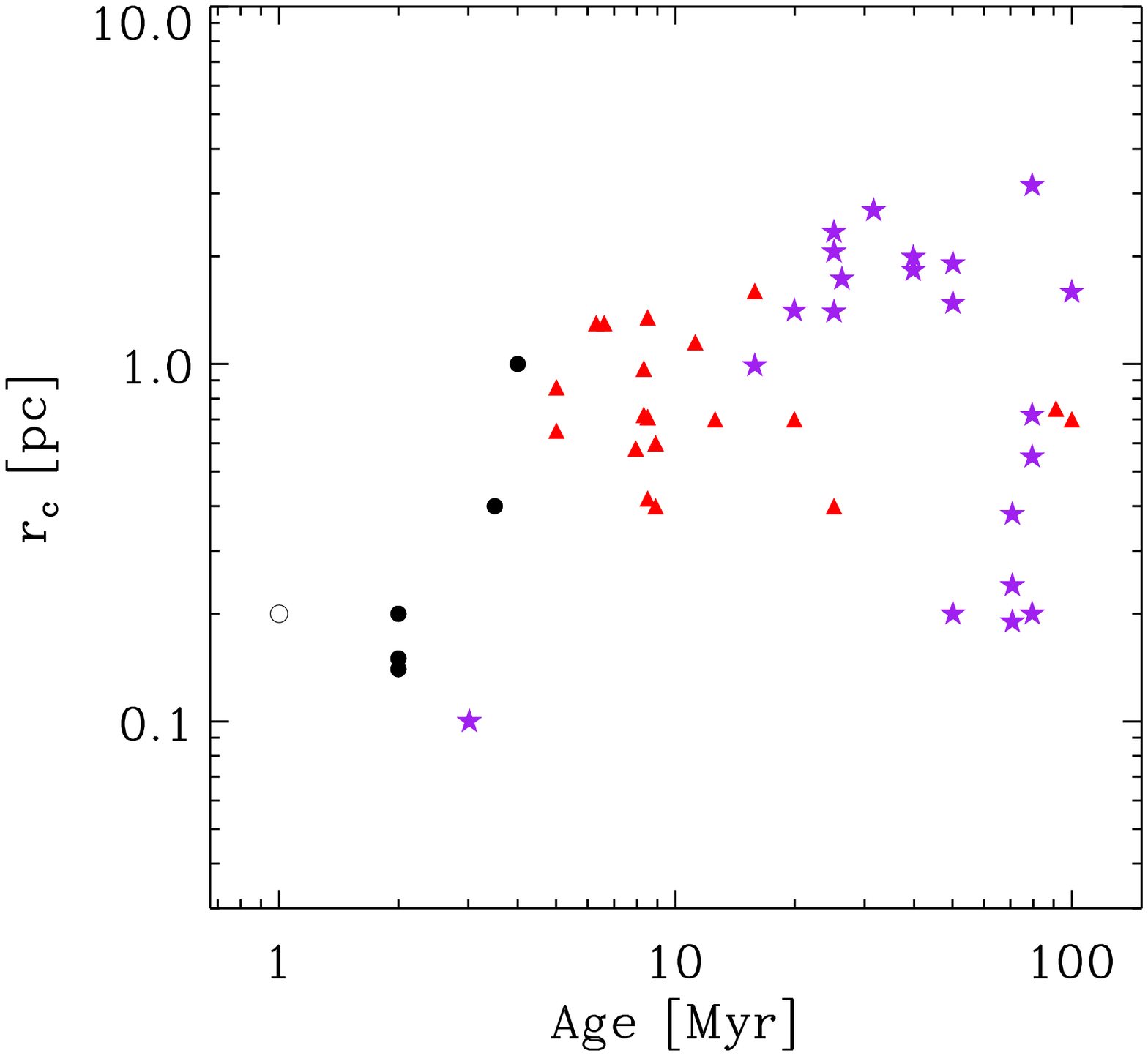,width=0.5\columnwidth,clip=} &
\hspace{-0.5cm}  \psfig{figure=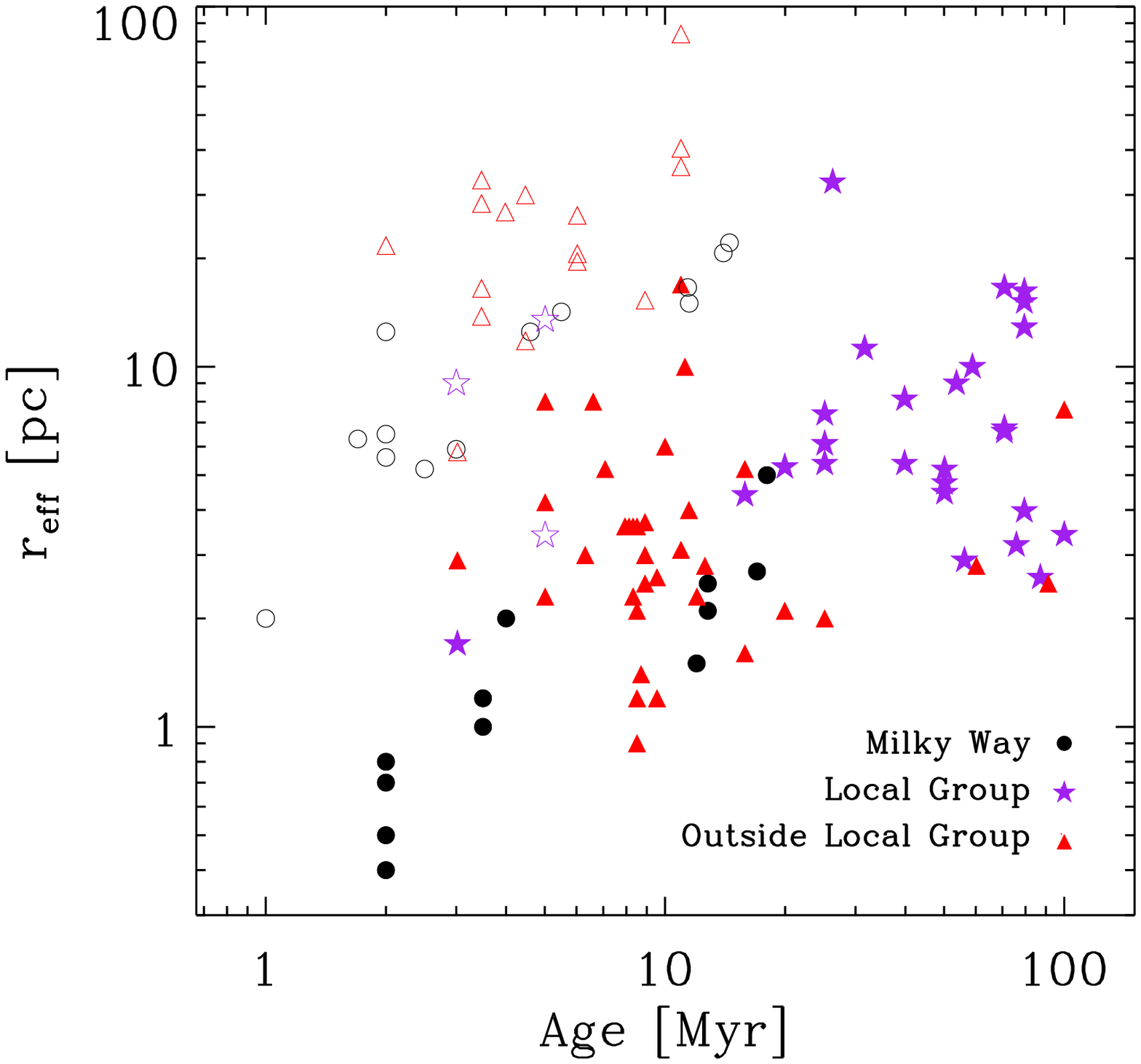,width=0.5\columnwidth}
\end{tabular}
\caption[] { Core radius ($\rcore$, left) and effective radius
  ($\reff$, right) as functions of age for the clusters in
  Tables~\ref{Tab:Galactic_Clusters}, \ref{Tab:Local_Group_Clusters}
  \& \ref{Tab:Extra_Galactic_Clusters}.  The filled symbols are
  clusters and the open symbols are associations.  }
\label{Fig:Radius_Evolution}
\end{figure}

Several studies have reported a striking increase of $\rcore$ with age
for LMC clusters \citep{1989ApJ...347L..69E, 2002MNRAS.337..597D,
  2003MNRAS.338...85M}. \citet{2008MNRAS.389..223B} finds a similar
increase for $\rcore$ of massive clusters in M51 and in a compilation
of literature data. \citet{2008A&A...478..137B} have shown that
$\reff$ of YMCs in the Milky Way increases with age.
Fig.~\ref{Fig:Radius_Evolution} shows the increase of $\rcore$ and
$\reff$ with age for all the clusters in
Tables~\ref{Tab:Galactic_Clusters}--\ref{Tab:Extra_Galactic_Clusters}.
There may be a rather strong selection bias toward dense objects in
the studies that consider $\rcore$, i.e., there may be young clusters
with large $\rcore$ which simply have not been classified as star
clusters or do not have a $\rcore$ measurement available.  This
becomes clear immediately when we look at $\reff$, where we have many
more values available for the associations (open symbols).  The values
of $\rcore$ are not determined for these objects, resulting in a
depletion of points at the top left of
Fig.~\ref{Fig:Radius_Evolution}.

At an age of 100\,Myr there is a large spread in $\rcore$.
\citet{1989ApJ...347L..69E} show that mass loss due to stellar
evolution from stellar populations with different IMF slopes gives
rise to different expansion rates. \citet{2008MNRAS.386...65M} invoke
different degrees of (primordial) mass segregation and retention
fractions of black holes to explain a range of growth rates for
$\rcore$ (see \S\ref{Sect:BHHeating} for further discussion). From the
lower envelope of points in the right panel of
Fig.~\ref{Fig:Radius_Evolution} it seems that $\reff$ rises
monotonically with age, with no obvious increase in the spread with
age.

An increase of a factor of 5--10 in radius has dramatic implications
for the evolution of these clusters, since it implies that in a very
short time the densities drop by two or three orders of magnitudes.
However, if very young clusters are mass segregated
(\S\ref{Sect:Mass_Segregation}), the observed $\rcore$ and $\reff$
(measured from the projected luminosity profile, see
\S\ref{Sect:TheRadius}) could be considerably smaller than the true
values \citep{2006MNRAS.369.1392F,2008MNRAS.391..190G}. We will return
to this topic, and the various physical mechanisms that drive cluster
expansion at these ages, in more detail in \S\ref{Sec:Dynamics} and
\S\ref{Sect:SCSurvival}.

Several studies have discussed the lack of any clear correlation
between the size of a cluster and its mass or luminosity
\citep{1999AJ....118..752Z, 2004A&A...416..537L,
  2007A&A...469..925S}. If this is how clusters form, it places
important constraints on models of cluster formation, since molecular
clouds and dense cores do follow a (virial) scaling relation between
mass and radius \citep{1981MNRAS.194..809L}. A variable star formation
efficiency that is higher for massive clumps could weaken or erase the
molecular cloud mass-radius relation \citep{2001AJ....122.1888A}.

\begin{figure}[!t]
\begin{center}
 \hspace{0.0\columnwidth}\psfig{figure=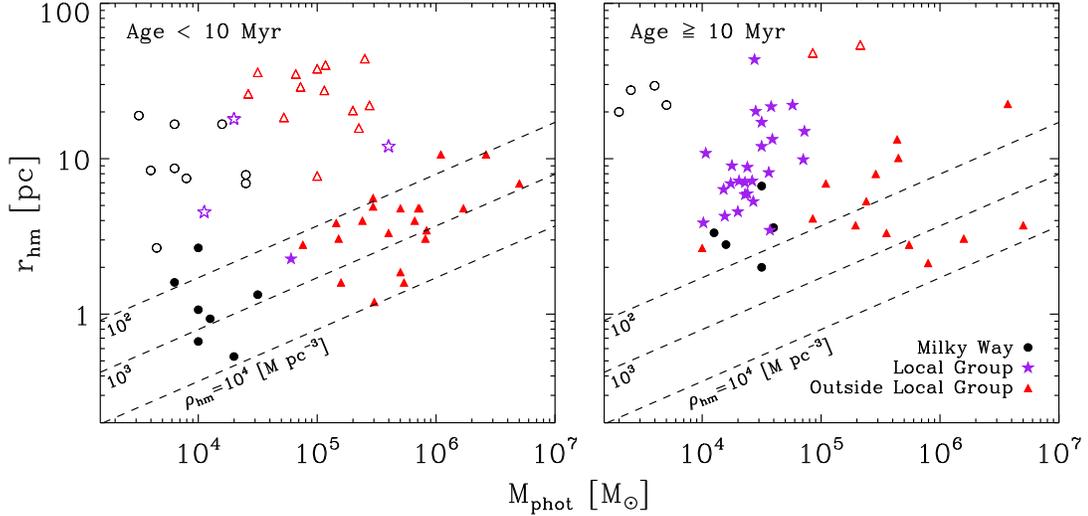,width=0.99\columnwidth}
\end{center}
\caption{ Mass-radius relation for all clusters (filled symbols) and
  associations (open symbols) from Tables~\ref{Tab:Galactic_Clusters},
  \ref{Tab:Local_Group_Clusters} \& \ref{Tab:Extra_Galactic_Clusters},
  using $\rhm=(4/3)\reff$. Lines of constant half-mass density,
  $\rhohm\equiv3M/(8\pi\rhm^3)$, are overplotted. The values of the
  densities shown correspond to $0.07 {\rm
    Myr}\lesssim\tdyn\lesssim0.7 {\rm Myr}$ (Eq.~\ref{Eq:tdyn}).  The
  clusters are subdivided into two groups: younger than 10\,Myr (left)
  and older than 10\,Myr (right). }
 \label{Fig:mass_radius_relation}
\end{figure}

Fig.~\ref{Fig:mass_radius_relation} shows the masses and radii of all
clusters and associations listed in
Tables~\ref{Tab:Galactic_Clusters}--\ref{Tab:Extra_Galactic_Clusters}. The
sample is divided into two age bins: less than $10\,$Myr (left panel)
and $10-100\,$Myr (right panel). We tentatively overplot lines of
constant density, which might indicate a trend with
$\rhohm\approx10^{3\pm1}\,\msun$ for the younger cluster sample.  This
mass-radius correlation seems to contradict earlier findings, but it
is important to bear in mind that we have limited ourselves to a
narrow range of (young) ages, and show the data for the clusters
(age$>3\,\tdyn$) with different (and more distinct) symbols, whereas
most literature studies do not separate the data in this way. The
distinction between clusters and assocations is of course a density
cut (Eq.~\ref{eq:dynamical}), corresponding to
$\rhohm\approx10\,\msunpc$ at an age of 10\,Myr, causing the upper
envelope of cluster points (filled) to line up with the constant
density lines. However, the lower envelope of points in the left panel
seems more consistent with constant density than with constant
radius. For the older clusters (right panel) the lower envelope seems
quite consistent with constant radius.  A correlation between mass and
radius, or indeed the lack of one, would have important implications
for the cluster's long-term survival, as we discuss in more detail in
\S\ref{Sect:SCSurvival}.

\subsection{Global properties}

The YMCs in the Magellanic Clouds and other nearby galaxies, such as
M31
\citep{2008ApJS..177..174N,2009AJ....138.1667B,2009AJ....137.4884M,2009ApJ...703.1872V,2009AJ....138..770H,2009AJ....137...94C,
  2010MNRAS.tmp...28P} and M33
\citep[e.g.][]{1999PASP..111..794C,2009ApJ...700..103P,2007AJ....134..447S,2009ApJ...699..839S}
provide opportunities for population studies. Observational studies of
these systems suffer less from the problematic distance and extinction
effects that challenge studies of Milky Way star clusters. They
provide a more global view of cluster populations, and are better
targets for studies of cluster age distributions
\citep{1985ApJ...299..211E,1987PASP...99..724H, 1995A&A...298...87G,
  2006MNRAS.366..295D,2008MNRAS.383.1103P} and luminosity functions
\citep{1985PASP...97..692E}.  

A comparison between Table~\ref{Tab:Galactic_Clusters} and
Table~\ref{Tab:Local_Group_Clusters} reveals the striking absence of
Milky Way YMCs with ages between 10 and 100 Myr, whereas in the SMC
and LMC, all YMCs except R\,136 have ages spanning this range. This
may well be an observational effect: Due to extinction and foreground
and background stars in the Milky Way, it is hard to discover clusters
without nebular emission and the luminous evolved stars found in very
young systems.  There might be a population of slightly older
(10--100\,Myr) YMCs in the Milky Way still awaiting discovery.

\subsubsection{The cluster luminosity function}
\label{Sect:LF}
YMCs have been observed and identified well beyond the Local Group,
providing exciting new opportunities for studies of star formation and
population synthesis.  They are found in abundance in galaxies with
high star formation rates, such as merging and interacting galaxies
\citep[e.g.][]{1992AJ....103..691H, 1997AJ....114.2381M,
  1999AJ....118.1551W}. However, they are also found in quiescent
spirals
\citep[e.g.][]{2000A&A...354..836L,2004A&A...416..537L,2009A&A...503...87C},
and there are many similarities between the young cluster populations
in these different environments. For example, the luminosity function
(LF), defined as the number of clusters per unit luminosity ($\dndl$)
is well described by a power-law $\dndl\sim L^{-\alpha}$ with
($\alpha\approx2$)
\citep[e.g.][]{1995AJ....109..960W,1997AJ....114.2381M,2002AJ....124.1393L,
  2003MNRAS.342..259D}, and tends to be slightly steeper at the bright
end \citep{1999AJ....118.1551W,2002AJ....124.1393L,
  2006A&A...450..129G}.

An appealing property of an $\alpha=2$ power-law is the fact that the
luminosity of the most luminous object, $\lmax$, increases linearly
with the total number of clusters,
$\ncl$. \citet[][]{2003dhst.symp..153W} showed that the value of
$\lmax$ for clusters in different galaxies scales as
$\lmax\propto\ncl^{\eta}$, with $\eta\approx0.75$ \citep[see
also][]{2002AJ....124.1393L}. The index $\eta$ is a proxy for the
shape of the bright end of the LF, since for a pure power-law
$\eta=1/(\alpha-1)$ \citep{2003AJ....126.1836H}.  A value of
$\eta=0.75$ corresponds to $\alpha=2.4$, supporting the finding that
the bright end of the LF is steeper than $L^{-2}$, since the $\lmax$
method traces the brightest clusters.

If a universal correlation between $\lmax$ and $\ncl$ exists, it means
that $\lmax$ is the result of the size of the sample.  It is therefore
determined by statistics, and no special physical conditions are
needed to form brighter clusters. A similar scaling between $\lmax$
and the star formation rate (SFR) was found by
\citet{2004MNRAS.350.1503W} and \citet{2008MNRAS.390..759B}.  This has
been interpreted in various ways. \citet{2008MNRAS.390..759B} uses the
similarity between the $\lmax-\ncl$ and $\lmax-\,$SFR relations to
conclude that the cluster formation efficiency is roughly constant
($\sim10\%$) over a large range of SFR.  \citet{2004MNRAS.350.1503W}
interpret it as an increase of the most massive cluster mass, $\mmax$,
with SFR and they conclude that $\mmax$ is set by the SFR of the host
galaxy, much like the \citet{2006MNRAS.365.1333W} relation between the
most massive star and the mass of the parent
cluster. \citet{2009A&A...494..539L} showed that the $\lmax$ clusters
have a large range of ages, with the brightest ones on average being
younger than the fainter ones. This implies a dependence of $M/L$ on
$\lmax$, such that the mass of the most {\em massive} cluster
($\mmax$) increases more slowly with $\ncl$ (and the SFR) than
$\lmax$, and hence that $\mmax$ does not follow the size of sample
prediction of the pure power-law LF.

\subsubsection{The cluster initial mass function}\label{Sect:CIMF}
It is tempting to interpret the LF as the underlying cluster mass
function.  As discussed above, it is not trivial to gain information
about the mass function from the LF, since the LF consists of clusters
with different ages, and clusters fade rapidly during their first
$\sim 1$ Gyr due to stellar evolution.  Determinations of cluster
initial mass functions (CIMFs) are rare, since it is hard to acquire
the cluster ages needed to select the youngest and to convert
luminosity into mass.  Several CIMF determinations have also found
power-law functions with indices close to $-2$
\citep{1999ApJ...527L..81Z,2007ApJ...663..844M,2003A&A...397..473B},
and other studies have found evidence for a truncation of this
power-law at the high-mass end \citep{2006A&A...446L...9G,
  2008MNRAS.390..759B, 2009A&A...494..539L, 2009ApJ...703.1872V}.  The
functional form of the initial mass function for young star clusters
is well represented by a \cite{1976ApJ...203..297S} distribution
\begin{equation}
  \phi(M) \equiv \frac{dN}{dM} =  A\,M^{-\beta}\,\exp(-M/M_*).
  \label{Eq:schechter}
\end{equation}
Here $\beta \simeq 2$ and the Schechter mass $M_*$ is equivalent to
the more familiar $L_*$ for the galaxy luminosity function.  For
Milky-Way type spiral galaxies $M_*\approx2\times10^5\,\msun$
\citep{2006A&A...446L...9G, 2009A&A...494..539L}.  For interacting
galaxies and luminous infrared galaxies, \citet{2008MNRAS.390..759B}
obtains $M_*\gtrsim10^6\,\msun$.

\begin{figure}
  \hspace{0.2\columnwidth}
  \psfig{figure=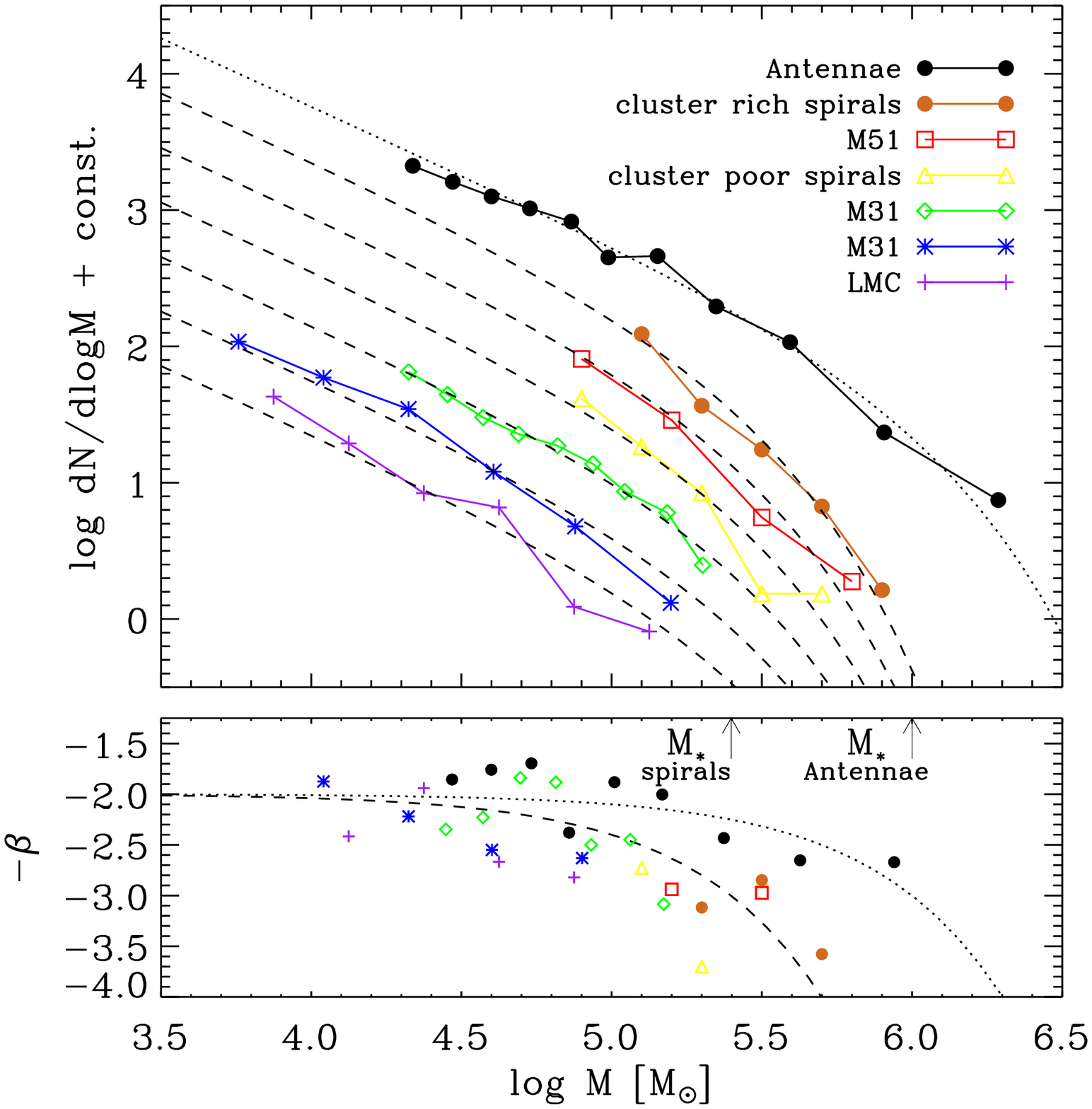,width=0.6\columnwidth}
  \caption[]{Top: Comparison of mass functions of clusters younger
    than $\sim1$\,Gyr in different galaxies. The results are taken
    from \citet{2009A&A...494..539L} (LMC, cluster rich spirals and
    cluster poor spirals); \citet{2009MNRAS.394.2113G} (M51);
    \citet{1999ApJ...527L..81Z} (the Antennae galaxies) and
    \citet{2009ApJ...703.1872V} (two versions of the M31 cluster mass
    function).  The cluster mass functions in spirals are compared to
    a Schechter function (Eq.~\ref{Eq:schechter}) with
    $M_*=2.5\times10^5\,\msun$ (dashed curves); for the Antennae,
    $M_*=10^6\,\msun$ (dotted curve) is used.  Bottom: the
    corresponding logarithmic slopes of the mass functions. The dotted
    and dashed curves are the logarithmic slopes of the functions
    shown in the top panel. }
\label{Fig:cluster_mass_functions}
\end{figure}

The upper panel of Fig.~\ref{Fig:cluster_mass_functions} compares
cluster mass functions for several galaxies with the Schechter
function, Eq.~\ref{Eq:schechter}.  The lower panel compares the
corresponding logarithmic slopes of the data with that of the
Schechter function.  Clearly the mass function of the Antennae
clusters extends to higher masses, and this is not only due to the
larger number of clusters, since the slope is also flatter at large
masses.  Thus the value of $M_*$ seems to depend on the local galactic
environment---it is possible to form more massive clusters in the
Antennae galaxies than in more quiescent environments.

Indirect indications for a truncation of the cluster mass function
also come from statistical arguments.  If we temporarily ignore the
(exponential) truncation, i.e. we adopt a power law with $\beta = 2$
without the exponential factor in Eq.\,\ref{Eq:schechter}, we can
relate the cluster formation rate to the masses of the most massive
clusters observed.  For an overall star formation rate of
$5000\,\msmyrkpc$ in the solar neighborhood
\citep{1979ApJS...41..513M}, comparable to the average values found in
external Milky-Way-type spirals \citep{1998ARA&A..36..189K}, and
assuming that $\sim10\%$ of this mass ends up in bound star clusters,
we find that the total mass formed in 10\,Myr in clusters within
4\,kpc of the Sun is $\sim 2 \times 10^5\,\msun$.  For our assumed
power-law mass function, the most massive cluster contains $\sim 10$\%
of the total mass, so the mass of the most massive cluster is a few
$\times10^4\,\msun$.  Within a 4\,kpc circle we find Westerlund~1,
with a mass of $\sim 6\times10^4\,\msun$ (see
Tab.\,\ref{Tab:Galactic_Clusters}), in reasonable agreement with
expectations.

Assuming the same star formation rate out to a distance of $\sim
8\,$kpc from the sun, which is a conservative assumption since the star formation
rate toward the Galactic center is probably higher than in the solar
neighborhood, the expected most massive cluster becomes a factor of
$8^2/4^2$ higher, or about $10^5\,\msun$.  Over a time span of 1\,Gyr,
clusters with masses of $\sim 10^7\,\msun$ should have formed within
that same distance, but if such a cluster existed it would most likely
already have been discovered, unless it has been disrupted, which seems unlikely. This
implies that for a quiescent environment like the Milky Way Galaxy,
the truncation of the cluster mass function must occur at considerably
lower mass than in a starburst environment. Similar arguments hold for
external galaxies, where the entire disk can be seen, and it is found
that the high-mass end of the cluster mass function falls off more
steeply than a power-law with exponent $-2$ (see
Fig.\,\ref{Fig:cluster_mass_functions}).

\subsubsection{Cluster formation efficiency}
\begin{figure}
\begin{tabular}{c}
  \hspace{0.1\columnwidth}\psfig{figure=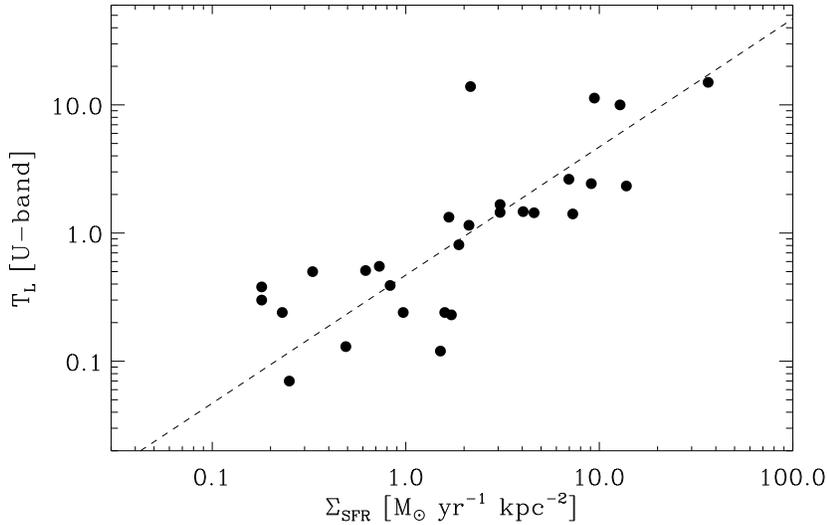,width=0.7\columnwidth} 
\end{tabular}
\caption[]{Specific $U$-band luminosity ($T_L$) for various cluster
  populations, as a function of $\sfrarea$ of the parent galaxy.  The
  dashed line shows a linear relation ($T_L\propto\sfrarea$). Data from
  \citet{2000A&A...354..836L}. }
\label{Fig:TLU}
\end{figure}

The number of globular clusters in a galaxy is often expressed in
terms of the specific frequency, the number of GCs per unit luminosity
of the host. For young clusters, this is probably not a very
meaningful quantity, since these clusters form with a power-law mass
function (or Schechter function, as in Eq.\,\ref{Eq:schechter}), and
the luminosity of the host galaxy depends strongly on the age of the
field star population. For this reason, \citet{2000A&A...354..836L}
introduce the {\it specific luminosity}, $T_L=100\,L_{\rm
  clusters}/L_{\rm galaxy}$ for samples of cluster populations in
different galaxies.  They show that, in the $U$-band, $T_L$ increases
strongly with the star formation rate per unit area ($\sfrarea$, see
Fig.~\ref{Fig:TLU}). This suggests that in galaxies with a higher
$\sfrarea$ a larger fraction of the newly formed stars end up in star
clusters.  The light in this blue filter is dominated by (short-lived)
hot stars, making $T_L$ a tracer of current star and star cluster
formation.  A compelling aspect of Fig.~\ref{Fig:TLU} is that both
axes are independent of distance. More recently,
\citet{2008MNRAS.390..759B} has used the $\lmax$--SFR relation to
derive a cluster formation efficiency of $\sim8\%$.  Both these
estimates suffer from (unknown) extinction effects, and the best way
to derive the cluster formation efficiency would be to compare the
fraction of the mass that forms in clusters to $\sfrarea$ for a sample
of galaxies.

The cluster formation efficiency is important for understanding the
extent to which YMCs can be used as tracers of star formation. An
interesting example is the lack of clusters with ages between 4 Gyr
and 12 Gyr in the LMC \citep[e.g.][]{1991ApJ...369....1V}. This ``age
gap'' could be the result of cluster disruption processes, or due to a
global pause in the SFR of the whole LMC. From comparison of the SFR
history and the age--metallicity relation of LMC field stars, the
latter scenario seems preferred, i.e. the LMC had a low SFR between 4
and 12 Gyr ago, and this is reflected in both the field stars and the
star clusters \citep{2009AJ....138.1243H}.  This suggests that YMCs
and their age distributions can be powerful tools in determining star
formation histories in more distant galaxies where individual stars
cannot be resolved.  We return to the interpretation of the age
distribution and its consequences for cluster lifetimes in
\S\ref{Sect:SCSurvival}.

\section{Dynamical processes in star clusters}
\label{Sec:Dynamics}

Studies of the evolution of a young star cluster split naturally into
three phases: (1) the first few megayears, during which stars are
still forming and the cluster contains significant amounts of ambient
gas, (2) a subsequent period when the cluster is largely gas-free, but
stellar mass loss plays an important role in the overall dynamics, and
(3) a late stage, during which purely stellar dynamical processes
dominate the long-term evolution of the cluster.  An upper limit on
the dividing line between phase 1 and phase 2 is the time of the first
supernovae in the cluster, some 3 Myr after formation
\citep{2006epbm.book.....E}, since these expel any remaining gas not
already ejected by winds and radiation from OB stars.  The dividing
line between phase 2 and phase 3 may be anywhere between 100\,Myr and
1\,Gyr, depending on the initial mass, radius, and density profile of
the cluster and the stellar mass function.

The evolution of the cluster during the first phase is a complex mix
of gas dynamics, stellar dynamics, stellar evolution, and radiative
transfer, and is currently incompletely understood
\citep[see][]{2007ApJ...668.1064E,2009MNRAS.398...33P}.  Unfortunately
this leaves uncertain many basic (and critical) cluster properties,
such as the duration and efficiency of the star-formation process, and
hence the cluster survival probability and the stellar mass function
at the beginning of phase 2 (see \S\ref{Sect:SCSurvival}).

Together, phases 2 aqnd 3 span the ``$N$-body simulation'' stage
familiar to many theorists.  Phase 3 is the domain of traditional
dynamical simulations; phase 2 is the ``kitchen sink'' phase, during
which stellar interactions, stellar evolution, and large-scale stellar
dynamics all contribute (see \S\ref{Sec:Simulations} and Appendix A).
As discussed in more detail below, the processes driving the dynamical
evolution during these phases are mostly well known and readily
modeled, allowing significant inroads to be made into the task of
interpreting cluster observations.  However, since the outcome of
phase 1 provides the initial conditions for phase 2, the proper
starting configuration for these simulations remains largely a matter
of conjecture.  Theoretical studies generally consist of throughput
experiments, mapping a set of assumed initial conditions into the
subsequent observable state of the cluster at a later age.

Setting aside the many uncertainties surrounding the early (phase 1)
evolution of the cluster, in this section we mainly describe the
assumed state of the cluster at the start of phase 2 and the physical
processes driving its subsequent evolution.  For better or worse,
$N$-body simulations generally assume quite idealized initial conditions
(summarized in Tab.\,\ref{Tab:InitialConditions}), with a spherically
symmetric, gas-free cluster in virial equilibrium, with all stars
already on the zero-age main-sequence.

\subsection{Initial conditions}\label{Sect:Init}

In the absence of a self-consistent understanding of cluster evolution
during phase 1, assumptions must be made about the following key
cluster properties before a phase 2 calculation can begin
\citep{2008LNP...760..181K}.  It must be noted that, in almost all
cases, the choices are poorly constrained by observations.

\begin{table}[htbp!]
\caption{Commonly adopted initial conditions for particle-based
  simulations of YMCs.
}
\medskip
\begin{center}
\begin{tabular}{llllllr} 
\hline
cluster		& parameter			& min		& max \\
property \\
\hline
number of stars	& N				&$10^3$		& $10^6$ \\
mass function	& single power law: $\phi(m) \propto m^{-2.35}$
						& 0.1\msun	& 100\msun \\
equilibrium	& $Q = -T/U$		        & \multicolumn{2}{c}{0.5} \\
density	distribution	& Plummer, King \\
tidal field	& $\rtide/\rJ$			& 0.25		& 1 \\
concentration	& $W_0$ 			& 1		& 16 \\
binary fraction & $f_b$			        & 0 		& 1 \\
mass ratio	& $\psi(q) = 1$			& 0 		& 1 \\
eccentricity	& $\Xi(e) = 2e$			& 0 		& 1 \\
orbital period	& $\Gamma(P) \propto 1/P$ 	& RLOF 		& hard \\
\hline		
\end{tabular}
\label{Tab:InitialConditions} 
\end{center}
\end{table} 

\begin{itemize}

\item The {\em stellar mass function} $\phi(m)=dN/dm$ is typically
  taken to be a ``standard'' distribution derived from studies of the
  solar neighborhood \citep[e.g.][]{1955ApJ...121..161S,
    1979ApJS...41..513M, 2001MNRAS.322..231K} although it has been
  suggested that the mass functions of some YMCs may be deficient in
  low-mass stars and/or ``top-heavy'' \citep{2001MNRAS.326.1027S}, in
  the sense that the slope $d\log N/d\log m$ of the mass function at
  the high-mass end is shallower (i.e.~less negative) than the
  standard Salpeter value of $-2.35$.

  In addition to the mass function, in many cases minimum and maximum
  stellar masses are imposed.  This is necessary for a pure power law,
  since the total mass would in general otherwise diverge, and is
  often desirable for other distributions, for which convergence is
  not an issue.  Often the minimum mass $m_{\rm min}$ is chosen to be
  a relatively high, $m_{\rm min}\apgt 1 \msun$, to emulate a more
  massive cluster simply by ignoring the low-mass stars.  This may
  suffice if one is interested in clusters younger than 100\,Myr
  (phase 2), but for older clusters the lower-mass stellar population
  is important, e.g. to reproduce the proper relaxation time.  The
  maximum mass $m_{\rm max}$ rarely poses a practical problem,
  although there may be some interesting correlations between the
  total cluster mass and the mass of the most massive star
  \citep{2004MNRAS.348..187W}

\item {\em Mass Segregation.}~~Traditionally, dynamical simulations
  have begun without initial mass segregation---that is, the local
  stellar mass distribution is assumed not to vary systematically with
  location in the cluster.  There is no good reason for this, other
  than simplicity.  Evidence for initial mass segregation can be found
  in some young clusters \citep[e.g.][]{1998ApJ...492..540H,
    2008AJ....135..173S}, simulations of star formation
  \citep[e.g.][]{2001ApJ...556..837K, 2006MNRAS.370..488B}, and
  dynamical evolution during phase 1 \citep{2007ApJ...655L..45M,
    2009ApJ...700L..99A}.  Several prescriptions have been used
  recently for initial mass segregation \citep{2008MNRAS.385.1673S,
    2008ApJ...685..247B,2009ApJ...698..615V}.  They differ in detail,
  but lead to similar conclusions, namely that initial mass
  segregation may be critical to cluster survival
  \citep{2009ApJ...698..615V}, since mass loss
  from centrally concentrated massive stars can be much more
  destructive than the same mass loss distributed throughout the body
  of the cluster (see \S\ref{Sect:SCSurvival}).

\item {\em Virial Ratio.}~~Simulations generally begin with a cluster
  in virial (dynamical) equilibrium, with virial ratio $Q
  \equiv -T/U=1/2$.  As with most of the other simplifying assumptions
  described here, the principal reason for this choice is reduction of
  the dimensionality of the initial parameter space, but there is no
  compelling physical reason for it.  The gas expulsion that marks the
  end of phase 1 is expected to leave the cluster significantly out of
  equilibrium, and quite possibly unbound
  \citep[e.g.][]{1980ApJ...235..986H}.  The time scale for a cluster
  to return to virial equilibrium may be comparable to the time scale
  on which mass loss due to stellar evolution subsequently modifies
  the cluster structure (see \S\ref{sec:TheFirstMyr}).

\item {\em Spatial Density and Velocity Distributions.}~~The initial
  density profiles of young star clusters are poorly constrained.
  Several standard models are used to model the stellar distribution:
  \citet{1911MNRAS..71..460P} and truncated Maxwellian
  \citep{1966AJ.....71...64K} models are the most common.  Other
  distributions, such as isothermal and homogeneous spheres, are also
  used \citep[e.g.][]{2005MNRAS.358..742S}.  King models provide good
  fits to many observed GCs, although their relation to YMCs is
  unclear (see \S\ref{sec:observations}).  In the absence of strong
  observational constraints, the stellar velocity distribution is
  normally taken to be non-rotating and isotropic, with (for a Plummer
  model) the dispersion following the local potential with the assumed
  virial ratio (see \S\ref{Sect:TheRadius}).

\item {\em Tidal Field.}~~Clusters do not exist in isolation, but
  rather are influenced by the local tidal field of their parent
  galaxy.  In many cases (see \S\ref{Sect:TheRadius}), the field is
  modeled explicitly as an external potential or its quadrupole moment
  relative to the cluster center; however simple ``stripping radius''
  prescriptions are also widely used.  In practice, incorporating a
  simple stripping radius instead of a self-consistent tidal field
  reduces the cluster lifetime by about a factor of two if the
  stripping radius is taken as $\rJ$. This effect can be mitigated by
  adopting a larger cut-off radius (for example $2\rJ$).  With few
  exceptions \citep[e.g.][]{2003MNRAS.340..227B,2003MNRAS.343.1025W}
  the parameters of the tidal field are held fixed in time,
  corresponding (for spherical or axisymmetric potentials) to an orbit
  at fixed galactocentric radius.  For a given orbit and cluster mass,
  the initial ratio of the cluster limiting radius \cite[e.g. the
    truncation radius of a][model, see
    \S\ref{Sect:TheRadius}]{1966AJ.....71...64K} to the Jacobi radius
  of the cluster in the local tidal field is a free parameter, often
  taken to be of order unity.  For a Plummer sphere, which extends to
  infinity, the tidal radius is often implemented as a simple cutoff
  at some relatively large distance.

\item {\em Binary Fraction.}~~Binary stars are critical to cluster
  evolution during phase 2 and especially phase 3
  \citep[e.g.][]{2007ApJ...665..707H,2007MNRAS.374...95P}. It is not
  so clear how important they are during phase 1, when major
  structural changes are induced by mass loss and mass segregation
  \citep{2000prpl.conf..151C,2003MNRAS.339..577B}.  There are few, if
  any, observational constraints on the overall binary fraction in
  YMCs.  Open clusters in the field typically have high binary
  fractions, approaching 100\% in some cases
  \citep[e.g.][]{2008MNRAS.386..447S,
    2009AJ....137.3358M,2009AJ....137.3437B}.  On the other hand, most
  recent studies of binaries in globular clusters suggest binary
  fractions of between $\sim6$\% and $\sim 15$\%
  \citep{2002AJ....123.1509B,2007MNRAS.380..781S,2008MmSAI..79..623M,2009A&A...493..947S}.

\item {\em Binary Secondary Masses.}~~The mass of the secondary star
  in a binary is typically selected uniformly between some minimum
  mass and the mass of the primary \citep{1991A&A...248..485D}.  With
  this choice, a binary tends to be more massive than the average
  cluster star, resulting in additional mass segregation of the binary
  population. This effect can be removed if desired by randomly
  selecting primary stars and splitting them into primary and
  secondary components, in which case adding binaries does not affect
  the mass function of cluster members (single stars and binaries),
  but it does introduce a deviation from the initial stellar mass
  function among binary components \citep{1995MNRAS.277.1491K}.

\item {\em Orbital Elements of the Binaries.}~~In general, the choices
  made for binary orbital elements tend to be defensive, given the
  lack of observational guidance.  Apart from the introduction of a
  whole new set of initial parameters, the presence of primordial
  binaries also introduces new mass, length, and time scales to the
  problem, greatly complicating direct comparison between runs having
  different initial conditions.

  The initial distribution of binary periods is unknown, but is often
  assumed to follow the observed distribution in the solar
  neighborhood \citep{1991A&A...248..485D}, which is approximately a
  Gaussian in $\log P$ with mean $\log P = 4.8$ and dispersion
  $\sigma_{\log P} = 2.3$, with $P$ in days.  However, distributions
  flat in $\log P$ \citep{1983ARA&A..21..343A} and $\log a$
  \citep{1990ApJ...362..522M} have also been adopted.  Observations of
  YMCs are of little help, as there are at most a handful of binaries
  with measured orbital parameters in YMCs, and those have very short
  orbital periods and high-mass components.  The eccentricity
  distribution is usually taken to be thermal
  \citep{1991A&A...248..485D}.  Binary orbital orientation and initial
  phase are chosen randomly.

\item {\em Higher order multiples.}~~Primordial multiple stars are
  rarely included in phase-2/3 dynamical simulations. There are a few
  examples of calculations with primordial triples
  \citep{2007MNRAS.379..111V} or hierarchical planetary systems
  \citep{2009ApJ...697..458S}.  The complications of adding primordial
  multiples greatly increase the already significant challenges of
  including binary dynamics and evolution.

\end{itemize}

\subsection{Multiple stellar populations}
The discovery of multiple populations of main-sequence stars and
giants in an increasing number of globular clusters
\citep{2005ApJ...621..777P,2008MmSAI..79..334P} has led to the
realization that star clusters are not idealized entities with single
well defined stellar populations.  In some clusters, the observed
stellar populations appear to be separated by less than $\sim 10^8$
years, well within our age range for young clusters.  The existence of
multiple populations indicates that a second epoch of star formation
must have taken place early in the cluster's lifetime (see
\S\ref{Sect:BlueStragglers}).  The differences in light-element
abundances suggest that the second-generation (SG) stars formed out of
gas containing matter processed through high-temperature CNO cycle
reactions in first-generation (FG) stars.

The two main candidates currently suggested as possible sources of
enriched gas for SG formation are rapidly rotating massive stars
\citep{2006A&A...458..135P, 2007A&A...464.1029D} and massive (4--9
\msun) Asymptotic Giant Branch (AGB) stars \citep{2001ApJ...550L..65V,
  2007PASA...24..103K}.  In order to form the large fraction of SG
stars suggested by observations (50\% or more of the current mass of
multiple-population clusters, \citealt{2008arXiv0811.3591C}), both
scenarios require that either the IMF of the FG stars was highly
anomalous, with an unusually large fraction of massive stars, or the
FG population had a normal IMF but was initially at least ten times
more massive than is now observed.  \citet{2007MNRAS.380.1589B} have
studied the subsequent evolution and mixing of the two-component
cluster in the first scenario.  \citet{2008MNRAS.391..825D} have
presented simulations of the second, in which the SG stars form deep
in the potential well of a FG cluster destabilized by early mass loss.
Most of the FG cluster dissolves, leaving a mixed FG/SG system after a
few gigayears.

If similar processes are operating today, multiple populations should
be expected in at least some young star clusters, but it is currently
not known whether, or to what extent, this phenomenon occurs in
observed YMCs.  For the unresolved extragalactic clusters, multiple
populations will be hard to confirm, but for clusters in the local
group this should be possible.  At present, however, only one known
cluster, Sandage 96, exhibits a young (10--16\,Myr) population
together with a relatively old (32--100\,Myr) population
\citep{2009ApJ...695..619V} (see also \S\ref{Sect:SNe}).  Spreads in
the main-sequence turn-off have been reported found for intermediate
age clusters in the LMC, which has been interpreted as an age spread
of $\sim300\,$Myr \citep{2007MNRAS.379..151M, 2009A&A...497..755M}.
If true, this has exciting implications for theories of the formation
of star clusters and the general assumption that clusters are simple
stellar populations.  However, \citet{2009MNRAS.398L..11B} find that
the observed main sequence turn-off spread could be explained by
stellar evolutionary effects induced by rotation.  Obviously if
multiple populations are common in YMCs, they significantly impact the
assumptions made for simulations of phase 2 and the long-term cluster
evolution during phase 3.  Except where noted, we will not explicitly
address the possibility of delayed SG star formation in this review.

\subsection{Overview of  cluster dynamical evolution}
\label{Sect:DynEvolution}

Most numerical studies start with initial conditions as described in
\S\ref{Sect:Init}.  To the extent that stellar mass loss can be
neglected, we can understand the dynamical evolution of a star cluster
from the fundamental physics of self-gravitating systems, driven by
relaxation.

\subsubsection{Evaporation}
\label{sect:Relaxation.Theory}

The relaxation time (Eq.\,\ref{Eq:trtd}) is the time scale on which
stars tend to establish a Maxwellian velocity distribution. A fraction
$\xie$ of the stars in the tail of that distribution have velocities
larger than $\vesc$ and consequently escape.  Assuming that this
high-velocity tail is refilled every $\trh$, the dissolution time
scale is $\tdis=\trh/\xie$.  For isolated clusters,
$\vesc=2\,\vrms$. For a Maxwellian velocity distribution, a fraction
$\xie=0.0074$ has $v>2\,\vrms$, and hence $\tdis=137\,\trh$. For
tidally limited cluster $\xie$ is higher since $\vesc$ is lower. For a
typical cluster density profile $\xie\approx0.033$, implying
$\tdis\approx30\,\trh$ \citep{1987degc.book.....S}.

The escape fraction $\xie$ is often taken to be constant
\citep[e.g.][]{1997ApJ...474..223G}, althouh it depends on $\rhm$
(through $\vrms$) and also on the strength of the tidal field, or
$\rj$ (through $\vesc$).  Effectively, $\xie$ depends on the ratio
$\rhm/\rj$ \citep[e.g.][]{1973ApJ...183..565S, 1988IAUS..126..393W}.
\citet{2008MNRAS.389L..28G} show that $\xie\propto(\rhm/\rj)^{3/2}$
for $\rhm/\rj\apgt0.05$ (the so-called {\it tidal regime}).  From
Eq.~\ref{Eq:trtd} we then find, for clusters on circular orbits in the
tidal regime, that $\tdis\propto N/\omega$ (neglecting the slowly
varying Coulomb logarithm). For a flat rotation curve, $\tdis\propto
R_G$ for a cluster of given mass \citep[e.g.][]{1990ApJ...351..121C,
  1997MNRAS.289..898V}.  This linear dependence of $\tdis$ on $R_G$
makes it difficult to explain the universality of the globular cluster
mass function via dynamical evolution of a power-law initial cluster
mass function \citep{2003ApJ...593..760V}, but this will not be
discussed further here.

\citet{2001MNRAS.325.1323B} showed that $\tcr$ also enters into the
escape rate and found, for equal-mass stars,
$\tdis\propto\trh^{3/4}\tcr^{1/4}$.  The non-linear scaling of the
dissolution time with the relaxation time results from the fact that a
star with sufficient energy to escape may orbit the system many times
before ``finding'' one of the Lagrangian points, through which escape
actually occurs \citep{2000MNRAS.318..753F}.
\citet{2003MNRAS.340..227B} found that this scaling also holds for
models of clusters with a stellar mass spectrum, stellar evolution,
and for different types of orbits in a logarithmic potential.  Their
result for $\tdis$ can be summarized as
\begin{equation}
\tdis\approx2\,{\rm Myr} 
            \left(\frac{N}{\ln\Lambda}\right)^{3/4}
            \left(\frac{R_G}{{\rm kpc}}\right)
            \left(\frac{V_G}{220\,\kms}\right)^{-1}(1-\varepsilon),
\label{eq:tdis3}
\end{equation}
where $\varepsilon$ is the eccentricity of the orbit.  For an
eccentric orbit, $\varepsilon > 0$ the distance to the galactic enter
$R_G$ is taken as the apogalcticon, whereas $V_G$ is circular
velocity, which is constant in a logarithmic potential.  If the
Coulomb logarithm is taken into account, the scaling is approximately
$\tdis\propto N^{0.65}$ in the range of about $10^3$ to
$10^6$\,\Msun\, \citep{2005A&A...429..173L}.

\subsubsection{Core collapse}
\label{sect:CoreCollapse.Theory}

Self-gravitating systems are inherently unstable, and no final
equilibrium state exists for a star cluster.  The evaporation of
high-velocity stars and the internal effects of two-body relaxation,
which transfers energy from the inner to the outer regions of the
cluster, result in core collapse
\citep{1962spss.book.....A,1968MNRAS.138..495L,1980ApJ...242..765C,1980MNRAS.191..483L,1996ApJ...471..796M}. During
this phase, the central portions of the cluster accelerate toward
infinite density while the outer regions expand.  The process is
readily understood by recognizing that, according to the virial
theorem, a self-gravitating system has negative specific
heat---reducing its energy causes it to heat up.  Hence, as relaxation
transports energy from the (dynamically) warmer central core to the
cooler outer regions, the core contracts and heats up as it loses
energy.  The time scale for the process to go to completion (i.e. a
core of zero size and formally infinite density) is $\tcc\sim 15
\trlx$ for an initial Plummer sphere of identical masses.  Starting
with a more concentrated \citet{1966AJ.....71...64K} distribution
shortens the time of core collapse considerably
\citep{1996NewA....1..255Q}, as does a broad spectrum of masses
\citep{1985ApJ...292..339I}.

In systems with a mass spectrum, two-body interactions accelerate the
dynamical evolution by driving the system toward energy equipartition,
in which the velocity dispersions of stars of different masses would
have $\langle mv^2\rangle\sim {\rm constant}$.  The result is mass
segregation, where more massive stars slow down and sink toward the
center of the cluster on a time scale \citep{1969ApJ...158L.139S}
\begin{equation}
	\ts \sim \frac{\langle m\rangle}{m}\,\trl \,.
\label{eq:ts}
\end{equation}
\citet{2002ApJ...576..899P} and \citet{2004ApJ...604..632G} find that,
for a typical \citet{2001MNRAS.322..231K} mass function, the time
scale for the most massive stars to reach the center and form a well
defined high-density core is $\sim0.2 \trl$, where $\trl$ is the
relaxation time (see Eq.\,\ref{Eq:LocalRelaxationTime}) of the region
of interest containing a significant number of massive stars---the
core of a massive cluster, or the half-mass radius of a smaller one
(in which case $\trl = \trlx$, see\,Eq.\,\ref{Eq:trtd}).  For dense
clusters, $\ts$ may be shorter than the time scale for stellar
evolution, or for the first supernovae to occur
\citep{1999A&A...348..117P}.

Thus, a collisional stellar system inevitably evolves toward a state
in which the most massive objects become concentrated in the
high-density central core (see \S\ref{Sect:Collisions}).  Dynamical
evolution provides a natural and effective mechanism for concentrating
astrophysically interesting objects in regions of high stellar
density.

\subsection{Internal Heating}
On longer time scales, the evolution of clusters that survive the
early phases of mass loss is driven by the competition between
relaxation and a variety of internal heating mechanisms.  High central
densities lead to interactions among stars and binaries.  Many of
these interactions can act as energy sources to the cluster on larger
scales, satisfying the relaxation-driven demands of the halo and
temporarily stabilizing the core against collapse
\citep{1989Natur.339...40G,1991ApJ...370..567G,
  1990ApJ...362..522M,1991ApJ...372..111M,
  1992MNRAS.257..513H,2003ApJ...593..772F}. On long time scales, these
processes lead to a slow (relaxation time) overall expansion of the
cluster, with $\rvir\propto t^{2/3}$, a result that follows from
simple considerations of the energy flux through the half-mass radius
\citep{1965AnAp...28...62H}.

While these processes are important to the long-term dynamical
evolution (phase 3), their relevance is somewhat different during the
first 100 Myr \citep{2007MNRAS.374...95P}, which is largely dominated
by stellar mass loss (phase 2) and the segregation of the most massive
stars.  Often, their major effect is to enhance the rate of collisions
and the formation of stellar exotica.  We now consider in turn the
following processes: binary heating (\S\ref{Sect:BinaryInteractions}),
stellar collisions (\S\ref{Sect:Collisions}), and black-hole heating
(\S\ref{Sect:BHHeating}).

\subsubsection{Binary Interactions}\label{Sect:BinaryInteractions}
Irrespective of the way they form, binaries are often described by
dynamicists as either ``hard'' or ``soft.''  A hard binary has binding
energy greater than the mean stellar kinetic energy in the cluster
\citep{1975MNRAS.173..729H}: $|E_b| > \frac12\langle mv^2\rangle
\approx \frac12\mmean\vrmssq$, where {\mmean} and $\vrms$ are the
local mean stellar mass and velocity dispersion. 
A binary with mass $m_b=m_1+m_2$ and semi-major axis $a_b$ has energy
$E_b = -Gm_1m_2/2a_b$, so hard binaries have $a_b<a_{\rm hard}$, where
\begin{equation}
	a_{\rm hard} ~=~ {G m_b^2 \over 4 \mmean \vrmssq} 
                    ~\approx~  9.5 \times 10^{4} \,\Rsun 
                            \left(\frac{m_b}{\msun}\right)^2  
                            \left(\frac{\vrms}{\kms}\right)^{-2}.
\label{Eq:hard}\end{equation}
Here we implicitely assumed that $m_1=m_2=\mmean$ to derive the
right-hand expression.  The hard--soft distinction is helpful when
discussing dynamical interactions between binaries and other cluster
members.  However, we note the definition of hardness depends on local
cluster properties, so the nomenclature changes with environment, and
even within the same cluster a binary that is hard in the halo could
be soft in the core.

The dynamical significance of ``hard'' binaries (see
Eq.\,\ref{Eq:hard}) has been understood since the 1970s
\citep{1975MNRAS.173..729H,1975AJ.....80..809H,1983ApJ...268..319H}
When a hard binary interacts with another cluster star, the resultant
binary (which may or may not have the same components as the original
binary) tends, on average, to be harder than the original binary,
making binary interactions a net heat source to the cluster.  Soft
binaries tend to be destroyed by encounters.  For equal-mass systems,
the mean energy liberated in a hard-binary encounter is proportional
to $E_b$: $\langle\Delta E_b\rangle = \gamma E_b$, where $\gamma =
0.4$ for ``resonant'' interactions \citep{1975MNRAS.173..729H}, and
$\gamma\sim0.2$ when wider ``flybys'' are taken into account
\citep{1987degc.book.....S}.

The liberated energy goes into the recoil of the binary and single
star after the interaction.  Adopting terminology commonly used in
this field, we write the binary energy as $E_b=-hkT$, where
$\frac32kT=\langle\frac12mv^2\rangle$ and $h\gg1$, so the total recoil
energy, in the center of mass frame of the interaction, is $\gamma
hkT$.  In the center of mass frame, a fraction ${m_b \over m_b + m}$
of this energy goes to the single star (of mass $m$) and ${m \over m_b
  + m}$ to the binary.  For equal-mass stars, these reduce to
$\frac23$ for the single star and $\frac13$ for the binary.
Neglecting the thermal motion of the center of mass frame, we identify
three regimes:
\begin{enumerate}
\item If $\frac23\gamma hkT < \frac12m\vesc^2 = 2m\vrmssq = 6kT$,
  i.e. $h<45$, neither the binary nor the single star acquires enough
  energy to escape the cluster. Binaries in this stage are kicked out
  of the core, then sink back by dynamical friction, in a process that
  we call ``binary convection.''
\item If $\frac23\gamma hkT > 6kT$ but $\frac13\gamma hkT < 4m\vrmssq
  = 12kT$, i.e. $45<h<180$, the single star escapes, but the binary is
  retained.  We refer to such a binary as a ``bully.''
\item If $h > 36/\gamma = 180$, both the binary and the single star are
  ejected.  Such a binary is a ``self-ejecter.''
\end{enumerate}
These numbers are only illustrative.  For a binary with components
more massive than average, as is often the case, the threshold for
bullying behavior drops, while that for self-ejection increases.

Tab.\,\ref{Tab:SHBSI} places these considerations in a more physical
context.  Note that, since the closest approach between particles in a
resonant interaction may be as little as a few percent of the binary
semi-major axis \citep{1985ApJ...298..502H}, the hardest binaries may
well experience physical stellar collisions rather than hardening to
the point of ejection; and a collision tends to soften the surviving
binary.  Alternatively, before their next interaction, they may enter
the regime in which internal processes, such as tidal circularization
and/or Roche-lobe overflow, become important.  The future of such a
binary may be determined by the internal evolution of its component
stars, rather than by further encounters.

\begin{table}[htbp!]
\caption{ Terminology (first column) and characterization (second and
  third columns) for the various stages of a binary (see text).  The
  subsequent columns give the orbital separation $a$ of a binary (in
  units of AU) with a total mass $m_1+m_2 \equiv m_b = 10\mmean$ or
  $m_b = 100\mmean$, in a cluster with a mass of $\mtot =
  10^5$\,\Msun\, and virial radius $\rvir = 1$\,pc and
  $\rvir=10$\,pc.
}
\medskip
\begin{tabular}{llrllllr} \hline
Binary 	  & relation  &$E_b$& \multicolumn{4}{|c|}{$M = 10^5$\,\msun} & Unit \\
          &           &[kT]& \multicolumn{2}{|c}{$\rvir =1$pc} 
                      & \multicolumn{2}{|c|}{$\rvir =10$pc} \\
          &           &&$m_b=10$ &$m_b=100$ &$m_b=10$&$m_b=100$ & \mmean \\
\hline
hard	  & $E_{\rm b} > \frac32 kT$     &1&$7.2\times10^4$&$7.2\times10^6$
                                   &$7.2\times10^5$& $7.2\times10^7$& AU\\
bully	  & $\vrec > \vesc m/m_b$  &10&$1.7\times10^3$&$1.7\times10^6$&$1.7\times10^4$&$1.7\times10^7$& AU\\
tenured   & $\tenc > \trlx$        &100&52     &58  & 53 & 5.8& AU\\
self-eject& $\vrec > \vesc m_b/m$  &100&0.016  & 1.6 &33  &$3.3\times10^3$& AU\\
\hline		
\end{tabular}
\label{Tab:SHBSI}
\end{table} 

The binary encounter time scale is $\tenc = (n \sigma \vrms)^{-1}$,
where $n$ is the local stellar density and $\sigma$ is the encounter
cross section (see Eq.\,\ref{Eq:CrossSection}).  If we arbitrarily
compute the binary interaction cross section as that for a flyby
within 3 binary semi-major axes, consistent with the encounters
contributing to the \cite{1987degc.book.....S} value $\gamma=0.2$, and
again assume equal masses ($m_b = 2m$), we find
\begin{equation}
  \tenc \sim 8h\trl,
\end{equation}
where we have used Eq. \ref{Eq:LocalRelaxationTime} and taken
$\ln\Lambda=10$.  Thus the net local heating rate per binary during
the 100\% efficient phase (\#1 above), when the recoil energy remains
in the cluster due to ``binary convection'' is
\begin{equation}
    {\cal E}_{bin} = \gamma hkT\,\tenc^{-1} \sim 0.1kT/\trl,
\end{equation}
that is, {\em on average}, each binary heats the cluster at a roughly
constant rate.  During the ``bully'' phase, the heating rate drops to
just over one-third of this value.  The limiting value of one-third is
not reached since the ejected single stars still heat the cluster
indirectly by reducing its binding energy by a few $kT$.  For
``self-ejecting'' binaries, the heating rate drops almost to zero,
with only indirect heating contributing.

Binary--binary interactions also heat the cluster, although the extra
degrees of freedom complicate somewhat the above discussion.  If the
binaries differ widely in semi-major axes, the interaction can be
handled in the three-body approximation, with the harder binary
considered a point mass.  If the semi-major axes are more comparable,
as a rule of thumb the harder binary tends to disrupt the wider one
\citep{1996MNRAS.281..830B}.

Numerical experiments over the past three decades have unambiguously
shown how initial binaries segregate to the cluster core, interact,
and support the core against further collapse
\citep{1990ApJ...362..522M,1992MNRAS.257..513H}.  The respite is only
temporary, however.  Sufficiently hard binaries are ejected from the
cluster by the recoil from their last interaction (self-ejection, see
Tab.\,\ref{Tab:SHBSI}), and binaries may be destroyed, either by
interactions with harder binaries, or when two or more stars collide
during the interaction.  For large initial binary fractions, this
binary-supported phase may exceed the age of the universe or the
lifetime of the cluster against tidal dissolution.  However, for low
initial binary fractions, as appears to have been the case for the GCs
observed today \citep{2008MmSAI..79..623M}, the binaries can be
depleted before the cluster dissolves, and core collapse resumes
\citep{2003ApJ...593..772F}.

Thus binary dynamics drives the evolution of the cluster while,
simultaneously, the combination of cluster dynamics and internal
stellar processes determine the internal evolution of each binary.
This interplay between stellar evolution and stellar dynamics is
sometimes referred to as {\em star-cluster ecology}
\citep{1992Natur.359..772H} or the {\em binary zoo}
\citep{2006NewA...12..201D} or {\em stellar promiscuouity}
\citep{2002ApJ...570..184H}.
 
\subsubsection{Stellar Collisions}\label{Sect:Collisions}

In systems without significant binary fractions---either initially or
following the depletion of core binaries---core collapse may continue
to densities at which actual stellar collisions occur.  In young
clusters, the density increase may be enhanced by rapid segregation of
the most massive stars in the system to the cluster core.  Since the
escape velocity from the stellar surface greatly exceeds the rms speed
of cluster stars ($\theta < 100$ in Eq.\,\ref{Eq:Safronov}),
collisions lead to mergers of the stars involved, with only small
fractional mass loss \citep{1987ApJ...323..614B,2001A&A...375..711F}.
If the merger products did not evolve, the effect of collisions would
be to dissipate kinetic energy, and hence cool the system,
accelerating core collapse \citep{1999A&A...348..117P}.  However, when
accelerated stellar evolution is taken into account, the (time
averaged) enhanced mass loss can result in a net heating effect
\citep{2008IAUS..246..151C}.

The cross section for an encounter between two objects of masses $m_1$
and $m_2$ and radii $r_1$ and $r_2$, respectively, is
\begin{equation}
	\sigma = \pi r^2 \left[ 1 + {2G(m_1+m_2) \over r v^2} \right]
\label{Eq:CrossSection}
\end{equation}
\citep{1976ApL....17...87H}, where $v$ is the relative velocity at
infinity and $r = r_1+r_2$.  For $r \ll G(m_1+m_2)/v^2$, as is usually
the case for the systems discussed in this review, the encounter is
dominated by the second term (gravitational focusing), and
Eq.\,\ref{Eq:CrossSection} reduces to
\begin{equation}
	\sigma \approx 2 \pi r {G m \over v^2},
\label{Eq:CrossSection2}
\end{equation}
which is nearly independent of the properties of the other stars.

Collisions between single stars are unlikely unless one (or both) of
the stars is very large and/or very massive, or the local density is
very high.  Consider a large, massive star of mass $m$ and radius $r$
moving through a field of smaller stars, so the collision cross
section is dominated by the properties of the massive star.  The rate
of increase of the star's mass due to collisions is
\begin{eqnarray}
    \frac{dM}{dt} &~\approx~& \rhocore \sigma v
		  ~~\approx~~ 2\pi G m r \rhocore / v \nonumber\\
		  &~=~& 0.6\, \left( {m \over 100 \Msun} \right) 
		         \left( {r \over 100 \Rsun} \right) 
		         \left( {\rhocore \over 10^6 \Msun/{\rm pc}^3} \right) 
		         \left( {10 \kms \over v} \right) 
                          ~~~~ M_\odot/{\rm Myr}\,.
\label{Fig:dmdt}
\end{eqnarray}
Thus a massive star ($m/\Msun \sim r/\Rsun \sim 1$) in a dense stellar
core ($\rhocore \sim 10^6$\,\Msun/pc$^3$) will experience numerous
collisions during its $\sim3-5$ Myr lifetime.

The presence of primoridal binaries can significantly increase the
chance of a traffic accident.  Hard binaries (see
\S\ref{Sect:BinaryInteractions}) are in the gravitational focusing
regime, so the binary interaction cross section may be obtained by
setting $r=a$ in Eq. \ref{Eq:CrossSection2}.  Such an encounter will
likely lead to the hardening of the binary and possibly the ejection
of the single star and also the binary, as just discussed.  However,
it may also lead to a hydrodynamical encounter, i.e. a physical
collision between two of the stars.  It is quite likely that the third
star will also be engulfed in the collision product
\citep{2004MNRAS.352....1F}.  Since binaries generally have semi-major
axes much greater than the radii of the component stars, such {\em
  binary-mediated collisions} play important roles in determining the
stellar collision rate in YMCs \citep{2002ApJ...576..899P}, leading to
significant numbers of mergers in lower-density, binary rich
environments.  Massive binaries in young dense clusters tend to be
collision targets rather than heat sources
\citep{2004ApJ...604..632G}.

In a sufficiently dense system, repeated stellar collisions can lead
to a so-called ``collision runaway'' \citep{1999A&A...348..117P}, in
which a massive star or collision product suffers repeated mergers and
grows enormously in mass before exploding as a supernova
\citep{2002ApJ...576..899P,2004Natur.428..724P,2004ApJ...604..632G}.
This has frequently been cited as a possible mechanism for producing
intermediate-mass black holes (IMBHs) in star clusters.  However,
while the dynamics is simple, numerous uncertainties in the stellar
evolution and mass loss of the resultant merger product have been
pointed out in the recent literature, suggesting that the net growth
rate, and hence the final mass of the resulting IMBH, may be much
lower than suggested by purely dynamical simulations---perhaps as
little as a few hundred solar masses
\citep{2008A&A...477..223Y,2009A&A...497..255G,2009Ap&SS.tmp..114V}.
Alternative formation mechanisms for more massive IMBHs involve gas
accretion onto a seed $\sim100$\,{\Msun} black hole during a second
round of star formation early in the cluster's lifetime
\citep{2009AAS...21333105V}, or repeated collisions between
stellar-mass black holes during the phase-3 evolution of the cluster
\citep{2002MNRAS.330..232C}.

\subsubsection{Black Hole Heating}\label{Sect:BHHeating}

An IMBH in a star cluster can be an efficient source of energy to the
stellar system.  Stars diffuse by two-body relaxation deeper and
deeper into the IMBH's potential well, and eventually are tidally
disrupted and consumed \citep{1976ApJ...209..214B}.
The energy lost during the process heats the system.  The heating rate
for an IMBH of mass $M_{BH}$ in a cluster core of density $\rho_c$ and
velocity dispersion $v_c$ is
\begin{equation}
    {\cal E}_{bh} \sim \frac{G^5\mmean\rho_c^2M_{BH}^3\ln\Lambda}{v_c^7}.
\end{equation}

Although cores are promising environments for the formation of IMBHs,
they may not be the best place to look for evidence of massive black
holes today.  Dynamical heating by even a modest IMBH is likely to
lead to a cluster containing a fairly extended core
\citep{2005ApJ...620..238B}.  Comparing the outward energy flux from
stars relaxing inward in the (Bahcall--Wolf) cusp surrounding the IMBH
to the outward flux implied by two-body relaxation at the cluster
half-mass radius, \citet{2007PASJ...59L..11H} estimate the equilibrium
ratio of the half-mass ($\rh$) to the core ($\rcore$) radius in a
cluster of mass $M$.  Calibrating to simulations, they conclude that
for systems with equal mass, except of course the black holes
\begin{equation}
  \frac{\rhm}{\rcore} ~\sim~ 0.23 \left(\frac{M}{M_{BH}}\right)^{3/4}.
\label{Eq:rcrh}
\end{equation}

\citet{2007MNRAS.374..857T} has suggested that the imprint of this
process can be seen in his ``isolated and relaxed'' sample of
simulated open clusters having relaxation times less than 1 Gyr, a
half-mass to tidal radius ratio $\rhm/\rt < 0.1$, and an orbital
ellipticity of less than 0.1.  Roughly half of the clusters in this
sample have core radii substantially larger than would be expected on
the basis of simple stellar dynamics and binary heating.  However,
\citet{2007MNRAS.379...93H} has argued that such anomalously large
core to half-mass ratios may also be explained by the presence of a
stellar-mass BH binaries heating the cores of these clusters
\citep[see
also][]{2004ApJ...608L..25M,2007MNRAS.379L..40M,2008MNRAS.386...65M}.
Many of the results discussed above and much of our physical
understanding of the dynamical evolution of star clusters have been
developed and calibrated by means of simulations.


\section{The survival of star clusters}
\label{Sect:SCSurvival}

The realization that the majority of star formation occurs in embedded
clusters, whereas only a small fraction of stars in the Galactic disk
currently reside in clusters (see \S\ref{Sect:Introduction}),
indicates that most clusters and associations are relatively short
lived; they dissolve on time scales comparable to the median age of
open clusters in the solar neighborhood \citep{2005A&A...440..403K},
which is about 250\,Myr (see \S\ref{Sect:ClusterProperties}).

Historically, studies of the lifetimes of star clusters have focused
on open clusters in the Milky Way.  The scarcity of open clusters with
ages $\apgt 1$~Gyr was reported independently in several studies
\citep{1958ZA.....44..221V,1957ApJ...125..445V,1958MNRAS.118..379O},
and has been attributed to their short median lifetimes \citep[about
  250 Myr;][]{1971A&A....13..309W}, rather than, say, a variation in
the formation history or a detection bias toward young objects.  These
short cluster lifetimes have been explained as due to the destructive
effects of encounters with giant molecular clouds (GMCs)
\citep{1958ApJ...127...17S}.  A typical Galactic cluster with a
density of $\sim1\,\msunpc$ can survive the heating due to passing
GMCs for about 250\,Myr.  The remarkable agreement between the
inferred mean lifetime and the expected survival time in the Galactic
disk implicated GMCs as responsible for the destruction of open
clusters (see \S\ref{Sect:ExternalPerturbations}). The argument is
further supported by the radial offset of the old (few Gyrs) open
clusters toward the anticenter of the Galactic disk and away from the
plane of the disk, where the density of GMCs is low
\citep{1980A&A....88..360V}, as is illustrated in
Fig.~\ref{Fig:distribution_open_globular_clusters}.  The galactic
bulge and spiral structure also contribute, though in lesser extend,
to the destruction of open clusters
\citep{1994AJ....108.1403W, 2007MNRAS.376..809G}.

By comparing the age distributions of clusters in the Magellanic
clouds with those in the Milky Way Galaxy, the former population is
found to be on average older and also more massive than the local
population \citep{1985ApJ...299..211E,1987PASP...99..724H}.  The
higher average cluster mass in the sample of Magellanic cloud clusters
is a consequence of the difficulty in detecting low mass clusters.
The apparent longer lifetimes of the Magellanic clusters could imply
that more massive clusters tend to live longer, although the longer
lifetimes could also be explained by the lower density of GMCs, the
absence of bulges and spiral structures and the overall weaker tidal
fields in the Magellanic clouds.  However, in
\S\ref{Sect:Observational.constraints} we argue that GMCs are unlikely
to play an important role in the early evolution of a YMC, due to
their high initial density. The mechanism leading to the destruction
of star clusters is therefore of major importance for understanding
the evolution of star clusters from youth to old age.

\subsection{Simulating star clusters}
\label{Sec:Simulations}

The formation, evolution and death of star clusters is a complex
problem combining stellar dynamics, gas dynamics, stellar evolution
and the evolution of the potential of the parent galaxy, all of which
contribute to the cluster's appearance over the entire lifetime of the
cluster (see also \S\ref{Sec:Dynamics}).  Over the past decade,
significant progress has been made in modeling many of these processes
simultaneously in numerical simulations of clusters during phase 2 and
phase 3.  A striking omission is the self-consistent treatment of the
interaction between stars and gas during phase 1.  We focus here on
simulations of phase-2 and phase-3 clusters, first describing
treatments of stellar dynamics (see below), then turning to the
inclusion of other physical processes into the mix.

A broad spectrum of numerical methodologies is available for
simulating the dynamical evolution of young star clusters.  In
approximate order of increasing algorithmic and physical complexity,
but not necessarily in increasing numerical complexity, the various
methods may be summarized as follows.

\begin{itemize}
\item {\em Static Models} are self-consistent potential--density pairs
  for specific choices of phase-space distribution functions
  \citep{1911MNRAS..71..460P,1966AJ.....71...64K,2008gady.book.....B}.
  They have been instrumental in furthering our understanding of
  cluster structure, and provide a framework for semi-analytical
  treatments of cluster dynamics.  However, they do not lend
  themselves to detailed study of star cluster evolution, and we will
  not discuss them further here, instead referring the reader to the
  discussion in \S\ref{Sect:TheRadius}, or to
  \citep{1987degc.book.....S}.

\item {\em ``Continuum'' Models} treat the cluster as a quasi-static
  continuous fluid whose phase-space distribution function evolves
  under the influence of two-body relaxation and other energy sources
  (such as binary heating) that operate on relaxation time scales (see
  Eq.\,\ref{Eq:Trlx}).

\item {\em Monte Carlo Models} treat some or all components of the
  cluster as pseudo-particles whose statistical properties represent
  the continuum properties of the system, and whose randomly chosen
  interactions model relaxation and other processes driving the
  long-term evolution.

\item {\em Direct $N$-body Models} follow the individual orbits of all
  stars in the system, automatically including dynamical and
  relaxation processes, and modeling other physical processes on a
  star-by-star basis.
\end{itemize}

Much of our current understanding of the evolution of star clusters
comes from detailed numerical simulations, and the above techniques
are used for the vast majority if simulations. In order to appreciate
some of the details presented in \S\ref{sec:TheFirstMyr} to
\S\ref{Sect:Observational.constraints} it may be important to have
some rudimentary understanding of this range of techniques.  However,
since this is a rather technical discussion, we present it in an
appendix (\S\ref{Sec:AppendixA}).

\subsection{Early violent gas expulsion}
\label{sec:TheFirstMyr}
Possibly the greatest discrepancy between star cluster simulations and
observations lies in the first few million years of the evolution
(phase 1 in \S\ref{Sec:Dynamics}).  Real star clusters are formed in a
complicated interaction between gas and gravity, which is imperfectly
understood. Once a primordial gas cloud starts to condense into stars
dynamical evolution also begins \citep{2003MNRAS.339..577B}. At the
end of the star formation process, probably brought about by the
developing winds of the most massive stars or the first supernovae,
the residual gas is ejected from the protostellar cluster.  The gas
expulsion phase is expected to be short---on the order of several
dynamical times (Eq.\,\ref{eq:dynamical})---and places the remaining
stellar population in a super-virial state, making the young cluster
vulnerable to dissolution. The sharp decrease in the number of young
and embedded star clusters at an age of a few megayears is thought to
be a consequence of this early process, and is often referred to as
``infant mortality'' \citep{2003ARA&A..41...57L}.

\subsubsection{Theoretical considerations}\label{sect:TheFirstMyr.Theory}

The total mass $\mtot = \mg + \ms$ of the primordial cluster contains
contributions from stars $\ms$ and gas $\mg$.  In virial equilibrium,
the rms velocity of the cluster is
\begin{equation}
  \vrmssq = {G\mtot \over 2\rvir}.
\label{eq:vrms}
\end{equation}
which in observable quantities becomes
\begin{equation}
  \sigmaoned^2 = {G\mtot \over \eta\reff},
\label{eq:sigmaoned}
\end{equation}
with $\eta\approx10$ (see \S\ref{Sect:TheRadius}).

The response of the cluster to the loss of the residual gas depends on
the gas expulsion time scale {\trem} relative to the dynamical time
scale {\tdyn} of the cluster (see Eq.\,\ref{Eq:tdyn}).  For many young
clusters $\trem \ll \tdyn$, in which case removing the gas shocks the
cluster.  In the extreme case, the positions and velocities of the
stars remains fixed during the gas expulsion phase, and the response
of the cluster to losing a fraction of gas, $f_e \equiv \ms/\mtot$,
can be calculated under the assumption that the mass loss is
instantaneous.  The rms velocity of the stars immediately before and
after gas loss is then given by Eq.~\ref{eq:vrms}; the stellar
positions are also unchanged.  As a consequence, the cluster expands
and re-establishes equilibrium at a new virial radius given by
\citep{1980ApJ...235..986H}
\begin{equation}
  {\rvir \over \rvir(t=0)} = {f_e \over 2 f_e-1}.
\label{eq:expimp}
\end{equation}
For high star formation efficiency ($f_e \apgt 0.5$) the arguments
leading to Eq.\,\ref{eq:expimp} may be reasonable, as in this case a
relatively small fraction of the stars is lost after the gas is
expelled.  However, the energy argument is too simple to determine the
survival probability if $f_e \aplt 0.5$.  Losing more than half of the
total mass ($f_e \le 0.5$) by explosive gas expulsion is devastating
for the cluster, leading to its complete dissolution in a few
dynamical time scales.  A low star formation efficiency may explain
the majority of disrupted young and embedded clusters, but even for
$f_e \aplt 0.5$ some small portion of the cluster can in practice
remain bound.  There are several independent arguments for the
survival of embedded clusters even for a star formation efficiency as
low as $f_e \aplt 0.1$.

The most important argument against the above simple analysis
(Eq.\,\ref{eq:expimp}) is the fact that the time scale for gas
expulsion, {\trem}, in practice is short but not instantaneous.  This
time scale should not be compared to the (global) half-mass crossing
time, but rather to the local dynamical time, which depends strongly
on the distance to the cluster center. For example, the dynamical time
scale in the cluster core has $t_{\rm core}/\tdyn = (\rho_{\rm
  hm}/\rho_{\rm c})^{1/2}$ and this fraction ranges from $\aplt
0.01$ for a concentrated cluster ($W_0 \apgt 12$, or $c > 2.7$) to
$\sim 0.2$ for a shallow potential ($c < 1$ or $W_0 \aplt 5$).  For
concentrated clusters ($c \apgt 2$), the gas expulsion occurs more or
less instantaneously for stars in the outskirts, but stars in the core
respond adiabatically to the loss of gas.  If $\trem > \tdyn$, the
cluster is likely to survive; in particular, the core may respond with
just a slight expansion (of at most a factor of 2), even in extreme
cases, $f_e \aplt 0.1$ \citep{2001MNRAS.323..988G}.

Further complications arise from the clumping of the gas and stellar
distributions in the embedded cluster \citep{2009MNRAS.397..954F}.  It
is likely that the radial dependence of the star formation process
causes the central part of the cluster to be depleted in gas, whereas
the outskirts are relatively gas rich \citep{2003MNRAS.343..413B}.
This radial variation of the star formation efficiency, combined with
the process of competitive star formation \citep{2003MNRAS.343..413B},
may actually render the cluster sub-virial after the gas is ejected,
for example if stars formed in the collapsing cloud are dynamically
cold \citep{1984ApJ...285..141L}, as suggested by the outcome of
turbulent self-gravitating simulations of cluster formation
\citep{2009ApJ...704L.124O}.  In that case, Eq.\,\ref{eq:expimp}
depends on the fractional deviation from virial equilibrium $q_{\rm
  vir}$, to become $\rvir/\rvir(t=0)=f_e/(2f_e-q_{\rm vir}^2)$,
i.e.~the condition for complete disruption becomes $f_e=q_{\rm
  vir}^2/2$ \citep{2009ApJ...697.1020P,2009Ap&SS.tmp..108G}. The
cluster survival probability then depends on the entire star formation
process, not just on its overall efficiency
\citep{2006MNRAS.373..752G}.  A further deviation from the simple
formulation comes from the effect of high-velocity escapers, which can
carry away a considerable fraction of the cluster's binding energy,
leaving the remaining stars more strongly bound
\citep{2007MNRAS.380.1589B}.

Thus the survival probability of an embedded cluster cannot be
determined by a single parameter, such as the star formation
efficiency, as the complete formation process of stars, clumps of
stars, and the entire cluster comes into play. The process whereby
stars form in massive star clusters is still poorly explored terrain
within astrophysics, and the formation of sub-clumps and clusters is
even less well charted.

\subsubsection{Observational constraints}
\label{Sect:ObservationalConstraints}

What sounds convincing from a theoretical standpoint is often very
hard to support with observations.  There are a number of interesting
observational indications of infant mortality, and of the associated
dissolution time scales, but since the embedded phase is 
short ($\sim1-2$\,Myr), many parameters are poorly constrained by
observations, and the interpretations of models tend to be sensitive
to assumptions made about the initial conditions.

Many observed young clusters, particularly the extragalactic
population in Tab.\,\ref{Tab:Extra_Galactic_Clusters}, appear to be
super-virial, which naively would lead to the early termination of the
cluster's existence. This is most easily seen in the higher value of
their dynamical mass \mdyn\, compared to the photometric mass \mphot.
The former is derived from measurements of the velocity dispersion and
the radius of the cluster and the use of Eq.~\ref{Eq:Mvirial}.  The
latter is derived from the total luminosity, calibrated to single
stellar population models (see \S\ref{sec:observations}).
\citet{2006A&A...448..881B} determine $\mphot$ and $\mdyn$ from a
compilation of 19 clusters and find that both independent mass
estimates are consistent for the somewhat older ($\apgt 50-100$\,Myr)
clusters, but for young ($\sim10\,$Myr) star clusters they find $\mdyn
> \mphot$.  \citet{2006MNRAS.373..752G} explain this as a signature of
the primordial gas expulsion, and of the process of infant mortality.

\begin{figure}[!t]
\begin{center}
\begin{tabular}{cc}
  \psfig{figure=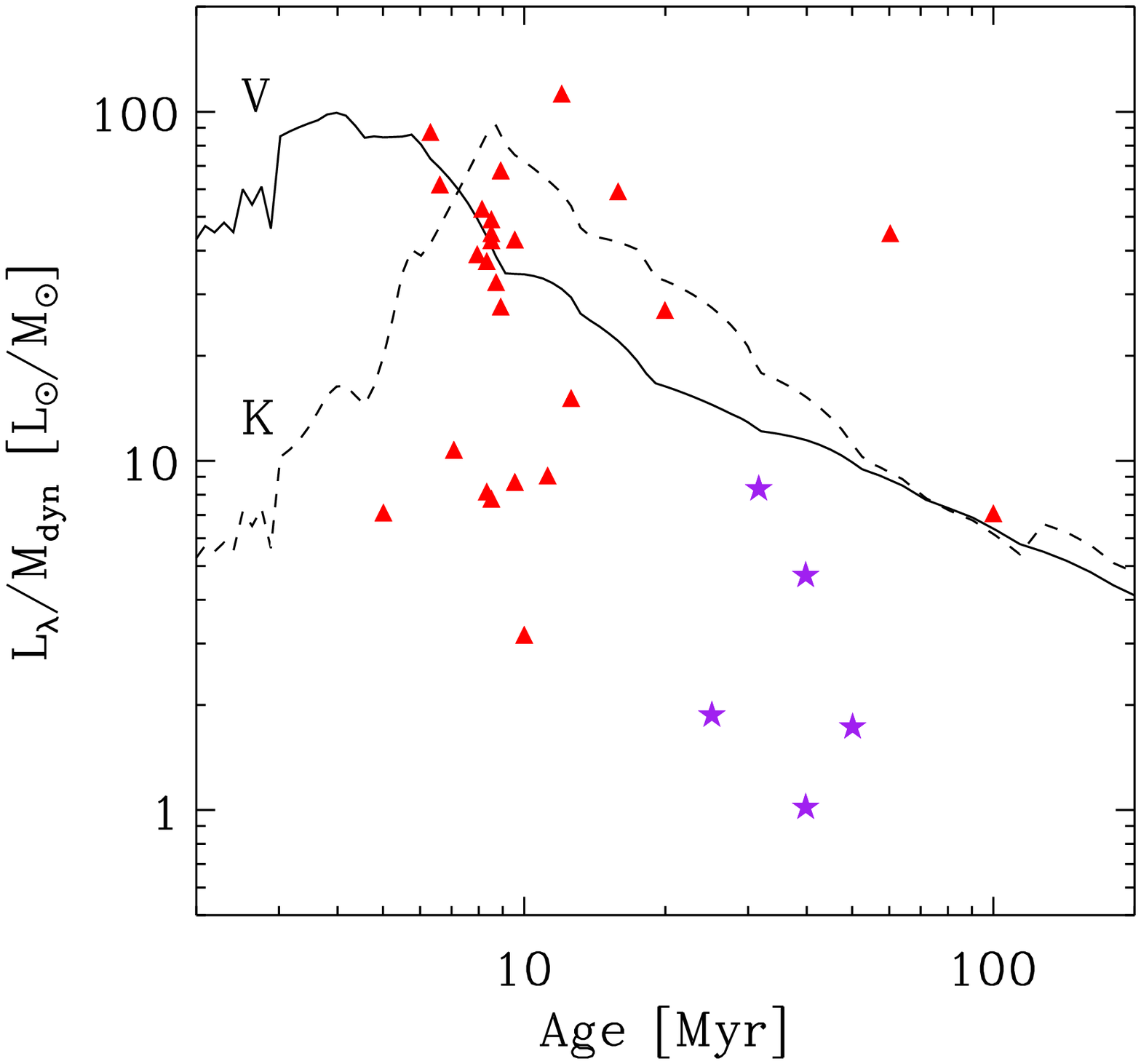,width=0.5\columnwidth}
  \psfig{figure=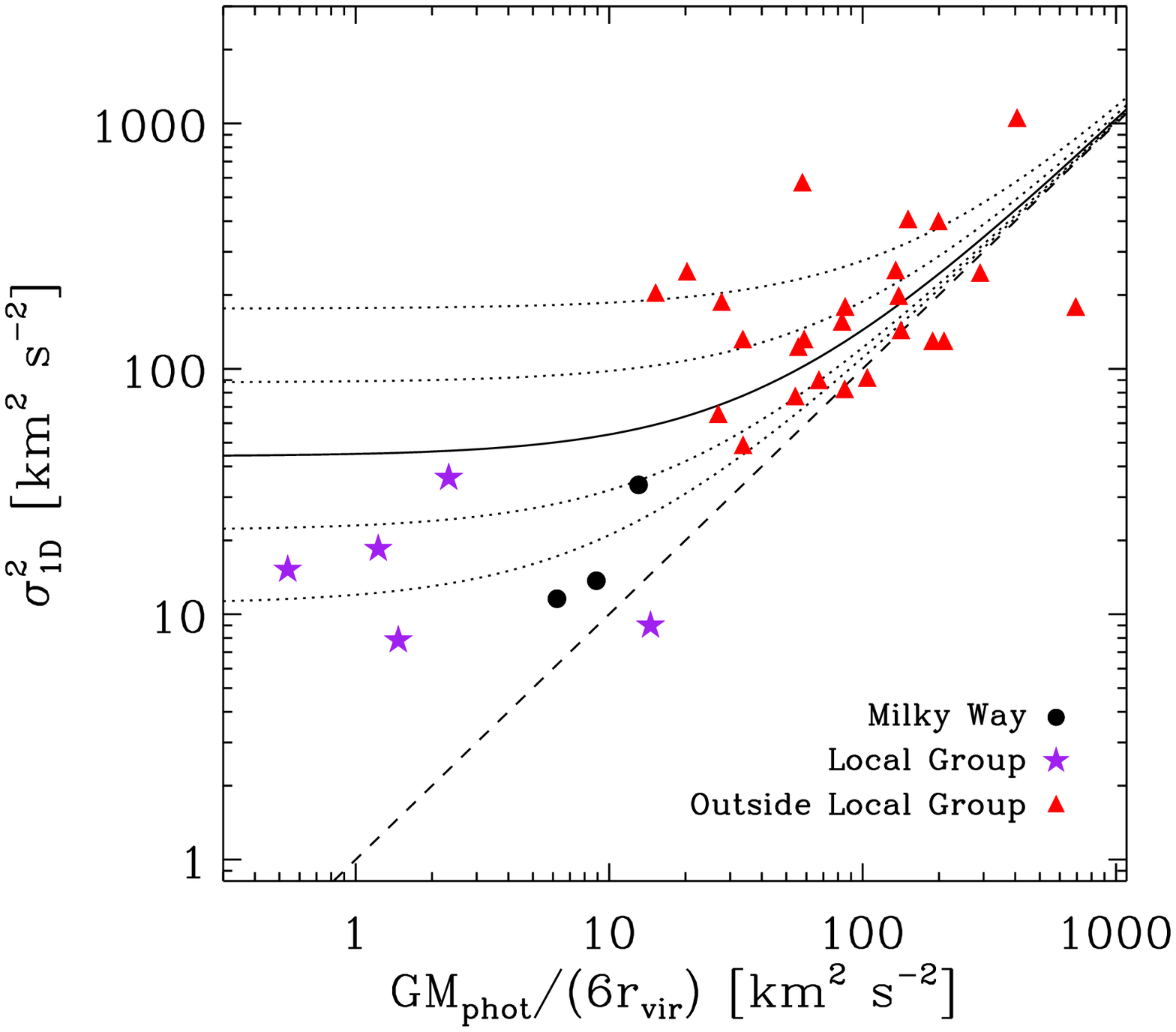,width=0.5\columnwidth}
\end{tabular}
\end{center}
 \caption{Left: Light to dynamical mass ratio for the clusters in
   Table~\ref{Tab:Extra_Galactic_Clusters} that have both quantities
   available.  Photometric evolution from the
   \citet{2003MNRAS.344.1000B} single stellar population models, using
   a Chabrier IMF, in the $V$ and $K$ bands is indicated by the full
   and dashed lines, respectively.
   Right: The measured velocity dispersion squared, presented as a
   function of a prediction for this quantity based on
   Eq.~\ref{Eq:Mvirial}.  The solid line is a prediction of the effect
   of binaries on $\mdyn$, with 1-$\sigma$ and 2-$\sigma$ variations
   due to stochastic fluctuations shown as dashed lines based on the
   results of \citet{2009arXiv0911.1557G}.
   }
   \label{Fig:light_to_mass}
\end{figure}

In Fig.~\ref{Fig:light_to_mass} we present an updated version of
Fig.~5 of \citet{2006A&A...448..881B}, showing the ratio of light to
dynamical mass of 24 clusters, taken from
Table~\ref{Tab:Extra_Galactic_Clusters}, each of which is about
10\,megayear old (see \S\ref{sec:observations}).  The age range is
quite narrow because the red supergiant phase, which starts around
10\,Myr, makes these clusters extremely bright (especially in the near
infrared), whereas younger clusters are still heavily obscured.

Many of the clusters in Fig.~\ref{Fig:light_to_mass} and
Tab.\,\ref{Tab:Extra_Galactic_Clusters} appear to have dynamical
masses too high for their luminosities, and the intuitive explanation
is that these clusters are expanding and are possibly even unbound
\citep{2006MNRAS.373..752G}.  The time needed for a cluster to
completely dissolve, or to find a new virial equilibrium after gas
expulsion, is about $\sim20-30\,\tdyn$ \citep[see for example Fig.~8
  in][]{2007MNRAS.380.1589B}, where $\tdyn$ is the initial dynamical
time, when the gas and the stars are still bound. As a result, to
`catch' an unbound or expanding cluster at 10 Myr, the initial $\tdyn$
should be $\gtrsim0.5\,$Myr. This corresponds to an half-mass density
of $\rhohm\lesssim300\,\msunpc$ (Eq.~\ref{eq:dynamical} and
$\rhohm\equiv3\mtot/8\pi\rh^3$). Clusters with shorter initial $\tdyn$
(higher density, see Eq.\,\ref{eq:dynamical}) have expanded into the
field, or found a new equilibrium a few megayears after gas expulsion,
and are not observable as super-virial clusters at
10\,Myr.\footnote{The models of \citet{2006MNRAS.373..752G} start with
  a density in the embedded phase of
  $\sim60\,\msunpc$($\tdyn\approx1\,$Myr), which is why their clusters
  are super-virial for at least 25\,Myr.}

The initial density of the clusters in Fig.~\ref{Fig:light_to_mass} is
unknown, but is likely to have been higher in the past that it is
today. In addition we may attemp to estimated their initial densities
from their current densities.  In Fig.~\ref{Fig:mass_radius_relation}
we show the radii and masses of the clusters under discussion,
together with lines of constant $\rhohm$. The present day densities of
the clusters younger than 10\,Myr are
$\rhohm\approx10^{3\pm1}\,\msunpc$. The densities in the embedded
phase were $O(1/f_e^4)$ higher --- where the reduction in cluster mass
accounts for a factor $1/f_e$ and the consequent adiabatic expansion a
factor $1/f_e^3$ (see~Eq.~\ref{Eq:adiabatic_expansion}).

The initial dynamical times were therefore a factor of $1/f_e^2$
shorter than {\tdyn} after phase 1.  Based on their physical ages of
$\sim 10\,$\,megayears and the fact that the gas ejection phase does
not last beyond the moment of the first supernova (within $\sim
3$\,Myr, see \S\ref{Sec:Simulations}), these clusters have evolved for
at least $10/f_e^2$ to $100/f_e^2$ initial dynamical times, and
therefore must be bound
\citep{2003ApJ...596..240M,2007ApJ...663..844M}, and the observed
discrepancy between \mphot\, and \mdyn\, cannot originate from the
overall expansion of the cluster. 

\citet{2006MNRAS.369.1392F} showed that the constant $\eta$ that
relates $\sigmaoned$ and $\reff$ to $\mdyn$ (Eq.~\ref{eq:sigmaoned})
is up to a factor of three higher when the cluster is mass
segregated. This results in a value of $\mdyn$ that is too low
compared to the true mass when this is not taken into account.  So
mass segregation is not a plausible explanation for the large downward
spread of points in Fig.~\ref{Fig:light_to_mass}.

We are still left with the question why the observed velocity
dispersions in some of these clusters are higher than would be
expected from the virial theorem.  Several independent and implicit
assumptions enter the derivation of $\mdyn$ and $\mphot$, and each of
them could be wrong. The stellar mass function, for example, could be
bottom-heavy, i.e.~steeper than Salpeter or with an excess of low-mass
stars.  Such a mass function would result in a velocity dispersion in
virial equilibrium higher than that of a cluster with a Salpeter IMF,
with little effect on $\mphot$.  However, to explain the observed
discrepancy, the cluster mass function must deviate substantially from
the canonical mass functions.  We do not favor this conjecture, since
the only star cluster for which the mass function was once anticipated
to be deficient in low-mass stars, the Arches
\citep{2005ApJ...628L.113S}, turns out to have a rather normal mass
function at least down to 1\,\Msun\, \citep{2006ApJ...653L.113K}.

Another interesting possibility is provided by the binarity of red
supergiants, which dominate the observed luminosity.  A relatively
high fraction of hard binaries (see \S\ref{Sect:BinaryInteractions})
leads to an overestimate of the cluster velocity dispersion due to the
contribution from their internal orbital motion.  This leads to an
overestimate of \sigmaoned, and therefore of the mass of the cluster
\citep{2008A&A...480..103K}.  For typical open clusters, with
$\sigmaoned\approx1\,\kms$, this can only account for a factor of
$\sim2$ increase of $\mdyn$ \citep{2008A&A...480..103K}.  However,
young star clusters are dominated by $\sim 13-22\,\msun$ red
supergiants, and a binary fraction of $\sim$25\% among these stars
could explain an apparent dynamical mass of up to an order of
magnitude more than the photometric mass \citep{2009arXiv0911.1557G}.
As a consequence the discrepancy between $\mdyn$ and $\mphot$ is
larger for clusters with high stellar velocity dispersion, or with a
small ratios of $\mphot/\reff$ (see Eq.~\ref{Eq:Mvirial}), which is
consistent with the observations.  The effect of binaries on
$\sigmaoned$ and the ratio of \mdyn/{\mphot} is presented in
Fig.~\ref{Fig:light_to_mass}.

Star clusters with $\rhohm\gtrsim100\,\msunpc$ at $10\,$Myr, such as
those listed in Tab.\,\ref{Tab:Extra_Galactic_Clusters} and shown in
Fig.~\ref{Fig:light_to_mass}, have survived the primordial gas
expulsion phase and should be considered bound, stable, and likely to
survive for a long time (see Eq.\,\ref{eq:tdis3}).  They enter the
next evolutionary phase, described in \S\ref{ssec:StellarMassLoss},
during which stellar mass loss dominates the evolution of the cluster
(phase 2).

\subsection{Stellar mass loss}
\label{ssec:StellarMassLoss}

Clusters that survive phase~1 (the embedded phase,
\S\ref{sec:TheFirstMyr}) continue to lose mass through stellar
evolution.  During phase~2, the most massive ($\apgt 50$\,\msun) stars
leave the main-sequence within $\aplt 4.0$\,Myr, and lose about 90\%
of their mass by the time they collapse to a black hole following a
supernova.  A $5\msun$ star loses $\apgt 80$\% of its mass by the time
its core forms a white dwarf at about 100\,Myr. For a
\citet{2001MNRAS.322..231K} IMF between $0.1\,\msun$ and $100\,\msun$
the total cluster mass decreases by roughly 10\%, 20\%, and 30\%
during the first 10, 100, and 500\,Myr.  The impact of this will be
discussed below.

\subsubsection{Theoretical considerations}
\label{sect:StellarMassLoss.Theory}

The time scale for mass loss depends on the mode in which it is
achieved; supernova explosions, Wolf-Rayet winds and AGB expulsion
result in high mass loss rates, whereas the general mass loss for an
older stellar population is relatively slow.  When the time scales for
mass loss by stellar evolution is considerably longer than $\tdyn$,
the cluster responds adiabatically, expanding through a series of
virial equilibria. For small $f_e$, Eq.~\ref{eq:expimp} reduces to
\begin{equation}
  \frac{\delta\rvir}{\rvir} = \frac{\delta M}{M},
  \label{Eq:infinitesimal}
\end{equation}
and therefore
\begin{equation}
  \frac{\rvir}{\rvir(t=0)} = \frac{M(t=0)}{M}.
  \label{Eq:adiabatic_expansion}
\end{equation}
Even losing half the mass by slow stellar evolution (which for the
canonical IMF would not occur within a Hubble time), the cluster would
expand by only a factor of two.  In the instantaneous approximation
(\S\ref{sec:TheFirstMyr} and Eq.\,\ref{eq:expimp}), such mass loss
would lead to the dissolution of the cluster.

In reality, the situation is more complicated, in particular because
of the connection between dynamical evolution and stellar mass loss.
For real clusters the expansion due to stellar mass loss is
considerably more severe than suggested above, and can even result in
complete disruption if the cluster is mass segregated before the bulk
of the stellar evolution takes place \citep{2009ApJ...698..615V}.
Even an initially unsegregated cluster can still undergo mass
segregation during the period when the residual gas is being ejected,
and certainly during the early evolution of its stars
\citep{1986ApJ...301..132A}, which can also lead to enhanced expansion
at later times.  The expansion of a mass segregated cluster, however,
will not be homologous, as the massive (segregated) core stellar
population tends to lose relatively more mass than the lower-mass halo
stars. The result is a more dramatic expansion of the cluster core,
with less severe effects farther out.

\begin{figure}[!t]
\begin{center}
 \hspace{0.0\columnwidth}
   \psfig{figure=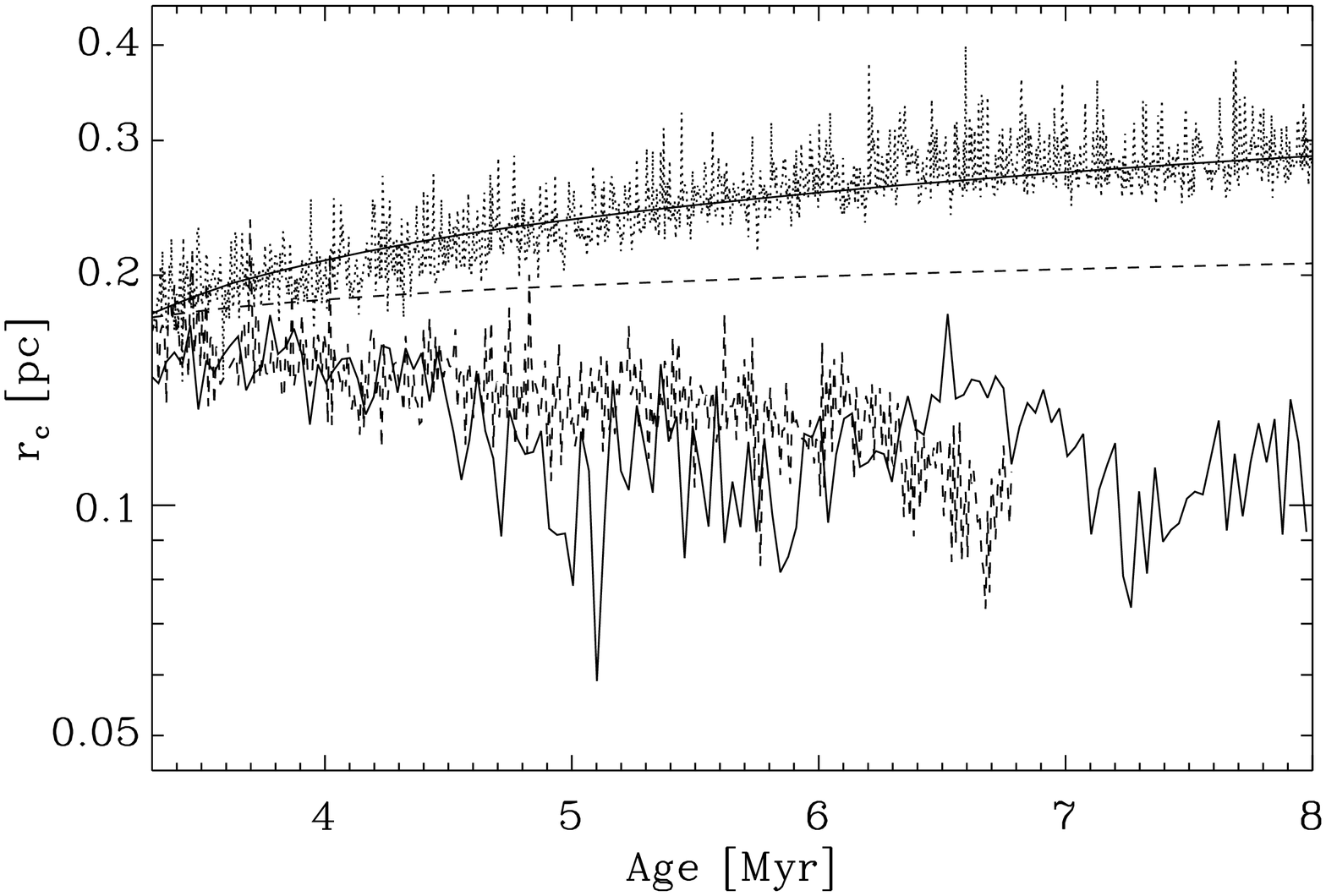,width=0.7\columnwidth}
\end{center}
   \caption[]{Evolution of the core radius during phase~2 of an
     $N$-body simulation \citep[$N=128k$, $\rvir = 3.2$\,pc, King
       $W_0=12$, Mass function is Salpeter between 1\,\msun\, and
       100\,\msun][]{2007MNRAS.374...95P}.  After about 3\,Myr stellar
     mass loss dominates the evolution of the cluster core radius.
     The dotted curve (top curve) is the result of the full
     simulation, with both stellar evolution and binary dynamics
     included; the irregular dashed curve is calculated with the same
     initial realization but without stellar evolution.  The wiggely
     solid curve (bottom) is calculated without stellar mass loss and
     without binary dynamics by collapsing the binaries in single
     objects. The smooth dashed line shows the expected expansion of
     the core, assuming adiabatic mass loss for a Salpeter initial
     mass.  The smooth solid curve is computed assuming mass loss by
     stellar evolution from a Salpeter mass function with a lower
     limit of 15 \msun, rather than the 0.1\,{\msun} used in the
     simulation.  }
   \label{fig:APhaseB}
\end{figure}

This effect is illustrated in Fig.\,\ref{fig:APhaseB}, which presents
the results of an isolated $N=128k$ body simulation, run on a GRAPE-4
\citep{1998sssp.book.....M} using the {\tt starlab} software
environment \citep{2001MNRAS.321..199P}.  The figure shows the
evolution of the core radius with and without stellar evolution.
Without stellar evolution the core tends to shrink, and eventually
reaches core collapse, whereas with stellar mass loss the core expands
\citep{2007MNRAS.374...95P}.

The combined effects of mass loss by stellar evolution and dynamical
evolution in the tidal field of the host galaxy have been extensively
studied
\citep{1995MNRAS.276..206F,2000ApJ...535..759T,2003MNRAS.340..227B}.
These studies show that when clusters expand to a radius of
$\sim$$0.5\,\rj$ they lose equilibrium and most of their stars
overflow $\rj$ (see \S\ref{Sect:TheRadius}) in a few crossing times.

We conclude that in phase~2 the overall evolution of the cluster is
completely dominated by expansion due to stellar mass loss.  However,
since most of the mass loss comes from massive stars in the core, the
core expansion is considerably larger than expected for the global
mass function.  Phase~2 lasts until the response of the cluster to
stellar mass loss deminishes and from that moment the cluster can
continue to expand until it completely dissolves or until the cluster
core starts to contract again due to internal dynamical effects, at
which point phase~3 begins.

\subsubsection{Observational constraints}
\label{sect:StellarMassLoss.Observations}

\begin{figure}[!t]
\begin{center}
 \hspace{0.0\columnwidth}
 \psfig{figure=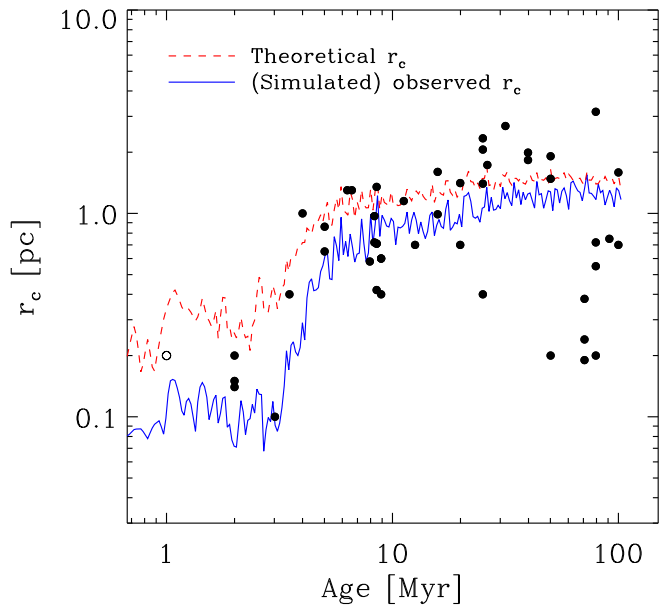,width=0.6\columnwidth} 
\end{center}
\caption{Evolution of observed core radii values of clusters, compared
  to the results of an $N$-body simulation including mass loss due to
  stellar evolution. For the simulation, the projected surface brightness
  profile was constructed and $\rcore$ was determined using
  and Eq.~\ref{Eq:EFF87} and Eq.~\ref{Eq:rcore} (full line). The dashed line shows the theoretical
  $\rcore$ from Eq.~\ref{Eq:theoryrcore2} discussed in \S\ref{Sect:TheRadius}.}
 \label{Fig:rc_nbody}
\end{figure}

Rapid dissolution due to stellar evolution mass loss occurs when
$\rhm/\rj \apgt 0.5$.  All the clusters in
Fig.~\ref{Fig:mass_radius_relation} have, within a factor of two,
$\rhm/\rj\approx0.03$, suggesting that they are probably all stable
against stellar mass loss.  \citet{2009A&A...498L..37P} recognizes two
evolutionary sequences in young Galactic star clusters, from which she
draws a similar conclusion.  The first sequence of dense ``starburst
clusters'', containing the Arches cluster, NGC~3603 and Trumpler 14,
starts at a density of $\sim10^5\,\msunpc$ at an age of a few
megayears. These clusters appear to expand at constant in mass, up to
an age of 10--20\,Myr, where we find the red supergiant clusters
RSGC01 and RSGC02. At that age the cluster density has dropped to
$\sim10^3\,\msunpc$.  The second sequence of ``leaky clusters'' (which
is identical to our definition of associations, see
\S\ref{Subsec:Populations}) starts at the same age but with much
lower densities of $\sim10\,\msunpc$, and expand while $M \propto
1/\rhm$ to densities comparable to the field star density.  The
clusters in Fig.~\ref{Fig:mass_radius_relation} may be compared with
the red supergiant clusters in the Milky Way, i.e.~the end point of
the dense cluster sequence. The associations discussed by \citet[][she
  refers to them as ``leaky'' clusters]{2009A&A...498L..37P} and
listed in Table~\ref{Tab:Galactic_Clusters} have dynamical times that
exceed the cluster age, and are expected to be unstable against mass
loss by stellar evolution.

In \S\ref{sec:observations} we discussed the remarkable increase in
the observed core radii of clusters with ages 1--50\,Myr
\citep{2003MNRAS.338...85M,2008MNRAS.389..223B}. Earlier studies
argued that this effect was the result of early gas expulsion
(phase~1), but as discussed in \S\ref{sec:observations} the time scale
for finding a new virial equilibrium after gas expulsion is too short
to affect the growth of {\rcore} over such a long period.  Mass loss
from the young stellar population (phase~2) can contribute to some
extent to the observed trend, but its effect is probably too weak to
explain it completely \citep{2007MNRAS.379L..40M}.  A more likely
solution is dynamical heating from relatively massive objects, such as
massive stars and stellar mass black holes, sinking to the cluster
center \citep{2004ApJ...608L..25M}.

An additional effect not yet discussed is the difference between the
observed $\rcore$, usually resulting from a fit to the surface
brightness profile (Eq.~\ref{Eq:EFF87}), and the 3D dynamical core
radius described in \S\ref{Sect:TheRadius}. If a young star cluster is
mass segregated, possibly already from its formation process, then the
light is dominated by the few massive stars in the core, which can
lead to an underestimate of $\rcore$
\citep{2005sf2a.conf..605F,2008MNRAS.391..190G}. \citet{2007MNRAS.379L..40M,
  2008MNRAS.386...65M} showed that when taking this into account, a
remarkable increase of $\rcore$ is ``observed," while the ``real" core
radius changes less.

We illustrate this in Fig.~\ref{Fig:rc_nbody}. The same data points as
in Fig.~\ref{Fig:Radius_Evolution} are shown, together with the
results of an $N$-body simulation (lines), where the theoretical value
of $\rcore$, as defined in \S\ref{Sect:Introduction} and the value of
$\rcore$ resulting from a fit of the EFF87 profile
(Eq.~\ref{Eq:EFF87}) to the projected light of the simulated cluster.
The cluster consists of $N=65\,536$ single stars initially distributed
according to a \citet{1911MNRAS..71..460P} profile with
$\rvir=2\,$pc. Before stellar evolution was turned on, we evolved the
cluster for 100 $\tdyn$, which corresponds to $\sim0.1\,\trlx$, to
mimic some degree of primordial mass segregation, in which stars with
masses $\gtrsim5\,\msun$ are more centrally concentrated than the less
massive stars (Eq.~\ref{eq:ts}). Since the massive stars dominate the
light, the observed $\rcore$ is almost a factor of three smaller than
the 3D version at $t=0$. The observed $\rcore$ increase by nearly a
factor of ten in a few tens of megayears, while the 3D core radius
expands only by a factor of three. After $\sim 30$\,Myr the two
quantities roughly agree. In the simulation $\rhm$ increased only by
60\% due to the stellar evolution mass loss.

Primordial mass segregation has a profound effect on the evolution of
a star cluster, but possibly much more relevant for this review is its
consequences for cluster observations. The assumption that a young
($\aplt 10$\,Myr) cluster is not mass segregated, when in reality it
is, can dramatically alter observationally derived quantities, such as
the cluster size, velocity dispersion, density profile, and central
density.

\subsection{External perturbations and evaporation}
\label{Sect:ExternalPerturbations}

\subsubsection{Theoretical considerations}\label{Sect:Perturbations.theory}

One important external disruptive factor, first considered by
\citet[][see \S\ref{sec:TheFirstMyr}]{1958ApJ...127...17S}, is
encounters between clusters and giant molecular clouds. Since GMCs are
typically more massive than clusters, the cluster is more affected by
an encounter than the cloud \citep{1991MmSAI..62..909T}. The cluster
lifetime due to heating by passing clouds is inversely proportional to
the volume density of molecular gas, $\rhogmc$, and proportional to
the density of the cluster:
\begin{equation}
\tdisgmc\approx1\,{\rm Gyr}
                \left(\frac{0.03\,\msunpc}{\rhogmc}\right)
                \left(\frac{\rhohm}{10\,\msunpc}\right).
\label{eq:gmc}
\end{equation}
This result is typical for disruption by external tidal perturbations
operating on short time scales ($\aplt \tdyn$), also known as tidal
shocks and can also be applied to passages through the disc
\citep[e.g.][]{1972ApJ...176L..51O}, bulge
\citep[e.g.][]{1997ApJ...474..223G} and spiral arms
\citep{2007MNRAS.376..809G}.  Here $0.03\,\msunpc$ is the molecular
gas density in the solar neighborhood and the constant is taken from
\citet{2006MNRAS.371..793G}, which is an update from the seminal
result by \citet{1958ApJ...127...17S}.  The dependence of $\tdis$ on
$\rhogmc$ indicates that the lifetimes of star clusters scales roughly
inversely with the observable surface density of molecular gas,
$\sigmagas$.

This result enables us to make order of magnitude estimates of the
lifetimes of clusters in other galaxies.  In spiral galaxies, GMC
encounters are especially frequent during the early stages of cluster
evolution, since clusters form in the thin gaseous disk where
$\rhogmc$ is high.  Older clusters are typically associated with the
thick disk, where $\rhogmc$ is low and GMC encounters are less
frequent.  Since young ($\lesssim$1\,Gyr) clusters in spiral galaxies
have only a small range in radii \citep[e.g.][]{2004A&A...416..537L},
more massive clusters tend to have higher densities, making them less
vulnerable to encounters with GMCs, which explains their longer
lifetimes compared to their lower-mass counterparts
\citep{2006MNRAS.371..793G}.  It is not clear whether the lack of a
mass-radius relation is a universal property imprinted at cluster
formation, or the result of evolution.

\subsubsection{Observational constraints}
\label{Sect:Observational.constraints}

If GMC encounters are a dominant disruption process, then it is
important to understand the mass-radius correlation of clusters, since
$\tdisgmc\propto\rhohm$ (\S\ref{Sect:Perturbations.theory}).  If
clusters form with a constant density, i.e. a mass-radius relation of
the form $\rhm\propto M^{1/3}$, then their destruction time due to GMC
encounters is independent of cluster mass.  Additional complications
arise from the time dependence of a mass-radius relation.  Clusters
older than those shown in Fig.~\ref{Fig:mass_radius_relation} For
young clusters ($\lesssim10\,$Myr) there seems to be some positive
correlation between mass and radius, roughly consistent with a density
of $10^{3\pm1}\,\msunpc$. Older clusters ($\gtrsim10\,$Myr) do not
seem to exhibit any correlation between mass and radius, consistent
with a near-constant radius, i.e. $\rhm =$~constant, which could be
the consequence of evolutionary effects, such as mass segregation and
stellar mass loss.  This would lead to a destruction time scale by
GMCs ($\tdisgmc$) that is dependent of the cluster mass:
$\tdisgmc\propto M/r^3_{\rm hm}$. In any case, the youngest YMCs
listed in Tab.\,\ref{Tab:Extra_Galactic_Clusters} are unlikely to be
rapidly destroyed by passing GMCs, since $\tdisgmc$ exceeds a Hubble
time due to their high densities (see left panel of
Fig.~\ref{Fig:mass_radius_relation}).

While considering mass loss from star clusters it is convenient to
distinguish between two fundamental processes: evaporation and tidal
stripping.  Evaporation is the steady loss of stars from the cluster
driven by the continuous repopulation by relaxation of the
high-velocity tail of the Maxwellian velocity distribution \citep[see
\S\ref{sect:Relaxation.Theory}, Eq.\,\ref{Eq:Trlx} and,
e.g.][]{amb38, 1940MNRAS.100..396S}.  This process has been the
subject of numerous comprehensive numerical studies
\citep{1987degc.book.....S,2003gnbs.book.....A,
  2003gmbp.book.....H,2003MNRAS.340..227B}.  Tidal stripping is the
prompt removal of stars that find themselves outside the cluster
Jacobi radius ($\rJ$, see \ref{Sect:TheRadius}) due to internal
processes such as stellar mass loss or a change in the external tidal
field, for example as the cluster approaches pericenter in its orbit
around its parent galaxy.  On a $\sim$100 Myr time scale, and for
clusters with masses $\apgt 10^4$\,{\Msun}, relaxation is unlikely to
be important, and so tidal stripping dominates the cluster mass loss.

\begin{figure}[!t]
\begin{center}
 \hspace{0.0\columnwidth}
 \psfig{figure=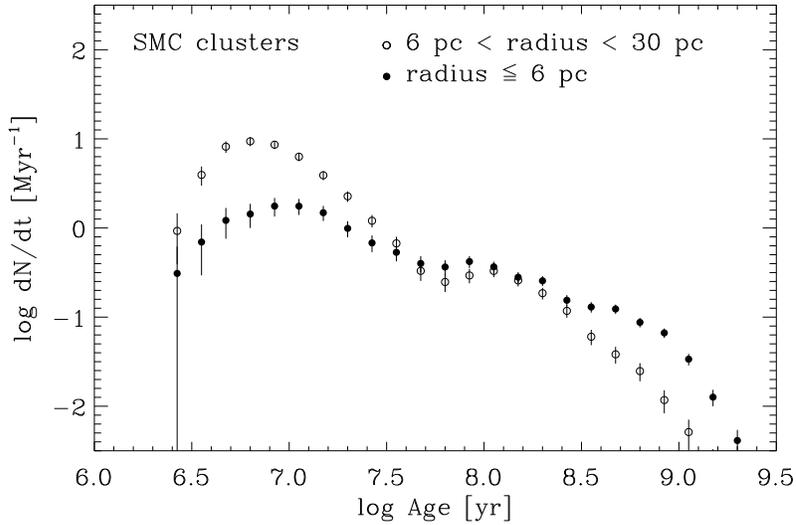,width=0.7\columnwidth} 
\end{center}
 \caption{Age distribution of SMC clusters based on the catalog of
   \citet{2006A&A...452..179C}. The sample is split into small and
   large clusters/associations, with the boundary at a radius of 6 pc.
   The histograms are made using a 0.5 dex bin width with different
   starting values (boxcar averaging).}
   \label{Fig:age_distribution_smc}
\end{figure}

Until the 1990s, the main targets of cluster disruption studies were
the open clusters in the Milky Way and the YMCs in the Magellanic
clouds.  Since then, HST observations have established the properties
of large populations containing more massive clusters in quiescent
spiral galaxies \citep[e.g.][]{2002AJ....124.1393L}, interacting
galaxies \citep[e.g.][]{1999AJ....118.1551W, 2005A&A...431..905B}, and
merger products \citep[e.g.][]{1997AJ....114.2381M}.  The primary tool
used in studies of cluster disruption is the cluster age
distribution. Different groups give different weights to the various
factors described in \S\ref{sec:TheFirstMyr} in interpreting the
results, but empirical cluster disruption studies follow one of two
basic models, in which the disruption is either {\em environment
  dependent} or {\it universal}.

In the environment dependent disruption model, the dissolution time
follows a simple scaling relation with cluster mass and environment,
following the age and mass distributions of luminosity-limited cluster
samples in different galaxies \citep{2003MNRAS.338..717B}. Variations
in dissolution time scales are explained by differences in the tidal
field strength \citep{2005A&A...429..173L} and the GMC density
\citep{2006MNRAS.371..793G}.

The universal disruption model assumes that internal processes
dominate the cluster disruption and that roughly 80--90\% of all
remaining clusters are destroyed during each decade in age, resulting
in a (mass limited) age distribution that declines inversely with time
($\propto t^{-1}$).  The main assumptions are that the majority
($80-90\%$) of clusters dissolve within a few hundred megayears and
that the disruption rate does not depend on mass.  This model however,
is calibrated against the cluster population in the Antennae galaxies,
as discussed in \S\ref{sec:TheFirstMyr}
\citep{2005ApJ...631L.133F,2007AJ....133.1067W}, and while it is quite
consistent with these observations, it is not clear how applicable it
is to other galactic environments.

The various cluster dissolution models have led to some controversy,
resulting in a number of spirited discussions at conferences and in
the literature. \citet{2006ApJ...650L.111C} demonstrated that the age
distribution of SMC clusters declines $\propto t^{-0.85}$, which is
consistent with their resuls for the Antennae.
\citet{2007ApJ...668..268G} are able to reproduce these results only
if they impose an incompleteness in the detection of clusters, which
is consistent with the arguments of \citet{2008MNRAS.383.1000D}.  As a
consequence \citet{2007ApJ...668..268G} conclude, using the same data
sample used by \citet{2006ApJ...650L.111C}, that the age distribution
of massive ($\apgt10^{3.5}\,\msun$) clusters in the LMC younger than a
few hundred megayears is not affected by internal processes,
contradicting the findings of \citet{2006ApJ...650L.111C}.

The conclusions of \citet{2007ApJ...668..268G} are supported by
\citet{2003MNRAS.338..717B}, who show that for a constant formation
rate of clusters and without disruption, the age distribution of a
luminosity-limited cluster sample declines $\propto
t^{-\zeta(\beta-1)}$, where $-\beta$ is the index of the cluster
initial mass function.  The parameter $\zeta$ describes how the
cluster fades with age, due to stellar evolution: $F_\lambda(t)\propto
t^{-\zeta}$, where $F_\lambda$ is the flux at wavelength $\lambda$ of
a cluster with constant mass.  For the U and V-bands $\zeta \approx
1.0$ and 0.7, respectively, which for $\beta = 2$ results in
$t^{-\kappa}$, with $0.7\lesssim\kappa\lesssim1.0$, due to fading
alone. If the distribution of cluster masses is described by a
Schechter function (see Eq.\,\ref{Eq:schechter} in \S\ref{Sect:CIMF})
the age distribution of a luminosity-limited sample that is not
affected by disruption is as steep as $0.9 \lesssim \kappa \lesssim
1.4$.

The internal disruption model for the age distribution relies on the
assumption that the cluster formation history is constant within the
age range considered (a few hundred megayears).
\citet{2009ApJ...701..607B} demonstrate that in the Antennae an
increase of the cluster formation rate as predicted by galaxy merger
models could alleviate the need to invoke longer duration cluster
disruption to explain the decline in the number of clusters with age.

The discussion of the distribution of the number of clusters with age
is complicated by the distinction between associations and dense
clusters (see \S\ref{sec:observations} and
\S\ref{Sect:ObservationalConstraints}), which may be hard to
distinghuish at large distances.  This is illustrated in
Fig.~\ref{Fig:age_distribution_smc}, which shows the age distribution
of clusters and associations in the central region of the SMC
\citep{2006A&A...452..179C}. The sample is divided into two subsamples
of $\sim 200$ clusters each, based on size.  The age distribution of
the large ($\reff > 6$pc) clusters falls off much more rapidly than
that of the more compact clusters. We note that this is consistent
with associations being disrupted by GMC encounters, since they have
(on average) much lower densities (see right panel of
Fig.~\ref{Fig:mass_radius_relation}) and are therefore destroyed much
quicker (Eq.~\ref{eq:gmc}). The median radii of the two samples are
4.5 pc and 9 pc, respectively, but at a distance of 20\,Mpc it would
be very difficult to tell these two groups apart.  We suggest that by
relaxing the cluster size in the observationally selected sample one
includes more short-lived associations, which if resolved should
probably not be considered genuine star clusters.

\subsection{Early evolution and the luminosity function of old globular cluster systems}

The LFs of old globular cluster systems is very different from the
(power-law) LFs of YMCs discussed in \S\ref{Sect:LF}. When
approximating the LF of old globular clusters in the Milky Way by a
Gaussian, the peak is at $M_V\approx-7.4\,$mag
\citep{2001stcl.conf..223H}. For a constant mass-to-light ratio of
$M/L\approx2$ the mass function, often referred to as globular cluster
mass function (GCMF), which in fact is a present-day mass function.
The GCMF is, like the LF, also peaked or bell-shaped (when using
logarithmic mass bins). The characteristic mass, or turn-over mass, is
$M_{\rm TO}\approx2\times10^5\,\msun$. It has been argued that old
globular clusters formed with this --bell-shaped-- mass function
\citep{1985ApJ...298...18F,2007MNRAS.377..352P}, which is completely
different than the power-law distribution we see for YMCs today.  The
discrepancy between the bell-shaped curve for old globular clusters
and the power-law mass function for YMCs then could originate from the
long timescale phase-3 dynamical evolution, this evolution could
severre the relatively low-mass ($M \aplt M_{\rm TO}$) clusters more
whereas it leaves the high-mass ($M \apgt M_{\rm TO}$) clusters
unaffected
\citep{1997ApJ...480..235E,1997MNRAS.287..915V,1998A&A...330..480B,
  2001ApJ...561..751F, 2008ApJ...679.1272M}.

The value of $M_{\rm TO}$, however, is remarkably constant between
galaxies \citep{2007ApJS..171..101J} and even within a single galaxy
it hardly varies with the distance to the galaxy center
\citep{2003ApJ...593..760V}. This is hard to reconsile with the models
that start with a power-law mass function and seek to explain the
present-day form of the GCMF by phase-3 dynamical evolution
\citep{2003ApJ...593..760V}.

For this review only those solutions to this conundrum are relevant
that play an important role in the early evolution (i.e. Phase~1 and
2). \citet{2008MNRAS.384.1231B} provide such a solution; they argue
that relatively low-mass clusters are more easily dissolved as a
result of phase-1 gas expulsion, leaving the most massive clusters
unaffected. They argue that the binding energy of embedded clusters
scales with the square of their total masses (gas and stars), whereas
the energy input from stellar winds and supernovae depends linearly on
the mass of all the stars in the cluster. If the star formation
efficiency ($f_e$) is independent of the mass of the primordial
molecular cloud, more massive clusters require more time to clear all
the residual gas from the natal cloud.  \citet{2008MNRAS.384.1231B}
find that clusters with $M \aplt 10^5\,\msun$ loose their residual gas
on time-scales much shorter than a crossing time, which effectively
shocks the cluster and drives it to dissolution (see
\S\ref{sec:TheFirstMyr}) More massive clusters respond adiabatically
to the loss of the mass of the cloud
(\S\ref{sect:StellarMassLoss.Theory}), resulting in a higher survival
probability. The weak dependence of $M_{\rm TO}$ on environmental
factors emerges quite naturally from this model
\citet{2008MNRAS.384.1231B}.

An alternative explanation for the evolution of the mass function of
YMCs compared to that of the GCMF is presented by
\citet{2003ApJ...587L..97V}.  They argue that phase-2 stellar mass
loss leads to a near universal dissolution rate of low-mass clusters.
Old globular clusters of low mass tend to be less concentrated than
high-mass clusters \citep{2000ApJ...539..618M}.  If this trend is
primordial, stellar mass loss would be more effective in destroying
low-mass clusters compared to high-mass clusters
\citep{1990ApJ...351..121C,2003ApJ...587L..97V}.  This mechanism is,
as the phase-1 outgassing, independent of environmental factors, like
the distance to the galactic center, and would therefore effectively
provide a satisfactory explanation of the weak dependence of $M_{\rm
  TO}$ to the environment.

Each of the three solution to this conundrum have their advantages and
disadvantages. It may be evident that this discussion has not yet
resulted in a satisfactory solution to the problem, but it promises to
continue to be an exciting discussion for the foreseeable future.


\section{Lusus Natur{\ae}}\label{Sect:Exotica}

Old globular clusters are of interest because of their old age, their
assumed relatively homogeneous populations, their relative isolation
in their parent galaxies, and because of their abundance of unusual
objects, such as blue stragglers, x-ray binaries, radio pulsars, etc.,
often referred to collectively as {\em stellar exotica}, or {\em lusus
  natur\ae}\footnote{Lusus Natur\ae\, is Latin for the freaks of
  nature, mutants or monsters.}. In the disk of the Milky Way, such
objects form through internal evolutionary processes in individual
stars or close binary systems.  In star clusters, these processes are
augmented by stellar interactions, mediated by the high encounter
frequency in dense cluster cores.

Stellar encounters generate new channels for the formation of exotic
objects, but can also catalyze existing channels.  For example, a
binary encounter may lead directly to a collision and the formation of
a blue straggler, or its effect may be indirect, perhaps resulting in
an exchange that eventually (billions of years later) leads to the
formation of a low-mass x-ray binary.  Repeated encounters can
transform binaries and multiple stellar systems, multiplying the
channels for the production of exotica
\citep{1995MNRAS.276..887D,2002ApJ...570..184H,2006NewA...12..201D}.
A clear understanding of the formation and evolution of these objects
can provide insight into the past dynamical evolution of the cluster
\citep{2009arXiv0903.0979D}.

Many of the stellar exotica observed in old globular clusters today
are the results of processes that began when the cluster was young.
In some cases, they are the products of the interplay between
dynamical and evolutionary processes involving stars and binaries
during the first $\sim 100$\,Myr of the cluster lifetime.  The
primordial seeds for many lusus natur{\ae} were planted during this
period \citep{2007MNRAS.374...95P}. Later stages (during phase 3, see
\S\,\ref{Sec:Dynamics}) of dynamical evolution, such as core collapse,
may produce additional generations of exotica.  Since the observed
YMCs will age to become old globular clusters (see
\S\ref{Sect:Introduction}), they provide a convenient testbed for the
study of the progenitors of stellar exotica.

The progenitors of observed lusus natur{\ae} in old globular clusters
are not necessarily easy to identify in the young cluster population,
although in some cases the evolutionary link is well established
\citep{2008A&A...488.1007G}.  There may well be entire populations of
peculiar objects in young star clusters that do not lead to observable
interesting objects at later stages, and some objects destined for
peculiarity may look perfectly ordinary at early times.  An example of
the latter is the dormant blue straggler population consisting of
stars that were rejuvenated by mass transfer or collisions while still
on the main sequence, and now lurk among their fellow main-sequence
stars until they remain behind after the others traverse the
Hertzsprung gap \citep{1997A&A...328..130P}.

\subsection{Binary Stars}\label{Sect:binaries}

Exotic objects in star clusters are closely related to binaries, as
they often form via internal binary evolution or during dynamical
interactions between binaries and other stars. Examples are the
formation of blue stragglers (\S\ref{Sect:BlueStragglers}), colliding
wind binaries (\S\ref{Sect:CWB}), and anomalous x-ray pulsars
(\S\ref{Sect:Magnetars}), all of which require the presence of binary
stars in the system.  In some cases, such as the slow evolution of an
accreting X-ray pulsar that leads from a low-mass x-ray binary to a
binary millisecond pulsar, the evolutionary track is readily
established \citep{1991PhR...203....1B}.  In others, however, the
punctuated equilibrium through which these objects evolve makes it
virtually impossible to catch the key transitions as they occur.
Examples are collisions between stars, or the common-envelope phase in
the Darwin--Riemann instability of a contact binary.

We will distinguish two fundamental types of binaries in star
clusters: (1)~``primordial'' binaries, which formed contemporaneously
with the stars in the cluster as a crucial part of the star-formation
process \citep{1989Natur.339...40G}, and (2)~``dynamical'' binaries,
which formed later via stellar interactions, often long after the
component stars reached the main sequence.  One can wonder to what
extend this limited terminology is still usable for binaries that
experiences one or more exchange interactions. It is for example,
quite possible that two stars that were initially single end-up in a
binary after two exchanges. As a practical matter we would still
consider such a binary primordial.  The second group of dynamical
binaries may be further divided into two sub-categories---binaries
formed by conservative three- and four-body stellar dynamical
interactions, and binaries formed by dissipative two-body tidal
capture.  The latter process was introduced by
\cite{1975MNRAS.172P..15F} to explain the relatively high specific
frequency of low-mass X-ray binaries in old globular clusters.  It has
fallen somewhat out of favor since the late 1980s, but it may be
entering a revival of sorts \citet{2004ApJ...610..477O}.  Many of the
curious objects discussed in this \S\, are related to binarity, either
primordial or tidal, although we tentatively will use this distinction
in the formation process.

\subsubsection{Blue Stragglers}\label{Sect:BlueStragglers}

A blue straggler is a star which exceeds the cluster main-sequence
turnoff in both temperature and luminosity, but which is not on the
horizontal branch.  Blue stragglers populate the region blueward of
the turnoff, as if they lagged behind on the cluster main sequence
while the other stars aged.  The first (34) blue stragglers were
discovered in the globular cluster M3 by \cite{1953AJ.....58...61S}.
At least 8 plausible explanations have been proposed for the formation
of blue stragglers \citep[see][]{1989AJ.....98..217L}.  Of these, two
are currently in favor:
\begin{itemize}
\item direct merger between two stars \citep{1976ApL....17...87H},
\item mass transfer in a close semi-detached binary star
  \citep{1964MNRAS.128..147M}.
\end{itemize}
The latter scenario is supported by the discovery of two blue
stragglers, in the young open star clusters NGC 663 and NGC 6649,
which have been found to be the donors in Be/X-ray binary systems
\citep{2007ASPC..367..645M}.  The discovery of a blue straggler in the
old open cluster M67, which appears to be about 2.5 times more massive
than the turn-off mass, favors the former view
\citep{1996ApJ...470..521L}.

Both favored mechanisms for blue straggler formation appear plausible
in YMCs.  However, no blue stragglers have yet been identified in any
observed YMC, although this may be explained by the absence of a clear
turn-off in the resolved clusters, which makes the identification (and
the definition) of a blue straggler impractical.  There are, however a
number of ``odd'' stars in YMCs that might possibly evolve to resemble
blue stragglers in the future.  Objects consistent with this broadened
definition include four O3~If/WN6-A stars in the star cluster R136 in
the 30 Doradus region of the LMC
\citep{1992AJ....104.1721C,1996ApJ...466..254B}.

Several interesting correlations exist between the numbers of blue
stragglers in old globular clusters and the numbers of red giants
\citep{1995A&A...294...80F}, and also with the binary fraction
\citep{2008A&A...481..701S,2009Natur.457..288K}.  In addition,
\cite{2004MNRAS.349..129D} find that the number of blue stragglers is
be independent of integrated absolute magnitude $M_V$ of the cluster,
and use this fact to argue that both production mechanisms are
relevant.

\subsubsection{Colliding wind binaries}\label{Sect:CWB}

Binaries containing two massive stars, such as Wolf-Rayet stars, with
strong stellar winds often exhibit intense radio and/or x-ray
emission.  Since this process requires copious stellar mass loss in a
fast wind, these sources do not occur in old globular clusters, but
YMCs appear to be excellent hosts for such systems.  Several young and
dense star clusters exhibit x-ray and radio emission from colliding
wind binaries.  In some relatively nearby cases, R136
\citep{2002ApJ...574..762P}, Wd1
\citep{2005A&A...434..949C,2006MNRAS.372.1407C}, and the Arches and
Quintuplet clusters \citep{2005AJ....130.2185L}, the counterparts of
the radio and x-ray sources have been identified.

\subsection{Compact objects}\label{Sect:XRayBinaries}

A compact object is generally produced in a supernova explosion (see
\S\,\ref{Sect:SNe}). During such event the compact object, either a
neutron star or a black hole, is likely to receive a high velocity,
often referred to as a 'kick'.  The direction of the kick is ill
constrained, but the velocity distribution of this kick is well
determined by observing the velocities of nearby young neutron stars
\citep{1994Natur.369..127L,1998ApJ...505..315C,2002ApJ...568..289A}
and the galactic scale-height of black hole x-ray transients
\citep{1996ApJ...473L..25W,2005ApJ...618..845G}. It seems that neutron
stars receive a considerably higher velocity kick than black holes. As
a result the majority ($\sim 95$\,\%) of neutron stars are likely
ejected from any YMC \citep{1996MNRAS.280..498D,2002ApJ...573..283P}
whereas most black holes are retained. However, since neutron stars
are considerably more massive than the mean mass of a star in a
cluster, once formed and retained they tend to stay in the cluster. At
a later age compact objects can subsequently dominate the cluster
evolution. For black holes this happens at an age of about 100\,Myr to
a Gyr \citep{1993Natur.364..421K,2007MNRAS.379L..40M}, whereas for
neutron stars the age of domination starts around a Hubble time
\citep{2007MNRAS.374...95P}. The evolution of compact objects in star
clusters deserves a separate review, but here we will present a very
brief summary of the current observational evidence for compact
objects in YMCs.

A number of YMCs have been observed in the radio and x-ray, revealing
a wealth of sources, even richer than found in old globular clusters
\citep{2008A&A...477..147C}.  Among the x-ray sources is a large
population of accreting neutron stars, stellar-mass black holes and
possibly intermediate-mass black holes.  Because of crowding in the
central regions of these clusters, where most of the x-ray sources are
found, very few sources have optical counterparts.

With the adopted age limit of 100\,Myr, only relatively few white
dwarfs have formed---about as many as neutron stars---and cataclysmic
variables are not expected.  The youth of these clusters seem to make
it unlikely that any low-mass x-ray binaries or millisecond pulsars
will be found.

\subsubsection{Magnetars}\label{Sect:Magnetars}

Shortly after a supernova (within $\sim10^5$\,years), a newly formed
neutron star may become observable as a magnetar, which can have a
magnetic field strength exceeding $\sim10^{15}$\,gauss
\citep{1999ApJ...510L.115K}.  The population of magnetars is
subdivided into two classes: soft gamma-ray repeaters (SGRs) and
anomalous x-ray pulsars (AXPs)\footnote{see {\tt
    http://www.physics.mcgill.ca/$\sim$pulsar/magnetar/main.html}.}.
As the products of supernovae, one might naively expect these objects
to reside mainly in YMCs, and indeed about half (3 of 8) of the known
SGRs and one-tenth (1 of 10) of the AXPs are known to reside in such
systems.  These are remarkably high fractions, given that only 0.05\%
of the stellar mass of the Galaxy resides in star clusters (see
\S\ref{Sect:Introduction}).

The single cluster AXP is CXOU J164710.2-455216
\citep{2006ApJ...636L..41M} in the Galactic young star cluster
Westerlund~1.  It exhibits a 20-ms burst with energy
$\sim10^{37}$\,erg (15-150 keV), and spins down at a rate of
$P/\dot{P} \simeq -10^{-4}$ \citep{2007MNRAS.378L..44M}, quite typical
of a magnetar.

Several interesting sources are hosted by young star clusters in the
LMC.  These include the microquasar LS I +61$^\circ$\,303, which may
have been ejected from the cluster IC~1805
\citep{2004A&A...422L..29M}.  The relatively low-density young star
cluster SL~463, too small for inclusion in this review, seems to be
associated with SGR 0526-66 \citep{2004ApJ...609L..13K} at a projected
distance of $\sim 30$\,pc.  Two other SGRs associated with relatively
low-density young star clusters are SGR 1806-20, at a projected
distance of $\sim 0.4$\,pc from the core of its parent cluster
\citep{2000AIPC..526..814M,2004A&A...419..191C}, and SGR 1900+14, at
$\sim 0.8$\,pc \citep{2000AIPC..526..814M,2000ApJ...533L..17V}.  This
latter SGR has a measured proper motion of 70 mas/yr away from the
cluster, suggesting that it indeed was ejected from the parent cluster
\citep{2009ApJ...692..158D}, which is relatively old ($14\pm1$\,Myr)
compared to the usual SGR-producing stars
\citep{2009arXiv0910.4859D}. The actual association between cluster
and SGR is hard to establish in this case since the distances to both
objects are ill constrained.

\subsubsection{Ultra-luminous X-ray sources}\label{Sect:ULX}

Old globular clusters are known to host an enormous excess of low-mass
x-ray binaries compared to the rest of the Galaxy.  Much of this
excess is attributed to the dynamical environment in dense cluster
cores \citep{1975MNRAS.172P..15F,2003ApJ...591L.131P}.  Young star
clusters are sites of intense dynamical activity, so it is not
surprising that YMCs also host many x-ray sources.
The majority of x-ray point sources in external galaxies appear to be
associated with young star clusters, as is the case for example in the
Antennae system (NGC4038/39) \cite{2002ApJ...577..710Z}.  The nature
of most of these x-ray sources is unknown, and we can only guess at
their origin.

We limit ourselves here to the most striking x-ray sources, the
subclass of ultra-luminous x-ray sources (ULXs), which are
characterized by x-ray luminosity $L_x \apgt 1.3 \times
10^{39}$\,erg/s, the maximum isotropic luminosity that can be produced
by a 10\,{\msun} black hole accreting pure hydrogen
\citep{2001ApJ...552L.109K}.  For practical reasons we round the
threshold up to $L_x \apgt 10^{40}$\,erg/s, mainly to ensure that such
luminosities are unlikely to be produced by stellar-mass black holes
accreting at the Eddington rate from a main-sequence companion star.
These ULXs are responsible for the brightest stellar x-ray sources in
the sky.

Several models attempt to explain the high x-ray luminosity of the
ULXs, but at present there is no consensus in the community on the
source of the x-rays. The current leading models are:
\begin{itemize}
\item anisotropic (collimated or beamed) emission from an accreting
  stellar-mass black hole \citep[although porosity, turbulence, and
    bubbles also provide interesting
    alternatives,][]{2001ApJ...552L.109K,2002MNRAS.335L..13K},
\item accretion from an evolved star onto a stellar-mass black hole,
  which can, in principle, lead to an accretion rate higher than from
  a main-sequence star, and therefore a higher x-ray luminosity
  \citep{2006ApJ...640..918M},
\item accretion from a companion star onto an ``intermediate-mass
  black hole'' (IMBH), with mass $\apgt 100$\,{\msun}
  \citep{2004Natur.428..724P}.
\end{itemize}

ULXs tend to be hosted by starburst and spiral galaxies
\citep{2000ApJ...535..632M}.  Some of the brightest are associated
with YMCs; a leading example is the ULX in the star cluster MGG~11 in
the starburst galaxy M82 \citep{2001MNRAS.321L..29K}.  The association
with YMCs argues in favor of an accreting black hole of $\sim
1000$\,{\msun} \citep{2004Natur.428..724P}.  The object in MGG~11 is
particularly interesting, as it shows a strong quasi-periodic
oscillation in the 50-100 mHz frequency range
\citep{2003ApJ...586L..61S}, providing a strong argument against
beamed emission, and supporting the hypothesis that the x-ray
luminosity comes from an accreting black hole of 200--5000\,\msun.
Further support is provided by the detected periodic variation of
$\sim 62$\,days, which can be explained if the black hole is orbited
by a 22--25\,{\msun} Roche-lobe filling donor star
\citep{2006MNRAS.370L...6P}.

ULXs have been associated with YMCs in NGC 5204
\citep{2004ApJ...602..249L}, the starburst galaxies M82
\citep{2001MNRAS.321L..29K} and NGC1313 \citep{2008A&A...486..151G},
the edge-on spiral NGC 4565 \citep{2002ApJ...576..738W}, the
interacting galaxies M51 \citep{2002ApJ...581L..93L}, NGC4038/39
\citep[Antennae][]{2001ApJ...554.1035F} and ESO 350-40
\citep[Cartwheel][]{2003ApJ...596L.171G}, and the type 1.5 Seyfert
galaxy NGC 1275 \citep{2006MNRAS.367.1132G}.  The higher abundance of
ULXs in active, starburst, and interacting galaxies may be related to
the empirical fact that YMCs tend to form in these environments; 60\%
of ULXs are associated with active star-forming regions
\citep[][although the definition of a ULX used here is somewhat
  faint]{2009ApJ...703..159S}.

If the counterpart of a ULX hosts an IMBHs, it is likely to be the the
acceptor from a windy or Roche-lobe overflowing massive star, as seems
to be the case in the $10^{39}$--$10^{41}$\,erg/s ULX in NGC 5204,
where the donor is identified as a B0~Ib supergiant with a 10-day
orbital period \citep{2004ApJ...602..249L}, and in ULX M51 X-7, which
has has an even shorter orbital period of only 2.1\,hr
\citep{2002ApJ...581L..93L}, although no stellar companion has been
identified.  The black holes that may be responsible for the ULXs NGC
1313 X-1 and X-2 (0.2-10.0 keV, assuming a distance of 3.7 Mpc) may
have masses in the range 100--1000\,{\msun}
\citep{2003ApJ...585L..37M}, and in these cases YMCs have been
identified as optical counterparts.

\subsection{Explosive events}

\subsubsection{Supernovae}\label{Sect:SNe}

Supernova are relatively rare events, occurring about once every 100
years in a galaxy like the Milky Way \citep{1999A&A...351..459C}.
Although type~I supernovae are unlikely to occur in star clusters
younger than 100\,Myr \citep{2009arXiv0903.1104P}, at these ages the
seeds may be planted for a rich future of type~I events
\citep{2002ApJ...571..830S}.

Since most massive stars tend to reside in clusters, as discussed in
\S\ref{Sect:Introduction} and \S\ref{sec:observations}, it is probable
that the majority of type~Ib/c and type~II supernovae occur in star
clusters.  However, since most supernovae occur in distant galaxies it
is hard to find optical counterparts, and very few associations of
supernovae with young star clusters have been reported.  Based on the
peculiar metallicity of SN 1987A, \cite{1991PAZh...17..404E} argue
that the star originated in the young LMC cluster MKM90.  SN~2006gy
\citep{2006CBET..695....1F}, which is a candidate for the formation of
an IMBH, may have been the result of a collision runaway
(\S\ref{Sect:Collisions}) in a YMC \citep{2007Natur.450..388P}.
However, perhaps the strongest case is the peculiar type~IIp supernova
SN~2004dj (probably produced by a 12--20\,\msun\, star) in the spiral
galaxy NGC~2403; as it faded, the star cluster Sandage-96 reappeared
\citep{2005ApJ...626L..89W}.

\subsubsection{Gamma-ray bursts}\label{Sect:GRB}

Any word written about gamma-ray bursts is likely to trigger its own
burst of e-mail, but we cannot resist the temptation to devote a few
lines to this fascinating transient phenomenon.  Gamma-ray bursts come
in two types, of short ($\aplt 2$\,s) and long duration, respectively
\cite{2002ARA&A..40..137M}.  Several theories exist which point to
either old globular or young star clusters as possible hosts for gamma-ray
bursts.  Colliding compact objects are often cited as sources for the
short bursts \citep{1991ApJ...379L..17N}. Long bursts are thought to
be hosted by massive star forming regions \citep{1998ApJ...494L..45P}.
Of particular interest is the elusive relation between the long bursts
and YMCs \citep{2000AstL...26..558E}.  The models for long-duration
gamma-ray bursts should be particularly applicable to YMCs, as they
require rapidly rotating high-mass stars \citep{2003ApJ...591..288H},
which could be achieved quite naturally by stellar collisions in a YMC
(see \S\ref{Sect:Collisions}).

\subsection{Summary of exotica}\label{Sect:OtherStellarExotica}

We have discussed several examples of exotica in YMCs, but other
curiosities remain.  Many of these exotic objects are well studied in
old globular clusters, but similar scrutiny is so far lacking in their
younger siblings.  Rather than providing a detailed description of
each of the oddities found, we simply mention a few developments and
recent discoveries of what today we call exotic objects, which one day
we may consider ``normal.''  The following list summarizes a number of
peculiar clusters, noteworthy because of a unique source or object.
The list is far from complete, but at the very least it indicates the
diversity of objects found in YMCs.
\begin{itemize}
\item {\bf R136} Contains some 13 colliding wind binaries
  \citep{2002ApJ...574..762P}, and possibly 3 blue stragglers (even
  though no clear turn-off can be distinguished).  In addition, the
  star cluster shows an OH (1720 MHz) Maser, which is probably related
  to the surrounding nebula rather than the star cluster itself
  \citep{2005AJ....129..805R}.
\item The {\bf Arches cluster} contains 10 radio point sources
  \citep{2005AJ....130.2185L} and several colliding wind binaries.
\item {\bf MGG11} is a YMC in M82 which may contain a ULX
  \citep{2001MNRAS.321L..29K}, alhtough the Chandra error box is
  slightly offset.
\item The {\bf Quintuplet cluster} contains the Pistol star
  \citep{1998ApJ...506..384F}, a candidate for the most massive star
  in the Galaxy, at a projected distance of about 1\,pc, as well as 9
  radio point sources \citep{2005AJ....130.2185L}.
\item {\bf Westerlund 1} hosts the anomalous x-ray pulsar CXOU
  J164710.2-455216 \citep{2006ApJ...636L..41M}, and a wealth of x-ray
  sources \citep{2008A&A...477..147C}. 
\item {\bf Westerlund 2} hosts the massive Wolf-Rayet binary WR 20a,
  containing two WN6ha stars, at a distance of 1.1\,pc from its center
  \citep{2005A&A...432..985R}.
\item {\bf NGC 663 and NGC 6649} contain blue stragglers which found
  to be the donors of Be/X-ray binary systems
  \citep{2007ASPC..367..645M}. 
\item {\bf Sandage-96} is a $\sim 96,000$\,{\msun} star cluster in
  NGC~2403 in which a type IIp supernova was detected
  \citep{2005ApJ...626L..89W}. This star cluster also exhibits
  multiple stellar populations \citep{2009ApJ...695..619V}.
\end{itemize}

\subsubsection{Planetary nebulae and supernova remnants}\label{Sect:Neblae}

The nuclear evolution of a sufficiently massive star ($\apgt
8$\,\msun) is associated with a supernova explosion (see
\S\ref{Sect:SNe}), while lower mass stars fizzle into the background
after a short, bright post-AGB phase. But following these lower mass
events a roughly spherical gas shell---a planetary nebula---remains
visible for a much longer time than either the supernova or the
post-AGB star, illuminated by the central stellar remnant and shocks
as the out-flowing gas encounters the interstellar medium.

Every star experiences either a supernova or a post AGB phase and a
cluster of $5 \times 10^4$ stars will experience some 350 supernovae,
leading to an equal number of remnants, while during the first
100\,Myr a similar number of planetary nebulae will form.  The
formation rate of planetary nebulae and supernova remnants is thus
$\sim7$/Myr.  With an observable lifetime for a nebula of about $10^4$
yr, we naively expect to see one nebula for every $\sim 14$ star
clusters.  Among 80 YMCs in the Milky Way Galaxy
\cite{2006A&A...459..103L} found 3 with a planetary nebula, consistent
with our naive estimate.  No planetary nebulae or supernova remnants
have so far been found in any of the YMCs in the Local Group. But
since only 16 of the clusters listed in
Tab.\,\ref{Tab:Local_Group_Clusters} are sufficiently old to have
formed white dwarfs from single stellar evolution, the lack of
planetary nebulae may be a statistical fluctuation.

\subsubsection{Brown dwarfs and planets}\label{Sect:Planets}

We will say little here about brown dwarfs and planets in YMCs since
the observations are sparse and the general topic deserves its own
review. However, a few words are in order.

Young dense star clusters are generally too distant for planets to be
detectable using current methods, and no planets have been found to
date \citep{2007ARA&A..45..397U}.  Old globular clusters seem to be
deficient in planets \citep[e.g.~47 Tuc,][]{2000ApJ...545L..47G},
possibly because of their low metallicities
\citep{2005ApJ...620.1043W}.  However, the recent HARPS discovery of a
planet around a metal poor star \citep{2009A&A...493..639M} makes this
argument less convincing.  The planet found orbiting the 11 ms pulsar
B1620-26 in the metal poor environment of the globular cluster M4
\citep{1993ASPC...36...11B,1993ApJ...412L..33T}, was formed by a
different mechanism than planets around solar-type stars, and we do
not (yet) expect such planets in YMCs. Star clusters in
high-metallicity environments, such as Westerlund~1 and NGC\,3603,
could host a large population of planets, but none have yet been
found.

We see no reason why YMCs should be deficient in planets.  However,
the high interaction rate in a dense cluster could make a planetary
system short-lived.  Disruption of a planetary system may leave the
planet separated from its parent star \citep{2009ApJ...697..458S}.
Several such free-floating objects have been found in the Orion
Trapezium cluster \citep{2000MNRAS.314..858L}. Once dislocated from
its parent star, a planet will easily escape the cluster, though.

\subsubsection{YMCs in a galactic context}

Due to supernovae and winds from massive stars, YMCs are the sources
of cluster winds \citep{2004ApJ...610..226S}, which may trigger
larger-scale winds and chimneys, as in the Perseus arm of the Milky
Way Galaxy \citep{1996Natur.380..687N}.  Although little studied from
the point of view of cluster evolution, this is an important topic
that provides a possible and rather natural link between the evolution
of a YMC and that of its parent galaxy.


\section{Concluding Remarks: Young Globular Clusters?}

The discovery of large numbers of young massive star clusters,
particularly in other galaxies, over the last decade has led to the
realization that such clusters are responsible for a significant
fraction of all current star formation in the local universe.  The
study of these systems, and especially their lifetimes, against
various stellar evolutionary and dynamical processes, is therefore of
critical importance to several branches of stellar and galactic
astrophysics.

Star clusters appear to form with a cluster mass function described by
a power-law with index $-2$.  This mass function seems to be the same
for both open and globular clusters, and does not depend significantly
on the local galactic environment or on the specific characteristics
of the giant molecular clouds from which the clusters formed.  An
important parameter for a young bound cluster appears to be its age
relative to its current dynamical time scale \tdyn.  For unbound
stellar agglomerates or associations, {\tdyn} exceeds the system age,
indicating that the cluster is either extremely young or dissolving
into the tidal field of its parent galaxy.  For the typical bound star
clusters discussed in this review, {\tdyn} is smaller than the current
age; for an association it is the other way around.

YMCs evolve and eventually dissolve due to the combined effects of a
number of physical processes, the most important (for clusters that
survive the early expulsion of their natal gas) being mass loss due to
stellar evolution.  The most massive clusters, such as those found in
the Antennae system, have expected lifetimes comparable to the age of
the universe, and we could well imagine that the Antennae will someday
be a medium-sized elliptical galaxy with an extended population of
intermediate-age clusters having overall properties quite comparable
to the old globular clusters seen in other ellipticals.

The seeds of the exotica observed in many present-day globular
clusters were initiated during the infancy of those systems, in the
strongly coupled mix of stellar dynamics and stellar evolution that
characterized their early evolution (in particular dring phase~2).
YMCs destined to survive to ages comparable to the globular clusters
appear to contain much richer populations of stellar exotica (per unit
mass) than are found in the field, and may provide important testbeds
for this unique period in cluster evolution.  Our limiting cluster age
of 100\,Myr is chosen in part to include the period when this
ecological interplay is strongest.

From an observational point of view, little is known of the formation
and evolution of stellar exotica in YMCs, mainly because such clusters
are relatively rare.  The Milky Way contains only half a dozen, and
the closest lies $\sim 4$\,kpc away, too distant for detailed study of
the individual stars in its central region.  However, a number of
studies have drawn connections between YMCs and exotic objects, such
as unusual supernovae, magenetars, x-ray binaries, and ultraluminous
x-ray sources, possibly placing YMCs in the same league as the old
GCs, which are rich in such objects.

Finally, we return to the term ``young globular cluster,'' which we
introduced in \S\ref{Sect:Introduction} but have deliberately avoided
throughout this review.  Is this an appropriate term for a YMC?  A
fairly common definition of a globular cluster, similar to that found
in many textbooks, appears in the Oxford English Dictionary (2009):
``{\em A roughly spherical cluster of stars, typically seen in
  galactic halos, containing large numbers of old, metal-poor
  stars}.''  This definition contains several qualifiers that would
seem to exclude YMCs, but in many ways the definition itself seems to
be a relic of a bygone era.

Our own Galaxy contains a significant population of bulge GCs, and
there may well be a disk component among GCs in other galaxies, such
as the LMC.  In any case, it is quite conceivable that YMCs will in
many cases come to populate the halos of their parent galaxies.  Since
there appears to be little dependence of YMC properties on current
galactic location, it seems unreasonable to exclude a YMC from the GC
definition based solely on this property.  The ``low metallicity''
qualifier in the definition also seems a side effect rather than a
requirement.  If we required simply that globular clusters be old---or
rather long-lived, as in our YMC definition---the metallicity becomes
redundant, merely reflecting the epoch at which they formed.
Furthermore, the LMC GCs display a broad range of ages, so any cutoff
on age or metallicity would be arbitrary.  Lastly, any massive cluster
more than a few tens of dynamical times old will necessarily have a
smooth, roughly spherical appearance, regardless of metallicity or
location.

Applying all of these arguments, we can strip away all of the
extraneous qualifying terms in the above GC definition until all that
remains is the word ``massive.''  In this view, globular clusters are
simply ``old massive clusters,'' the logical descendants of YMCs in
the early universe.


\section*{Acknowledgments}
We are particularly grateful to Douglas Heggie, S{\o}ren Larsen, Nate
Bastian and Fred Rasio for their careful reading of an earlier version
of this manuscript and their useful (and often quite critical)
comments. In addition we received very helpful comments from Holger
Baumgardt, Melvyn Davies, Richard de Grijs, Diederik Kruijssen and
Hans Zinnecker.  The results of the $N$-body simulation shown in
Fig.~\ref{Fig:rc_nbody} was done on one of the GRAPE-6~BLX64 boards of
the European Southern Observatory in Garching.
This work was made possible with the financial support of DFG cluster
of excellence Origin and Structure of the Universe
(www.universe-cluster.de), NASA (grant NNX07AG95G), the National
Science Foundation (NSF, via grant AST-0708299), Nederlandse
Onderzoekschool voor Astronomie (NOVA), het Leids-Kerkhoven Bosscha
Fonds (LKBF), the Kavli Institute for Theoretical Physics (KITP,
through their hosptality and NSF grant PHY05-51164), and in particular
of the Nederlandse Organizatie voor Wetenschappelijk Onderzoek (NWO,
via grants \#643.200.503 and \#639.073.803) and the Nederlandse
Computer Faciliteiten (NCF, via project \#SH-095-08).


\def\aa{\ {A\&A}\ }
\def\aap{\ {A\&A}\ }
\def\aarev{\ {A\&A Review}\ }
\def\aas{\ {A\&AS}\ }
\def\aaps{\ {A\&AS}\ }
\def\acta{\ {Acta Astron.}\ }
\def\advspres{\ {Adv. Space Res.}\ }
\def\aj{\ {AJ}\ }
\def\ajp{\ {Aust. J. Phys.}\ }
\def\ajpas{\ {Aust. J. Phys. Astr. Supp.}\ }
\def\annap{\ {Ann. d'Ap.}\ }
\def\apj{\ {ApJ}\ }
\def\apjl{\ {ApJL}\ }
\def\apjs{\ {ApJS}\ }
\def\aplett{Astrophys.~Lett.}
\def\apss{\ {Ap\&SS}\ }
\def\araa{\ {ARA\&A}\ }
\def\astrolett{\ {Astro. Lett. and Communications}\ }
\def\baas{\ {BAAS}\ }
\def\ban{\ {BAN}\ }
\def\cntphys{\ {Contemp. Phys.}\ }
\def\currsci{\ {Curr. Sci.}\ }
\def\jaa{\ {JA\&A}\ }
\def\jphys{\ {J. de Physique}\ }
\def\memsai{\ {Mem. Soc. Astron. Ital.}\ }
\def\mess{\ {Messenger}\ }
\def\mnras{\ {MNRAS}\ }
\def\nat{\ {Nat}\ }
\def\natps{\ {Nature Phys. Sci.}\ }
\def\newa{\ {New Astron.}\ }
\def\obs{\ {Observatory}\ }
\def\pasa{\ {Proc. Astron. Soc. Aust.}\ }
\def\pasp{\ {PASP}\ }
\def\pasj{\ {Publ. Astr. Soc. Japan}\ }
\def\pcps{\ {Proc. Cambridge Phil. Soc.}\ }
\def\pddo{\ {Pub. David Dunlap Obs.}\ }
\def\philtrans{\ {Phil. Trans. R. Soc. London A}\ }
\def\physrep{\ {PhysRep}\ }
\def\physrevl{\ {Phys. Rev. Lett.}\ }
\def\physscrip{\ {Physica Scripta}\ }
\def\phystod{\ {Physics Today}\ }
\def\pnas{\ {Proc. Nat. Acad. Sci.}\ }
\def\prd{\ {Phys. Rev. D.}\ }
\def\qjras{\ {QJRAS}\ }
\def\roa{\ {R. Obs. Annals}\ }
\def\rob{\ {R. Obs. Bull.}\ }
\def\sci{\ {Science}\ }
\def\sovast{\ {SvA}\ }
\def\ssr{\ {Space Sci. Rev.}\ }
\def\zfa{\ {Zeitschr. f. Astroph.}\ }
\def\zeiss{\ {Zeiss Inform. with Jena Rev.}\ }


\appendix
\section{Appendix: Dynamical algorithms}
\label{Sect:Algorithms}
\label{Sec:AppendixA}

Even though the fundamental physics is not hard to understand,
simulating star clusters is not a trivial matter.  Significant
complications arise due to the long-range nature of the gravitational
force, which means that every star in the cluster is effectively in
constant communication with every other.  The result is a potentially
enormous number of interactions that must be calculated as we follow
the time evolution of the system, leading to high computational cost.
Further complications arise from the enormous range in spatial and
temporal scales inherent in a star cluster.  Computers, by the way
they are constructed, have difficulty in resolving such wide ranges,
and many of the software problems in simulations of self-gravitating
systems arise from this basic limitation.  The combination of many
physical processes occurring on many scales, with high raw processing
requirements, makes numerical gravitational dynamics among the most
demanding and challenging areas of computational science.  In this
Appendix we discuss some of the issues involved in the numerical
modeling of massive star clusters.

The broad spectrum of numerical methodologies currently available is
summarized in \S\ref{Sec:Simulations}. Here we present some of the
details of these simulation techniques, starting with {\em
  ``Continuum'' Models}, followed by {\em Monte Carlo Models} and {\em
  Direct $N$-body Models}.  We end this section with brief discussions
of computer hardware and multi-core codes, and in particular {\em
  Kitchen Sink Models} and {\em Future Prospects}.

\subsection{Continuum methods}

The two leading classes of continuum models are gas-sphere
\citep{1980MNRAS.191..483L,1984MNRAS.208..493B,2001A&AT...20...47D}
and Fokker--Planck
\citep{1979ApJ...234.1036C,1985IAUS..113..373S,1990ApJ...351..121C,1992ApJ...386..106D,1996PASJ...48..691T,1997PASJ...49..547T,1998ApJ...503L..49T}
methods.  They have mainly been applied to spherically symmetric
systems, although axisymmetric extensions to rotating systems have
also been implemented
\citep{1999MNRAS.302...81E,2002MNRAS.334..310K,2004MNRAS.351..220K},
and some limited experiments with rudimentary binary treatments have
also been performed \citep{1991ApJ...370..567G}.

Both approaches start with the collisional Boltzmann equation as the
basic description for a stellar system, then simplify it by averaging
the distribution function $f({\bf x}, {\bf v})$ in different ways.
Gas-sphere methods proceed in a manner closely analogous to the
derivation of the equations of fluid motion, taking velocity averages
to construct the moments of the distribution: $\rho=\int\,d^3v\,f({\bf
  x}, {\bf v})$, ${\bf u}=\int\,d^3v\,{\bf v}f({\bf x}, {\bf v})$,
$\sigma^2= \frac13 \int\,d^3v\,v^2f({\bf x}, {\bf v})$, etc.  Application of a
closure condition leads to a set of equations identical to those of a
classical conducting fluid, in which the conductivity depends
inversely on the local relaxation time.  Fokker--Planck methods
transform the Boltzmann equation by orbit-averaging all quantities and
recasting the equation as a diffusion equation in $E-J$ space, where
$E$ is stellar energy and $J$ is angular momentum.  Since both $E$ and
$J$ are conserved orbital quantities in a static, spherically
symmetric system, two-body relaxation enters into the problem via the
diffusion coefficients. 

These methods have been of enormous value in developing and refining
theoretical insights into the fundamental physical processes driving
the dynamical evolution of stellar systems
\citep{1984MNRAS.208..493B}.  However, as the degree of realism
demanded of the simulation increases---adding a mass spectrum, stellar
evolution, binaries, etc.---the algorithms rapidly become cumbersome,
inefficient, and of questionable validity \citep{1999CeMDA..73..179P}.
As a result, they are generally not applied to the young stellar
systems of interest here.  The major approaches currently used for
simulating young massive clusters are particle-based Monte Carlo or
direct $N$-body codes.

\subsection{Monte Carlo methods}\label{Sect:MonteCarlo}

Depending on one's point of view, Monte Carlo methods can be regarded
as particle algorithms for solving the partial differential equations
arising from the continuum models, or approximate schemes for
determining the long-term average gravitational interactions of a
large collection of particles.  The early techniques developed in the
1970s and 1980s
\citep{1971ApJ...164..399S,1973dses.conf..183H,1975IAUS...69....3S,1982AcA....32...63S,1986AcA....36...19S}
fall into the former category, but recent studies, in particular
\citep{1998MNRAS.298.1239G,2000ApJ...540..969J,2001A&A...375..711F,2001MNRAS.324..218G,2003ApJ...593..772F,2006MNRAS.371..484G,2007ApJ...658.1047F,2008MNRAS.389.1858H,2009MNRAS.395.1173G},
tend to adopt the latter view.  The hybrid Monte Carlo scheme of
\citep{1998MNRAS.298.1239G,2001MNRAS.324..218G,2003MNRAS.343..781G}
combines a gas-sphere treatment of the ``background'' stellar
population with a Monte Carlo realization of the orbits and
interactions of binaries and other objects of interest.  These
approaches have allowed the first simulations of an entire globular
cluster, from a very early (although gas depleted) phase to complete
dissolution.

Monte Carlo methods are designed for efficient computation of
relaxation effects in collisional stellar systems, a task which they
accomplish by reducing stellar orbits to their orbital
elements---energy and angular momentum---effectively orbit averaging
the motion of each star.  Relaxation is modeled by randomly selecting
pairs of stars and applying interactions between them in such a way
that, on average, the correct rate is obtained.  This may be
implemented in a number of ways, but interactions are generally
realized on time scales comparable to the orbit-averaged relaxation
time.  As a result, Monte Carlo schemes can be orders of magnitude
faster than direct $N$-body codes.  For example,
\cite{2000ApJ...540..969J} report a CPU time scaling for their
Monte Carlo scheme of $O(N^{1.4})$ for core-collapse problems,
compared to $N^3$ for $N$-body methods, as discussed below.  To
achieve these speeds, however, the geometry of the system must be
simple enough that the orbital integrals can be computed from a star's
instantaneous energy and angular momentum.  In practice, this limits
the approach to spherically symmetric systems in virial equilibrium,
and global dynamical processes occurring on relaxation (or longer)
time scales.

\subsection{$N$-body methods}\label{Sect:DirectNbody}

$N$-body codes incorporate detailed descriptions of stellar dynamics
at all levels, using direct integration of the individual (Newtonian)
stellar equations of motion for all stars
\citep{2003gnbs.book.....A,2003gmbp.book.....H}.  Their major
attraction is that they are assumption-free, in the sense that all
stellar interactions are automatically included to all orders, without
the need for any simplifying approximations or the inclusion of
additional reaction rates to model particular physical processes of
interest.  Thus, problems inherent to Monte Carlo methods (see
\S\ref{Sect:MonteCarlo}), related to departures from virial
equilibrium, spherical symmetry, statistical fluctuations, the form of
(and indeed the existence of) phase space distribution functions, and
the possibility of interactions not explicitly coded in advance,
simply do not arise, and therefore do not require fine-tuning as in
the Monte Carlo models.

The price of all these advantages is computational expense.  Each of
the $N$ particles must interact with every other particle a few
hundred times over the course of every orbit, each interaction
requires $O(N)$ force calculations, and a typical (relaxation time)
run spans $O(N)$ orbits (see Eq.~\ref{Eq:trtd}).  The resulting
$O(N^3)$ scaling of the total CPU time means that, even with the best
time-step algorithms, integrating even a fairly small system of, say,
$N\sim10^5$ stars requires sustained teraflops speeds for several
months \citep{1988Natur.336...31H}.  Radically improved performance
can be achieved by writing better software, or by building faster
computers (or both).  In fact, the remarkable speed-up of $N$-body
codes over the last four decades has mainly been due to advances in
hardware, and in a lesser extend due to software.

Substantial performance improvements were realized by adopting better
(individual) time stepping schemes (as opposed to earlier shared time
step schemes), in which particles advance using steps appropriate to
their individual orbits, rather than a single step for all.  Further
gains were made by utilizing neighbor schemes
\citep{1973ApJ...179..885A}, which divide the force on every particle
into irregular (rapidly varying) and regular (slowly varying) parts,
due (loosely speaking) to nearby and more distant bodies.  By
recomputing the regular force at every particle step, but
extrapolating the more expensive $O(N)$ regular force for most time
steps, and recomputing it only on longer time scales, significant
improvements in efficiency have been realized.  A multi-level
generalization of this approach by \citet{2003JCoPh.185..484D} is
incorporated into the collisonal NBODY6++ \citep{1999JCoAM.109..407S}.

Another important algorithmic improvement was introduced in the
mid-1980s with the development of tree codes
\citep{1986Natur.324..446B}, which reduce the force calculation
complexity from $O(N)$ to $O(\log N)$.  Despite their algorithmic
efficiency, tree codes have not been widely used in modeling
collisional systems \citep[but see][]{1993ApJ...414..200M}.  This seems
principally to be because of lingering technical concerns about their
long-term accuracy in systems dominated by relaxation processes and
their performance in clusters with large dynamic ranges in densities
and time scales, even though these objections may no longer be well
founded \citep{1999MNRAS.310.1147M,2000ApJ...536L..39D}.  Very
promising direct--treecode methods have recently been developed to
model the dynamical interaction between a cluster and the surrounding
galactic population \citep{2007PASJ...59.1095F,2009NewA...14..369P}.

\subsection{Parallelization}

Individual time step schemes are generally hard to optimize on
parallel machines.  For those architectures, block time step schemes
\citep{1986LNP...267..156M,2006NewA...12..124M} offer substantially
better performance.  By rounding each star's ``natural'' step down to
the nearest negative integer power of two, such a scheme effectively
discretizes the time variable, allowing the possibility that large
blocks of stars will be ``next'' on the time step list, and so can be
efficiently integrated in parallel.  

The two most important parallel integration techniques are the {\em
  ring} and {\em copy} algorithms.  Both have advantages and
disadvantages, but the execution times for each, on computers with $p$
processors, scale as $N/p$, while the communication times scale as $p$
\citep{2007NewA...12..357H}.  Both algorithms are implemented in a
range of $N$-body codes, including NBODY6++
\citep{2003JCoPh.185..484D} and the {\tt kira} integrator in {\tt
  Starlab} \citep{2008NewA...13..285P}.  The two-dimensional lattice
parallelization for direct $N$-body kernels has comparable CPU time
scaling, but the communication has a weaker scaling ($\propto
1/\sqrt{p}$), enabling the code to maintain satisfactory performance
even on computers with $p \apgt 10^3$ processors
\citep{2002NewA....7..373M,2004Bisseling.book}.  So far, however, this
scheme has not been implemented in a production $N$-body code.

An interesting further step is to use a widely distributed grid of
computers \citep{FosterAndKesselman}.  In this extreme form of
parallel computing the computational bottleneck often shifts from the
$O(N^2)$ force calculation (see \S\ref{Sect:DirectNbody}) to
communication (latency and bandwidth).  However, even in the
worst-case scenario the communication costs scale $\propto N$, so, for
a sufficiently large number of stars even intercontinental grid
computing can be practical \citep{2008PCAA.book.....H}. In addition,
if (as seems likely---see \S\ref{KitchenSink}) future simulation
environments will combine a range of codes in addition to pure stellar
dynamics to address the evolution of YMCs in detail, grid computing
may provide the solution to the problem of limited supply of local
computer resources.  This is particularly relevant if the desired
algorithms for solving stellar dynamics, stellar evolution,
hydrodynamics, etc, require a diversity in computer architectures that
may not be locally available.

\subsection{Hardware acceleration}

A quantum leap in gravitational $N$-body simulation speed came from
the introduction of special purpose computers.  All $N$-body codes,
including neighbor schemes and treecodes, suffer from the cost of
computing inter-particle forces at every step along the orbit.  A
technological solution in widespread use is the ``GRAPE'' (short for
``GRAvity PipE'') series of machines developed by Sugimoto and
co-workers at Tokyo University \citep{1993PASJ...45..269E}.
Abandoning algorithmic sophistication in favor of simplicity and raw
computing power, these machines achieved high performance by mating a
fourth-order Hermite integration scheme \citep{1992PASJ...44..141M}
with special-purpose hardware in the form of highly parallel,
pipelined ``Newtonian force accelerators'' implementing the
computation of all inter-particle forces entirely in hardware.
Operationally, the GRAPE hardware is simple to program, as it merely
replaces the function that computes the (regular) force on a particle
by a call to the hardware interface libraries; the remainder of the
user's $N$-body code is unchanged.

The effect of GRAPE on simulations of stellar systems has been nothing
short of revolutionary.  Today, GRAPE-enabled code lies at the heart
of almost all detailed $N$-body simulations of star clusters and dense
stellar systems.  GRAPE-like optimizations of the innermost force
calculation operations using the IA-32 Streaming SIMD Extensions 2
(SSE2), suitable for use on any Intel/AMD processor, are described by
\cite{2006NewA...12..169N}.

Recently, Graphics Processing Units (GPUs) have achieved speeds and
price/performance levels previously attainable only by GRAPE systems
\citep[see][]{2007NewA...12..641P,2007astro.ph..3100H,
  2008NewA...13..103B,2009NewA...14..630G} for recent GPU
implementations of the GRAPE interface).  In addition, the programming
model for GPUs \citep[as well as the GRAPE-DR,][]{2005astro.ph..9278M},
means that many other kinds of algorithms can (in principle) be
accelerated, although, in practice, it currently seems that
CPU-intensive operations such as direct $N$-body force summation show
substantially better acceleration than, say, treecodes running on the
same hardware.  It appears that commodity components may be poised to
outpace special-purpose computers in this specialized area of
computational science, just as they have already done in
general-purpose computing.

\subsection{The kitchen sink}\label{KitchenSink}

Consistent with our growing understanding of the role of stellar and
binary interactions in collisional stellar systems, the leading
programs in this field are ``kitchen sink'' packages that combine
treatments of dynamics, stellar and binary evolution, and stellar
hydrodynamics within a single simulation.  Of these, the most widely
used are the $N$-body codes {\tt NBODY}
\citep{2001MNRAS.323..630H,2003gnbs.book.....A}, {\tt kira} which is
part of the {\tt starlab} package \citep[e.g.][]{2001MNRAS.321..199P},
and the Monte Carlo codes developed by Giersz
\citep{1998MNRAS.298.1239G,2008MNRAS.389.1858H,2009MNRAS.395.1173G},
Freitag \citep{2003ApJ...593..772F,2006MNRAS.368..121F}, and Fregeau
\citep{2007ApJ...658.1047F}.

Despite the differences in their handling of the large-scale dynamics,
as just outlined, these codes all employ conceptually similar
approaches to stellar and binary evolution and collisions.  All use
approximate descriptions of stellar evolution, generally derived from
look-up tables based on the detailed evolutionary models of
\citet{1989ApJ...347..998E} and \citet{2000MNRAS.315..543H}.  They
also rely on semi-analytic or heuristic rule-based treatments of
binary evolution \citep{1996A&A...309..179P,2002MNRAS.329..897H},
conceptually similar from code to code, but significantly different in
detail.

In most cases, collisions are implemented in the simple
``sticky-sphere'' approximation, where stars are taken to collide (and
merge) if they approach within the sum of their effective radii.  The
effective radii may be calibrated using hydrodynamical simulations,
and mass loss may be included in some approximate way.  Freitag's
Monte Carlo code, geared mainly to studies of galactic nuclei, uses a
more sophisticated approach, interpolating encounter outcomes from a
pre-computed grid of smoothed particles hydrodynamics (SPH)
simulations \citep{2005MNRAS.358.1133F}.  An interesting alternative,
though currently only operational in AMUSE (see
\S\ref{Sect:Future}), is the ``Make Me A Star'' package
\citep[MMAS;][]{2003MNRAS.345..762L}\footnote{See {\tt
    http://webpub.allegheny.edu/employee/j/jalombar/mmas/}} and its
extension ``Make Me a Massive Star''
\citep[MMAMS;][]{2008MNRAS.383L...5G}\footnote{See {\tt
    http://castle.strw.leidenuniv.nl/}}.  MMA(M)S constructs a merged
stellar model by sorting the fluid elements of the original stars by
entropy or density, then recomputing their equilibrium configuration,
using mass loss and shock heating data derived from SPH calculations.

Small-scale dynamics of multiple stellar encounters, such as binary
and higher-order encounters, are often handled by look-up from
pre-computed cross sections or---more commonly---by direct
integration, either in isolation or as part of a larger $N$-body
calculation.  Codes employing direct integration may also include
post-Newtonian terms in the interactions between compact objects
\citep{2006MNRAS.371L..45K}.

\subsection{Future prospects}\label{Sect:Future}

The very comprehensiveness of kitchen-sink codes gives them the great
advantage of applicability to complex stellar systems, but also the
significant disadvantage of inflexibility.  By selecting one of these
codes, one is implicitly choosing a particular hard-coded combination
of dynamical integrator, stellar and binary evolution schemes,
collision prescription, and treatment of multiple dynamics.  The
structure of these codes is such that implementing a different
algorithm within the larger framework is difficult at best, and
practically impossible.

However, studies of dense stellar systems force interactions between
programs that were never intended to interact with other programs, and
by extension require new communication channels between the
programmers responsible for them.  Closely related to this effort is
the ``MUSE'' (MUltiscale, MUltiphysics Software Environment)
project\footnote{{\tt http://muse.li}} \citep{2009NewA...14..369P},
and its successor AMUSE (Astrophysical MUltipurpose Software
Environment), two ambitious open-source efforts in code integration.
(A)MUSE aims at the self-consistent integration of dynamics,
collisions, stellar evolution, and other relevant physical processes,
thereby realizing one vision of the MODEST\footnote{MODEST stands for
  MOdeling DEnse STellar systems, and can be found at {\tt
    http://www.manybody.org/modest}.} community
\citep{2003NewA....8..337H,2003NewA....8..605S,2006NewA...12..201D}.
The long-term goal is a comprehensive environment for modeling dense
stellar systems, including multiphysics/legacy codes and flexible
interfaces to integrate existing software (written in many languages)
within this unifying environment.

Aside from future developments in modular simulation environments, the
recent appearance of programmable high-performance hardware has
spurred the development of new algorithms for implementing the
force-evaluation operations in N-body codes.
\citep{2008NewA...13..498N} have developed extensions of the standard
fourth-order Hermite scheme to higher (sixth and eighth) orders, and
Gaburov and Nitadori (2010, in preparation) have incorporated these
methods into the well-known \citet{AhmadCohen1973} neighbor scheme.
Other intriguing future prospects come from hybridization of direct
methods with hierarchical tree and particle-mesh algorithms, an
approach currently under development by Nitadori and Makino (2009,
private communication), but not yet operational.


\begin{thebibliography}{}
\expandafter\ifx\csname natexlab\endcsname\relax\def\natexlab#1{#1}\fi

\bibitem[{{Aarseth}(2003)}]{2003gnbs.book.....A}
{Aarseth} SJ. 2003.
\newblock \textit{{Gravitational N-Body Simulations}}.
\newblock Gravitational N-Body Simulations, by Sverre J.~Aarseth, pp.~430.~ISBN
  0521432723.~Cambridge, UK: Cambridge University Press, November 2003.

\bibitem[{{Abel}, {Bryan} \& {Norman}(2002)}]{2002Sci...295...93A}
{Abel} T, {Bryan} GL, {Norman} ML. 2002.
\newblock \textit{Science} 295:93--98

\bibitem[{{Abt}(1983)}]{1983ARA&A..21..343A}
{Abt} HA. 1983.
\newblock \textit{\araa} 21:343--372

\bibitem[{{Ahmad} \& {Cohen}(1973{\natexlab{a}})}]{AhmadCohen1973}
{Ahmad} A, {Cohen} L. 1973{\natexlab{a}}.
\newblock \textit{J. of Comp. Phys.} 12:289

\bibitem[{{Ahmad} \& {Cohen}(1973{\natexlab{b}})}]{1973ApJ...179..885A}
{Ahmad} A, {Cohen} L. 1973{\natexlab{b}}.
\newblock \textit{\apj} 179:885--896

\bibitem[{{Allison} et~al.(2009){Allison}, {Goodwin}, {Parker}, {de Grijs},
  {Portegies Zwart} \& {Kouwenhoven}}]{2009ApJ...700L..99A}
{Allison} RJ, {Goodwin} SP, {Parker} RJ, {de Grijs} R, {Portegies Zwart} SF,
  {Kouwenhoven} MBN. 2009.
\newblock \textit{\apjl} 700:L99--L103

\bibitem[{{Ambartsumian}(1938)}]{amb38}
{Ambartsumian} VA. 1938.
\newblock \textit{Sci. Mem. Leningrade State Univ. \#22, ser. Math. Sci.
  (astronomy)} 4:19

\bibitem[{{Anders} et~al.(2004){Anders}, {de Grijs}, {Fritze-v.~Alvensleben} \&
  {Bissantz}}]{2004MNRAS.347...17A}
{Anders} P, {de Grijs} R, {Fritze-v.~Alvensleben} U, {Bissantz} N. 2004.
\newblock \textit{\mnras} 347:17--28

\bibitem[{{Andersen} et~al.(2009){Andersen}, {Zinnecker}, {Moneti},
  {McCaughrean}, {Brandl} et~al.}]{2009ApJ...707.1347A}
{Andersen} M, {Zinnecker} H, {Moneti} A, {McCaughrean} MJ, {Brandl} B, et~al.
  2009.
\newblock \textit{\apj} 707:1347--1360

\bibitem[{{Antonov}(1962)}]{1962spss.book.....A}
{Antonov} VA. 1962.
\newblock \textit{{Solution of the problem of stability of stellar system
  Emden's density law and the spherical distribution of velocities}}

\bibitem[{{Applegate}(1986)}]{1986ApJ...301..132A}
{Applegate} JH. 1986.
\newblock \textit{\apj} 301:132--144

\bibitem[{{Arp} \& {Sandage}(1985)}]{1985AJ.....90.1163A}
{Arp} H, {Sandage} A. 1985.
\newblock \textit{\aj} 90:1163--1171

\bibitem[{{Arzoumanian}, {Chernoff} \& {Cordes}(2002)}]{2002ApJ...568..289A}
{Arzoumanian} Z, {Chernoff} DF, {Cordes} JM. 2002.
\newblock \textit{\apj} 568:289--301

\bibitem[{{Ascenso}, {Alves} \& {Lago}(2009)}]{2009A&A...495..147A}
{Ascenso} J, {Alves} J, {Lago} MTVT. 2009.
\newblock \textit{\aap} 495:147--155

\bibitem[{{Ascenso} et~al.(2007){Ascenso}, {Alves}, {Vicente} \&
  {Lago}}]{2007A&A...476..199A}
{Ascenso} J, {Alves} J, {Vicente} S, {Lago} MTVT. 2007.
\newblock \textit{\aap} 476:199--215

\bibitem[{{Ashman} \& {Zepf}(2001)}]{2001AJ....122.1888A}
{Ashman} KM, {Zepf} SE. 2001.
\newblock \textit{\aj} 122:1888--1895

\bibitem[{{Backer}(1993)}]{1993ASPC...36...11B}
{Backer} DC. 1993.
\newblock In \textit{Planets Around Pulsars}, ed. {J.~A.~Phillips,
  S.~E.~Thorsett, \& S.~R.~Kulkarni}, vol.~36 of \textit{Astronomical Society
  of the Pacific Conference Series}

\bibitem[{{Bacon}, {Sigurdsson} \& {Davies}(1996)}]{1996MNRAS.281..830B}
{Bacon} D, {Sigurdsson} S, {Davies} MB. 1996.
\newblock \textit{\mnras} 281:830--846

\bibitem[{{Bahcall} \& {Wolf}(1976)}]{1976ApJ...209..214B}
{Bahcall} JN, {Wolf} RA. 1976.
\newblock \textit{\apj} 209:214--232

\bibitem[{{Barmby} et~al.(2009){Barmby}, {Perina}, {Bellazzini}, {Cohen},
  {Hodge} et~al.}]{2009AJ....138.1667B}
{Barmby} P, {Perina} S, {Bellazzini} M, {Cohen} JG, {Hodge} PW, et~al. 2009.
\newblock \textit{\aj} 138:1667--1680

\bibitem[{{Barnes} \& {Hut}(1986)}]{1986Natur.324..446B}
{Barnes} J, {Hut} P. 1986.
\newblock \textit{\nat} 324:446--449

\bibitem[{{Bastian}(2008)}]{2008MNRAS.390..759B}
{Bastian} N. 2008.
\newblock \textit{\mnras} 390:759--768

\bibitem[{{Bastian}, {Covey} \& {Meyer}(2010)}]{2010arXiv1001.2965B}
{Bastian} N, {Covey} KR, {Meyer} MR. 2010.
\newblock \textit{\araa, in press (ArXiv:1001.2965)}

\bibitem[{{Bastian} \& {de Mink}(2009)}]{2009MNRAS.398L..11B}
{Bastian} N, {de Mink} SE. 2009.
\newblock \textit{\mnras} 398:L11--L15

\bibitem[{{Bastian} et~al.(2008){Bastian}, {Gieles}, {Goodwin}, {Trancho},
  {Smith} et~al.}]{2008MNRAS.389..223B}
{Bastian} N, {Gieles} M, {Goodwin} SP, {Trancho} G, {Smith} LJ, et~al. 2008.
\newblock \textit{\mnras} 389:223--230

\bibitem[{{Bastian} et~al.(2005){Bastian}, {Gieles}, {Lamers}, {Scheepmaker} \&
  {de Grijs}}]{2005A&A...431..905B}
{Bastian} N, {Gieles} M, {Lamers} HJGLM, {Scheepmaker} RA, {de Grijs} R. 2005.
\newblock \textit{\aap} 431:905--924

\bibitem[{{Bastian} et~al.(2007){Bastian}, {Konstantopoulos}, {Smith},
  {Trancho}, {Westmoquette} \& {Gallagher}}]{2007MNRAS.379.1333B}
{Bastian} N, {Konstantopoulos} I, {Smith} LJ, {Trancho} G, {Westmoquette} MS,
  {Gallagher} JS. 2007.
\newblock \textit{\mnras} 379:1333--1342

\bibitem[{{Bastian} et~al.(2006){Bastian}, {Saglia}, {Goudfrooij},
  {Kissler-Patig}, {Maraston} et~al.}]{2006A&A...448..881B}
{Bastian} N, {Saglia} RP, {Goudfrooij} P, {Kissler-Patig} M, {Maraston} C,
  et~al. 2006.
\newblock \textit{\aap} 448:881--891

\bibitem[{{Bastian} et~al.(2009){Bastian}, {Trancho}, {Konstantopoulos} \&
  {Miller}}]{2009ApJ...701..607B}
{Bastian} N, {Trancho} G, {Konstantopoulos} IS, {Miller} BW. 2009.
\newblock \textit{\apj} 701:607--619

\bibitem[{{Bate}, {Bonnell} \& {Bromm}(2003)}]{2003MNRAS.339..577B}
{Bate} MR, {Bonnell} IA, {Bromm} V. 2003.
\newblock \textit{\mnras} 339:577--599

\bibitem[{{Battinelli} \& {Capuzzo-Dolcetta}(1991)}]{1991MNRAS.249...76B}
{Battinelli} P, {Capuzzo-Dolcetta} R. 1991.
\newblock \textit{\mnras} 249:76--83

\bibitem[{{Baumgardt}(1998)}]{1998A&A...330..480B}
{Baumgardt} H. 1998.
\newblock \textit{\aap} 330:480--491

\bibitem[{{Baumgardt}(2001)}]{2001MNRAS.325.1323B}
{Baumgardt} H. 2001.
\newblock \textit{\mnras} 325:1323--1331

\bibitem[{{Baumgardt}, {De Marchi} \& {Kroupa}(2008)}]{2008ApJ...685..247B}
{Baumgardt} H, {De Marchi} G, {Kroupa} P. 2008.
\newblock \textit{\apj} 685:247--253

\bibitem[{{Baumgardt} \& {Kroupa}(2007)}]{2007MNRAS.380.1589B}
{Baumgardt} H, {Kroupa} P. 2007.
\newblock \textit{\mnras} 380:1589--1598

\bibitem[{{Baumgardt}, {Kroupa} \& {Parmentier}(2008)}]{2008MNRAS.384.1231B}
{Baumgardt} H, {Kroupa} P, {Parmentier} G. 2008.
\newblock \textit{\mnras} 384:1231--1241

\bibitem[{{Baumgardt} \& {Makino}(2003)}]{2003MNRAS.340..227B}
{Baumgardt} H, {Makino} J. 2003.
\newblock \textit{\mnras} 340:227--246

\bibitem[{{Baumgardt}, {Makino} \& {Hut}(2005)}]{2005ApJ...620..238B}
{Baumgardt} H, {Makino} J, {Hut} P. 2005.
\newblock \textit{\apj} 620:238--243

\bibitem[{{Bell} et~al.(2008){Bell}, {Zucker}, {Belokurov}, {Sharma},
  {Johnston} et~al.}]{2008ApJ...680..295B}
{Bell} EF, {Zucker} DB, {Belokurov} V, {Sharma} S, {Johnston} KV, et~al. 2008.
\newblock \textit{\apj} 680:295--311

\bibitem[{{Bellazzini} et~al.(2002){Bellazzini}, {Fusi Pecci}, {Messineo},
  {Monaco} \& {Rood}}]{2002AJ....123.1509B}
{Bellazzini} M, {Fusi Pecci} F, {Messineo} M, {Monaco} L, {Rood} RT. 2002.
\newblock \textit{\aj} 123:1509--1527

\bibitem[{{Belleman}, {B{\'e}dorf} \& {Portegies
  Zwart}(2008)}]{2008NewA...13..103B}
{Belleman} RG, {B{\'e}dorf} J, {Portegies Zwart} SF. 2008.
\newblock \textit{New Astronomy} 13:103--112

\bibitem[{{Benz} \& {Hills}(1987)}]{1987ApJ...323..614B}
{Benz} W, {Hills} JG. 1987.
\newblock \textit{\apj} 323:614--628

\bibitem[{{Bettwieser} \& {Sugimoto}(1984)}]{1984MNRAS.208..493B}
{Bettwieser} E, {Sugimoto} D. 1984.
\newblock \textit{\mnras} 208:493--509

\bibitem[{{Bhattacharya} \& {van den Heuvel}(1991)}]{1991PhR...203....1B}
{Bhattacharya} D, {van den Heuvel} EPJ. 1991.
\newblock \textit{\physrep} 203:1--124

\bibitem[{{Bik} et~al.(2003){Bik}, {Lamers}, {Bastian}, {Panagia} \&
  {Romaniello}}]{2003A&A...397..473B}
{Bik} A, {Lamers} HJGLM, {Bastian} N, {Panagia} N, {Romaniello} M. 2003.
\newblock \textit{\aap} 397:473--486

\bibitem[{{Binney} \& {Tremaine}(2008)}]{2008gady.book.....B}
{Binney} J, {Tremaine} S. 2008.
\newblock \textit{{Galactic Dynamics: Second Edition}}.
\newblock Princeton University Press

\bibitem[{{Bisseling}(2004)}]{2004Bisseling.book}
{Bisseling} RH. 2004.
\newblock \textit{{Parallel Scientific Computation: A Structured Approach using
  BSP and MPI}}.
\newblock Oxford University Press, pp.~324.~ISBN 0-19-852939-2, March 2004.

\bibitem[{{Bonnell} \& {Bate}(2006)}]{2006MNRAS.370..488B}
{Bonnell} IA, {Bate} MR. 2006.
\newblock \textit{\mnras} 370:488--494

\bibitem[{{Bonnell}, {Bate} \& {Vine}(2003)}]{2003MNRAS.343..413B}
{Bonnell} IA, {Bate} MR, {Vine} SG. 2003.
\newblock \textit{\mnras} 343:413--418

\bibitem[{{Bosch}, {Terlevich} \& {Terlevich}(2009)}]{2009AJ....137.3437B}
{Bosch} G, {Terlevich} E, {Terlevich} R. 2009.
\newblock \textit{\aj} 137:3437--3441

\bibitem[{{Boutloukos} \& {Lamers}(2003)}]{2003MNRAS.338..717B}
{Boutloukos} SG, {Lamers} HJGLM. 2003.
\newblock \textit{\mnras} 338:717--732

\bibitem[{{Brandl} et~al.(1996){Brandl}, {Sams}, {Bertoldi}, {Eckart}, {Genzel}
  et~al.}]{1996ApJ...466..254B}
{Brandl} B, {Sams} BJ, {Bertoldi} F, {Eckart} A, {Genzel} R, et~al. 1996.
\newblock \textit{\apj} 466:254

\bibitem[{{Brandner} et~al.(2008){Brandner}, {Clark}, {Stolte}, {Waters},
  {Negueruela} \& {Goodwin}}]{2008A&A...478..137B}
{Brandner} W, {Clark} JS, {Stolte} A, {Waters} R, {Negueruela} I, {Goodwin} SP.
  2008.
\newblock \textit{\aap} 478:137--149

\bibitem[{{Bruzual} \& {Charlot}(2003)}]{2003MNRAS.344.1000B}
{Bruzual} G, {Charlot} S. 2003.
\newblock \textit{\mnras} 344:1000--1028

\bibitem[{{Caldwell} et~al.(2009){Caldwell}, {Harding}, {Morrison}, {Rose},
  {Schiavon} \& {Kriessler}}]{2009AJ....137...94C}
{Caldwell} N, {Harding} P, {Morrison} H, {Rose} JA, {Schiavon} R, {Kriessler}
  J. 2009.
\newblock \textit{\aj} 137:94--110

\bibitem[{{Campbell} et~al.(1992){Campbell}, {Hunter}, {Holtzman}, {Lauer},
  {Shayer} et~al.}]{1992AJ....104.1721C}
{Campbell} B, {Hunter} DA, {Holtzman} JA, {Lauer} TR, {Shayer} EJ, et~al. 1992.
\newblock \textit{\aj} 104:1721--1742

\bibitem[{{Campbell} et~al.(2010){Campbell}, {Evans}, {Mackey}, {Gieles},
  {Alves} et~al.}]{2010arXiv1002.0288C}
{Campbell} MA, {Evans} CJ, {Mackey} AD, {Gieles} M, {Alves} J, et~al. 2010.
\newblock \textit{\mnras, in press (ArXiv:1002.0288)}

\bibitem[{{Cantiello}, {Brocato} \& {Blakeslee}(2009)}]{2009A&A...503...87C}
{Cantiello} M, {Brocato} E, {Blakeslee} JP. 2009.
\newblock \textit{\aap} 503:87--101

\bibitem[{{Cappellaro}, {Evans} \& {Turatto}(1999)}]{1999A&A...351..459C}
{Cappellaro} E, {Evans} R, {Turatto} M. 1999.
\newblock \textit{\aap} 351:459--466

\bibitem[{{Carretta} et~al.(2008){Carretta}, {Bragaglia}, {Gratton} \&
  {Lucatello}}]{2008arXiv0811.3591C}
{Carretta} E, {Bragaglia} A, {Gratton} RG, {Lucatello} S. 2008.
\newblock \textit{ArXiv e-prints}

\bibitem[{{Casertano} \& {Hut}(1985)}]{1985ApJ...298...80C}
{Casertano} S, {Hut} P. 1985.
\newblock \textit{\apj} 298:80--94

\bibitem[{{Chabrier}(2003)}]{2003PASP..115..763C}
{Chabrier} G. 2003.
\newblock \textit{\pasp} 115:763--795

\bibitem[{{Chandar}, {Bianchi} \& {Ford}(2000)}]{2000AJ....120.3088C}
{Chandar} R, {Bianchi} L, {Ford} HC. 2000.
\newblock \textit{\aj} 120:3088--3097

\bibitem[{{Chandar} et~al.(1999){Chandar}, {Bianchi}, {Ford} \&
  {Salasnich}}]{1999PASP..111..794C}
{Chandar} R, {Bianchi} L, {Ford} HC, {Salasnich} B. 1999.
\newblock \textit{\pasp} 111:794--800

\bibitem[{{Chandar}, {Fall} \& {Whitmore}(2006)}]{2006ApJ...650L.111C}
{Chandar} R, {Fall} SM, {Whitmore} BC. 2006.
\newblock \textit{\apjl} 650:L111--L114

\bibitem[{{Chatterjee}, {Fregeau} \& {Rasio}(2008)}]{2008IAUS..246..151C}
{Chatterjee} S, {Fregeau} JM, {Rasio} FA. 2008.
\newblock In \textit{IAU Symposium}, ed. {E.~Vesperini, M.~Giersz, \&
  A.~Sills}, vol. 246 of \textit{IAU Symposium}

\bibitem[{{Chernoff} \& {Weinberg}(1990)}]{1990ApJ...351..121C}
{Chernoff} DF, {Weinberg} MD. 1990.
\newblock \textit{\apj} 351:121--156

\bibitem[{{Chiosi} et~al.(2006){Chiosi}, {Vallenari}, {Held}, {Rizzi} \&
  {Moretti}}]{2006A&A...452..179C}
{Chiosi} E, {Vallenari} A, {Held} EV, {Rizzi} L, {Moretti} A. 2006.
\newblock \textit{\aap} 452:179--193

\bibitem[{{Clark} et~al.(2008){Clark}, {Muno}, {Negueruela}, {Dougherty},
  {Crowther} et~al.}]{2008A&A...477..147C}
{Clark} JS, {Muno} MP, {Negueruela} I, {Dougherty} SM, {Crowther} PA, et~al.
  2008.
\newblock \textit{\aap} 477:147--163

\bibitem[{{Clark} et~al.(2005){Clark}, {Negueruela}, {Crowther} \&
  {Goodwin}}]{2005A&A...434..949C}
{Clark} JS, {Negueruela} I, {Crowther} PA, {Goodwin} SP. 2005.
\newblock \textit{\aap} 434:949--969

\bibitem[{{Clark} et~al.(2009){Clark}, {Negueruela}, {Davies}, {Larionov},
  {Ritchie} et~al.}]{2009A&A...498..109C}
{Clark} JS, {Negueruela} I, {Davies} B, {Larionov} VM, {Ritchie} BW, et~al.
  2009.
\newblock \textit{\aap} 498:109--114

\bibitem[{{Clarke}, {Bonnell} \& {Hillenbrand}(2000)}]{2000prpl.conf..151C}
{Clarke} CJ, {Bonnell} IA, {Hillenbrand} LA. 2000.
\newblock \textit{Protostars and Planets IV} :151

\bibitem[{{Cohn}(1979)}]{1979ApJ...234.1036C}
{Cohn} H. 1979.
\newblock \textit{\apj} 234:1036--1053

\bibitem[{{Cohn}(1980)}]{1980ApJ...242..765C}
{Cohn} H. 1980.
\newblock \textit{\apj} 242:765--771

\bibitem[{{Corbel} \& {Eikenberry}(2004)}]{2004A&A...419..191C}
{Corbel} S, {Eikenberry} SS. 2004.
\newblock \textit{\aap} 419:191--201

\bibitem[{{Cordes} \& {Chernoff}(1998)}]{1998ApJ...505..315C}
{Cordes} JM, {Chernoff} DF. 1998.
\newblock \textit{\apj} 505:315--338

\bibitem[{{Crowther} et~al.(2006){Crowther}, {Hadfield}, {Clark}, {Negueruela}
  \& {Vacca}}]{2006MNRAS.372.1407C}
{Crowther} PA, {Hadfield} LJ, {Clark} JS, {Negueruela} I, {Vacca} WD. 2006.
\newblock \textit{\mnras} 372:1407--1424

\bibitem[{{Davies}(2009)}]{2009arXiv0903.0979D}
{Davies} B. 2009.
\newblock \textit{ArXiv e-prints}

\bibitem[{{Davies} et~al.(2009){Davies}, {Figer}, {Kudritzki}, {Trombley},
  {Kouveliotou} \& {Wachter}}]{2009arXiv0910.4859D}
{Davies} B, {Figer} DF, {Kudritzki} R, {Trombley} C, {Kouveliotou} C, {Wachter}
  S. 2009.
\newblock \textit{ArXiv e-prints}

\bibitem[{{Davies} et~al.(2007){Davies}, {Figer}, {Kudritzki}, {MacKenty},
  {Najarro} \& {Herrero}}]{2007ApJ...671..781D}
{Davies} B, {Figer} DF, {Kudritzki} RP, {MacKenty} J, {Najarro} F, {Herrero} A.
  2007.
\newblock \textit{\apj} 671:781--801

\bibitem[{{Davies} et~al.(2008){Davies}, {Figer}, {Law}, {Kudritzki}, {Najarro}
  et~al.}]{2008ApJ...676.1016D}
{Davies} B, {Figer} DF, {Law} CJ, {Kudritzki} R, {Najarro} F, et~al. 2008.
\newblock \textit{\apj} 676:1016--1028

\bibitem[{{Davies}(1995)}]{1995MNRAS.276..887D}
{Davies} MB. 1995.
\newblock \textit{\mnras} 276:887--905

\bibitem[{{Davies} et~al.(2006){Davies}, {Amaro-Seoane}, {Bassa}, {Dale}, {de
  Angeli} et~al.}]{2006NewA...12..201D}
{Davies} MB, {Amaro-Seoane} P, {Bassa} C, {Dale} J, {de Angeli} F, et~al. 2006.
\newblock \textit{New Astronomy} 12:201--214

\bibitem[{{Davies}, {Piotto} \& {de Angeli}(2004)}]{2004MNRAS.349..129D}
{Davies} MB, {Piotto} G, {de Angeli} F. 2004.
\newblock \textit{\mnras} 349:129--134

\bibitem[{{de Grijs} \& {Anders}(2006)}]{2006MNRAS.366..295D}
{de Grijs} R, {Anders} P. 2006.
\newblock \textit{\mnras} 366:295--307

\bibitem[{{de Grijs} et~al.(2003){de Grijs}, {Fritze-v.~Alvensleben}, {Anders},
  {Gallagher}, {Bastian} et~al.}]{2003MNRAS.342..259D}
{de Grijs} R, {Fritze-v.~Alvensleben} U, {Anders} P, {Gallagher} JS, {Bastian}
  N, et~al. 2003.
\newblock \textit{\mnras} 342:259--273

\bibitem[{{de Grijs} et~al.(2002{\natexlab{a}}){de Grijs}, {Gilmore}, {Johnson}
  \& {Mackey}}]{2002MNRAS.331..245D}
{de Grijs} R, {Gilmore} GF, {Johnson} RA, {Mackey} AD. 2002{\natexlab{a}}.
\newblock \textit{\mnras} 331:245--258

\bibitem[{{de Grijs} et~al.(2002{\natexlab{b}}){de Grijs}, {Gilmore}, {Mackey},
  {Wilkinson}, {Beaulieu} et~al.}]{2002MNRAS.337..597D}
{de Grijs} R, {Gilmore} GF, {Mackey} AD, {Wilkinson} MI, {Beaulieu} SF, et~al.
  2002{\natexlab{b}}.
\newblock \textit{\mnras} 337:597--608

\bibitem[{{de Grijs} \& {Goodwin}(2008)}]{2008MNRAS.383.1000D}
{de Grijs} R, {Goodwin} SP. 2008.
\newblock \textit{\mnras} 383:1000--1006

\bibitem[{{de Grijs} \& {Parmentier}(2007)}]{2007ChJAA...7..155D}
{de Grijs} R, {Parmentier} G. 2007.
\newblock \textit{Chinese Journal of Astronomy and Astrophysics} 7:155--186

\bibitem[{{de Wit} et~al.(2005){de Wit}, {Testi}, {Palla} \&
  {Zinnecker}}]{2005A&A...437..247D}
{de Wit} WJ, {Testi} L, {Palla} F, {Zinnecker} H. 2005.
\newblock \textit{\aap} 437:247--255

\bibitem[{{Decressin} et~al.(2007){Decressin}, {Meynet}, {Charbonnel},
  {Prantzos} \& {Ekstr{\"o}m}}]{2007A&A...464.1029D}
{Decressin} T, {Meynet} G, {Charbonnel} C, {Prantzos} N, {Ekstr{\"o}m} S. 2007.
\newblock \textit{\aap} 464:1029--1044

\bibitem[{{Dehnen}(2000)}]{2000ApJ...536L..39D}
{Dehnen} W. 2000.
\newblock \textit{\apjl} 536:L39--L42

\bibitem[{{Deiters} \& {Spurzem}(2001)}]{2001A&AT...20...47D}
{Deiters} S, {Spurzem} R. 2001.
\newblock \textit{Astronomical and Astrophysical Transactions} 20:47--50

\bibitem[{{DeLuca} et~al.(2009){DeLuca}, {Caraveo}, {Esposito} \&
  {Hurley}}]{2009ApJ...692..158D}
{DeLuca} A, {Caraveo} PA, {Esposito} P, {Hurley} K. 2009.
\newblock \textit{\apj} 692:158--161

\bibitem[{{D'Ercole} et~al.(2008){D'Ercole}, {Vesperini}, {D'Antona},
  {McMillan} \& {Recchi}}]{2008MNRAS.391..825D}
{D'Ercole} A, {Vesperini} E, {D'Antona} F, {McMillan} SLW, {Recchi} S. 2008.
\newblock \textit{\mnras} 391:825--843

\bibitem[{{Dias} et~al.(2002){Dias}, {Alessi}, {Moitinho} \&
  {L{\'e}pine}}]{2002A&A...389..871D}
{Dias} WS, {Alessi} BS, {Moitinho} A, {L{\'e}pine} JRD. 2002.
\newblock \textit{\aap} 389:871--873

\bibitem[{{Djorgovski} \& {Meylan}(1993)}]{1993ASPC...50.....D}
{Djorgovski} SG, {Meylan} G, eds. 1993.
\newblock \textit{{Structure and dynamics of globular clusters}}, vol.~50 of
  \textit{Astronomical Society of the Pacific Conference Series}

\bibitem[{{Dorband}, {Hemsendorf} \& {Merritt}(2003)}]{2003JCoPh.185..484D}
{Dorband} EN, {Hemsendorf} M, {Merritt} D. 2003.
\newblock \textit{Journal of Computational Physics} 185:484--511

\bibitem[{{Drukier}(1996)}]{1996MNRAS.280..498D}
{Drukier} GA. 1996.
\newblock \textit{\mnras} 280:498--514

\bibitem[{{Drukier}, {Fahlman} \& {Richer}(1992)}]{1992ApJ...386..106D}
{Drukier} GA, {Fahlman} GG, {Richer} HB. 1992.
\newblock \textit{\apj} 386:106--119

\bibitem[{{Duquennoy} \& {Mayor}(1991)}]{1991A&A...248..485D}
{Duquennoy} A, {Mayor} M. 1991.
\newblock \textit{\aap} 248:485--524

\bibitem[{{Ebisuzaki} et~al.(1993){Ebisuzaki}, {Makino}, {Fukushige}, {Taiji},
  {Sugimoto} et~al.}]{1993PASJ...45..269E}
{Ebisuzaki} T, {Makino} J, {Fukushige} T, {Taiji} M, {Sugimoto} D, et~al. 1993.
\newblock \textit{\pasj} 45:269--278

\bibitem[{{Efremov}(1991)}]{1991PAZh...17..404E}
{Efremov} YN. 1991.
\newblock \textit{Pis ma Astronomicheskii Zhurnal} 17:404--409

\bibitem[{{Efremov}(2000)}]{2000AstL...26..558E}
{Efremov} YN. 2000.
\newblock \textit{Astronomy Letters} 26:558--564

\bibitem[{{Eggleton}(2006)}]{2006epbm.book.....E}
{Eggleton} P. 2006.
\newblock \textit{{Evolutionary Processes in Binary and Multiple Stars}}

\bibitem[{{Eggleton}, {Fitchett} \& {Tout}(1989)}]{1989ApJ...347..998E}
{Eggleton} PP, {Fitchett} MJ, {Tout} CA. 1989.
\newblock \textit{\apj} 347:998--1011

\bibitem[{{Einsel} \& {Spurzem}(1999)}]{1999MNRAS.302...81E}
{Einsel} C, {Spurzem} R. 1999.
\newblock \textit{\mnras} 302:81--95

\bibitem[{{Elmegreen}(2007)}]{2007ApJ...668.1064E}
{Elmegreen} BG. 2007.
\newblock \textit{\apj} 668:1064--1082

\bibitem[{{Elmegreen} \& {Efremov}(1997)}]{1997ApJ...480..235E}
{Elmegreen} BG, {Efremov} YN. 1997.
\newblock \textit{\apj} 480:235

\bibitem[{{Elson} \& {Fall}(1985{\natexlab{a}})}]{1985PASP...97..692E}
{Elson} RAW, {Fall} SM. 1985{\natexlab{a}}.
\newblock \textit{\pasp} 97:692--696

\bibitem[{{Elson} \& {Fall}(1985{\natexlab{b}})}]{1985ApJ...299..211E}
{Elson} RAW, {Fall} SM. 1985{\natexlab{b}}.
\newblock \textit{\apj} 299:211--218

\bibitem[{{Elson}, {Fall} \& {Freeman}(1987)}]{1987ApJ...323...54E}
{Elson} RAW, {Fall} SM, {Freeman} KC. 1987.
\newblock \textit{\apj} 323:54--78

\bibitem[{{Elson}, {Freeman} \& {Lauer}(1989)}]{1989ApJ...347L..69E}
{Elson} RAW, {Freeman} KC, {Lauer} TR. 1989.
\newblock \textit{\apjl} 347:L69--L71

\bibitem[{{Espinoza}, {Selman} \& {Melnick}(2009)}]{2009arXiv0903.2222E}
{Espinoza} P, {Selman} FJ, {Melnick} J. 2009.
\newblock \textit{ArXiv e-prints}

\bibitem[{{Fabbiano}, {Zezas} \& {Murray}(2001)}]{2001ApJ...554.1035F}
{Fabbiano} G, {Zezas} A, {Murray} SS. 2001.
\newblock \textit{\apj} 554:1035--1043

\bibitem[{{Fabian}, {Pringle} \& {Rees}(1975)}]{1975MNRAS.172P..15F}
{Fabian} AC, {Pringle} JE, {Rees} MJ. 1975.
\newblock \textit{\mnras} 172:15P

\bibitem[{{Fall}, {Chandar} \& {Whitmore}(2005)}]{2005ApJ...631L.133F}
{Fall} SM, {Chandar} R, {Whitmore} BC. 2005.
\newblock \textit{\apjl} 631:L133--L136

\bibitem[{{Fall} \& {Rees}(1985)}]{1985ApJ...298...18F}
{Fall} SM, {Rees} MJ. 1985.
\newblock \textit{\apj} 298:18--26

\bibitem[{{Fall} \& {Zhang}(2001)}]{2001ApJ...561..751F}
{Fall} SM, {Zhang} Q. 2001.
\newblock \textit{\apj} 561:751--765

\bibitem[{{Fellhauer}, {Wilkinson} \& {Kroupa}(2009)}]{2009MNRAS.397..954F}
{Fellhauer} M, {Wilkinson} MI, {Kroupa} P. 2009.
\newblock \textit{\mnras} 397:954--962

\bibitem[{{Ferraro}, {Fusi Pecci} \& {Bellazzini}(1995)}]{1995A&A...294...80F}
{Ferraro} FR, {Fusi Pecci} F, {Bellazzini} M. 1995.
\newblock \textit{\aap} 294:80--88

\bibitem[{{Figer} et~al.(2006){Figer}, {MacKenty}, {Robberto}, {Smith},
  {Najarro} et~al.}]{2006ApJ...643.1166F}
{Figer} DF, {MacKenty} JW, {Robberto} M, {Smith} K, {Najarro} F, et~al. 2006.
\newblock \textit{\apj} 643:1166--1179

\bibitem[{{Figer}, {McLean} \& {Morris}(1999)}]{1999ApJ...514..202F}
{Figer} DF, {McLean} IS, {Morris} M. 1999.
\newblock \textit{\apj} 514:202--220

\bibitem[{{Figer} et~al.(2002){Figer}, {Najarro}, {Gilmore}, {Morris}, {Kim}
  et~al.}]{2002ApJ...581..258F}
{Figer} DF, {Najarro} F, {Gilmore} D, {Morris} M, {Kim} SS, et~al. 2002.
\newblock \textit{\apj} 581:258--275

\bibitem[{{Figer} et~al.(1998){Figer}, {Najarro}, {Morris}, {McLean}, {Geballe}
  et~al.}]{1998ApJ...506..384F}
{Figer} DF, {Najarro} F, {Morris} M, {McLean} IS, {Geballe} TR, et~al. 1998.
\newblock \textit{\apj} 506:384--404

\bibitem[{{Fleck} et~al.(2005){Fleck}, {Boily}, {Lan{\c c}on}, {Heggie} \&
  {Deiters}}]{2005sf2a.conf..605F}
{Fleck} J, {Boily} C, {Lan{\c c}on} A, {Heggie} D, {Deiters} S. 2005.
\newblock In \textit{SF2A-2005: Semaine de l'Astrophysique Francaise}, ed.
  {F.~Casoli, T.~Contini, J.~M.~Hameury, \& L.~Pagani}

\bibitem[{{Fleck} et~al.(2006){Fleck}, {Boily}, {Lan{\c c}on} \&
  {Deiters}}]{2006MNRAS.369.1392F}
{Fleck} JJ, {Boily} CM, {Lan{\c c}on} A, {Deiters} S. 2006.
\newblock \textit{\mnras} 369:1392--1406

\bibitem[{{Foley} et~al.(2006){Foley}, {Li}, {Moore}, {Wong}, {Pooley} \&
  {Filippenko}}]{2006CBET..695....1F}
{Foley} RJ, {Li} W, {Moore} M, {Wong} DS, {Pooley} D, {Filippenko} AV. 2006.
\newblock \textit{Central Bureau Electronic Telegrams} 695:1

\bibitem[{{Forte}, {Vega} \& {Faifer}(2009)}]{2009MNRAS.397.1003F}
{Forte} JC, {Vega} EI, {Faifer} F. 2009.
\newblock \textit{\mnras} 397:1003--1020

\bibitem[{Foster \& Kesselman(2004)}]{FosterAndKesselman}
Foster I, Kesselman C, eds. 2004.
\newblock \textit{The Grid: Blueprint for a New Computing Infrastructure}.
\newblock Published by Morgan Kaufmann

\bibitem[{{Fregeau} et~al.(2004){Fregeau}, {Cheung}, {Portegies Zwart} \&
  {Rasio}}]{2004MNRAS.352....1F}
{Fregeau} JM, {Cheung} P, {Portegies Zwart} SF, {Rasio} FA. 2004.
\newblock \textit{\mnras} 352:1--19

\bibitem[{{Fregeau} et~al.(2003){Fregeau}, {G{\"u}rkan}, {Joshi} \&
  {Rasio}}]{2003ApJ...593..772F}
{Fregeau} JM, {G{\"u}rkan} MA, {Joshi} KJ, {Rasio} FA. 2003.
\newblock \textit{\apj} 593:772--787

\bibitem[{{Fregeau} \& {Rasio}(2007)}]{2007ApJ...658.1047F}
{Fregeau} JM, {Rasio} FA. 2007.
\newblock \textit{\apj} 658:1047--1061

\bibitem[{{Freitag} \& {Benz}(2001)}]{2001A&A...375..711F}
{Freitag} M, {Benz} W. 2001.
\newblock \textit{\aap} 375:711--738

\bibitem[{{Freitag} \& {Benz}(2005)}]{2005MNRAS.358.1133F}
{Freitag} M, {Benz} W. 2005.
\newblock \textit{\mnras} 358:1133--1158

\bibitem[{{Freitag}, {Rasio} \& {Baumgardt}(2006)}]{2006MNRAS.368..121F}
{Freitag} M, {Rasio} FA, {Baumgardt} H. 2006.
\newblock \textit{\mnras} 368:121--140

\bibitem[{{Fujii} et~al.(2007){Fujii}, {Iwasawa}, {Funato} \&
  {Makino}}]{2007PASJ...59.1095F}
{Fujii} M, {Iwasawa} M, {Funato} Y, {Makino} J. 2007.
\newblock \textit{\pasj} 59:1095--

\bibitem[{{Fukushige} \& {Heggie}(1995)}]{1995MNRAS.276..206F}
{Fukushige} T, {Heggie} DC. 1995.
\newblock \textit{\mnras} 276:206--218

\bibitem[{{Fukushige} \& {Heggie}(2000)}]{2000MNRAS.318..753F}
{Fukushige} T, {Heggie} DC. 2000.
\newblock \textit{\mnras} 318:753--761

\bibitem[{{Gaburov} \& {Gieles}(2008)}]{2008MNRAS.391..190G}
{Gaburov} E, {Gieles} M. 2008.
\newblock \textit{\mnras} 391:190--196

\bibitem[{{Gaburov}, {Harfst} \& {Portegies Zwart}(2009)}]{2009NewA...14..630G}
{Gaburov} E, {Harfst} S, {Portegies Zwart} SF. 2009.
\newblock \textit{New Astronomy} 14:630--637

\bibitem[{{Gaburov}, {Lombardi} \& {Portegies
  Zwart}(2008)}]{2008MNRAS.383L...5G}
{Gaburov} E, {Lombardi} JC, {Portegies Zwart} S. 2008.
\newblock \textit{\mnras} 383:L5--L9

\bibitem[{{Gao} et~al.(1991){Gao}, {Goodman}, {Cohn} \&
  {Murphy}}]{1991ApJ...370..567G}
{Gao} B, {Goodman} J, {Cohn} H, {Murphy} B. 1991.
\newblock \textit{\apj} 370:567--582

\bibitem[{{Gao} et~al.(2003){Gao}, {Wang}, {Appleton} \&
  {Lucas}}]{2003ApJ...596L.171G}
{Gao} Y, {Wang} QD, {Appleton} PN, {Lucas} RA. 2003.
\newblock \textit{\apjl} 596:L171--L174

\bibitem[{{Geyer} \& {Burkert}(2001)}]{2001MNRAS.323..988G}
{Geyer} MP, {Burkert} A. 2001.
\newblock \textit{\mnras} 323:988--994

\bibitem[{{Gieles}(2009)}]{2009MNRAS.394.2113G}
{Gieles} M. 2009.
\newblock \textit{\mnras} 394:2113

\bibitem[{{Gieles}, {Athanassoula} \& {Portegies
  Zwart}(2007)}]{2007MNRAS.376..809G}
{Gieles} M, {Athanassoula} E, {Portegies Zwart} SF. 2007.
\newblock \textit{\mnras} 376:809--819

\bibitem[{{Gieles} \& {Baumgardt}(2008)}]{2008MNRAS.389L..28G}
{Gieles} M, {Baumgardt} H. 2008.
\newblock \textit{\mnras} 389:L28--L32

\bibitem[{{Gieles}, {Lamers} \& {Portegies Zwart}(2007)}]{2007ApJ...668..268G}
{Gieles} M, {Lamers} HJGLM, {Portegies Zwart} SF. 2007.
\newblock \textit{\apj} 668:268--274

\bibitem[{{Gieles} et~al.(2006{\natexlab{a}}){Gieles}, {Larsen}, {Bastian} \&
  {Stein}}]{2006A&A...450..129G}
{Gieles} M, {Larsen} SS, {Bastian} N, {Stein} IT. 2006{\natexlab{a}}.
\newblock \textit{\aap} 450:129--145

\bibitem[{{Gieles} et~al.(2006{\natexlab{b}}){Gieles}, {Larsen}, {Scheepmaker},
  {Bastian}, {Haas} \& {Lamers}}]{2006A&A...446L...9G}
{Gieles} M, {Larsen} SS, {Scheepmaker} RA, {Bastian} N, {Haas} MR, {Lamers}
  HJGLM. 2006{\natexlab{b}}.
\newblock \textit{\aap} 446:L9--L12

\bibitem[{{Gieles} et~al.(2006{\natexlab{c}}){Gieles}, {Portegies Zwart},
  {Baumgardt}, {Athanassoula}, {Lamers} et~al.}]{2006MNRAS.371..793G}
{Gieles} M, {Portegies Zwart} SF, {Baumgardt} H, {Athanassoula} E, {Lamers}
  HJGLM, et~al. 2006{\natexlab{c}}.
\newblock \textit{\mnras} 371:793--804

\bibitem[{{Gieles}, {Sana} \& {Portegies Zwart}(2009)}]{2009arXiv0911.1557G}
{Gieles} M, {Sana} H, {Portegies Zwart} SF. 2009.
\newblock \textit{MNRAS, in press (ArXiv:0911.1557)}

\bibitem[{{Giersz}(1998)}]{1998MNRAS.298.1239G}
{Giersz} M. 1998.
\newblock \textit{\mnras} 298:1239--1248

\bibitem[{{Giersz}(2001)}]{2001MNRAS.324..218G}
{Giersz} M. 2001.
\newblock \textit{\mnras} 324:218--230

\bibitem[{{Giersz}(2006)}]{2006MNRAS.371..484G}
{Giersz} M. 2006.
\newblock \textit{\mnras} 371:484--494

\bibitem[{{Giersz} \& {Heggie}(1994)}]{1994MNRAS.268..257G}
{Giersz} M, {Heggie} DC. 1994.
\newblock \textit{\mnras} 268:257

\bibitem[{{Giersz} \& {Heggie}(2009)}]{2009MNRAS.395.1173G}
{Giersz} M, {Heggie} DC. 2009.
\newblock \textit{\mnras} 395:1173--1183

\bibitem[{{Giersz} \& {Spurzem}(2003)}]{2003MNRAS.343..781G}
{Giersz} M, {Spurzem} R. 2003.
\newblock \textit{\mnras} 343:781--795

\bibitem[{{Gies}(1987)}]{1987ApJS...64..545G}
{Gies} DR. 1987.
\newblock \textit{\apjs} 64:545--563

\bibitem[{{Gilliland} et~al.(2000){Gilliland}, {Brown}, {Guhathakurta},
  {Sarajedini}, {Milone} et~al.}]{2000ApJ...545L..47G}
{Gilliland} RL, {Brown} TM, {Guhathakurta} P, {Sarajedini} A, {Milone} EF,
  et~al. 2000.
\newblock \textit{\apjl} 545:L47--L51

\bibitem[{{Girardi} et~al.(1995){Girardi}, {Chiosi}, {Bertelli} \&
  {Bressan}}]{1995A&A...298...87G}
{Girardi} L, {Chiosi} C, {Bertelli} G, {Bressan} A. 1995.
\newblock \textit{\aap} 298:87

\bibitem[{{Glebbeek} et~al.(2009){Glebbeek}, {Gaburov}, {de Mink}, {Pols} \&
  {Portegies Zwart}}]{2009A&A...497..255G}
{Glebbeek} E, {Gaburov} E, {de Mink} SE, {Pols} OR, {Portegies Zwart} SF. 2009.
\newblock \textit{\aap} 497:255--264

\bibitem[{{Glebbeek}, {Pols} \& {Hurley}(2008)}]{2008A&A...488.1007G}
{Glebbeek} E, {Pols} OR, {Hurley} JR. 2008.
\newblock \textit{\aap} 488:1007--1015

\bibitem[{{Gnedin} \& {Ostriker}(1997)}]{1997ApJ...474..223G}
{Gnedin} OY, {Ostriker} JP. 1997.
\newblock \textit{\apj} 474:223

\bibitem[{{Gonz{\'a}lez-Mart{\'{\i}}n}, {Fabian} \&
  {Sanders}(2006)}]{2006MNRAS.367.1132G}
{Gonz{\'a}lez-Mart{\'{\i}}n} O, {Fabian} AC, {Sanders} JS. 2006.
\newblock \textit{\mnras} 367:1132--1138

\bibitem[{{Goodman} \& {Hut}(1989)}]{1989Natur.339...40G}
{Goodman} J, {Hut} P. 1989.
\newblock \textit{\nat} 339:40--42

\bibitem[{{Goodwin}(2009)}]{2009Ap&SS.tmp..108G}
{Goodwin} SP. 2009.
\newblock \textit{\apss} :108

\bibitem[{{Goodwin} \& {Bastian}(2006)}]{2006MNRAS.373..752G}
{Goodwin} SP, {Bastian} N. 2006.
\newblock \textit{\mnras} 373:752--758

\bibitem[{{Gouliermis} et~al.(2004){Gouliermis}, {Keller}, {Kontizas},
  {Kontizas} \& {Bellas-Velidis}}]{2004A&A...416..137G}
{Gouliermis} D, {Keller} SC, {Kontizas} M, {Kontizas} E, {Bellas-Velidis} I.
  2004.
\newblock \textit{\aap} 416:137--155

\bibitem[{{Gris{\'e}} et~al.(2008){Gris{\'e}}, {Pakull}, {Soria}, {Motch},
  {Smith} et~al.}]{2008A&A...486..151G}
{Gris{\'e}} F, {Pakull} MW, {Soria} R, {Motch} C, {Smith} IA, et~al. 2008.
\newblock \textit{\aap} 486:151--163

\bibitem[{{Gualandris} et~al.(2005){Gualandris}, {Colpi}, {Portegies Zwart} \&
  {Possenti}}]{2005ApJ...618..845G}
{Gualandris} A, {Colpi} M, {Portegies Zwart} S, {Possenti} A. 2005.
\newblock \textit{\apj} 618:845--851

\bibitem[{{G{\"u}rkan}, {Freitag} \& {Rasio}(2004)}]{2004ApJ...604..632G}
{G{\"u}rkan} MA, {Freitag} M, {Rasio} FA. 2004.
\newblock \textit{\apj} 604:632--652

\bibitem[{{Gutermuth} et~al.(2005){Gutermuth}, {Megeath}, {Pipher}, {Williams},
  {Allen} et~al.}]{2005ApJ...632..397G}
{Gutermuth} RA, {Megeath} ST, {Pipher} JL, {Williams} JP, {Allen} LE, et~al.
  2005.
\newblock \textit{\apj} 632:397--420

\bibitem[{{Gvaramadze} \& {Bomans}(2008)}]{2008A&A...490.1071G}
{Gvaramadze} VV, {Bomans} DJ. 2008.
\newblock \textit{\aap} 490:1071--1077

\bibitem[{{Hamada} \& {Iitaka}(2007)}]{2007astro.ph..3100H}
{Hamada} T, {Iitaka} T. 2007.
\newblock \textit{ArXiv Astrophysics e-prints}

\bibitem[{{Harayama}, {Eisenhauer} \& {Martins}(2008)}]{2008ApJ...675.1319H}
{Harayama} Y, {Eisenhauer} F, {Martins} F. 2008.
\newblock \textit{\apj} 675:1319--1342

\bibitem[{{Harfst} et~al.(2007){Harfst}, {Gualandris}, {Merritt}, {Spurzem},
  {Portegies Zwart} \& {Berczik}}]{2007NewA...12..357H}
{Harfst} S, {Gualandris} A, {Merritt} D, {Spurzem} R, {Portegies Zwart} S,
  {Berczik} P. 2007.
\newblock \textit{New Astronomy} 12:357--377

\bibitem[{{Harfst}, {Portegies Zwart} \& {Stolte}(2009)}]{2009arXiv0911.3058H}
{Harfst} S, {Portegies Zwart} S, {Stolte} A. 2009.
\newblock \textit{ArXiv e-prints}

\bibitem[{{Harris} \& {Zaritsky}(2009)}]{2009AJ....138.1243H}
{Harris} J, {Zaritsky} D. 2009.
\newblock \textit{\aj} 138:1243--1260

\bibitem[{{Harris}(1996)}]{1996AJ....112.1487H}
{Harris} WE. 1996.
\newblock \textit{\aj} 112:1487

\bibitem[{{Harris}(2001)}]{2001stcl.conf..223H}
{Harris} WE. 2001.
\newblock In \textit{Saas-Fee Advanced Course 28: Star Clusters}, ed.
  {L.~Labhardt \& B.~Binggeli}

\bibitem[{{Heger} et~al.(2003){Heger}, {Fryer}, {Woosley}, {Langer} \&
  {Hartmann}}]{2003ApJ...591..288H}
{Heger} A, {Fryer} CL, {Woosley} SE, {Langer} N, {Hartmann} DH. 2003.
\newblock \textit{\apj} 591:288--300

\bibitem[{{Heggie} \& {Hut}(2003)}]{2003gmbp.book.....H}
{Heggie} D, {Hut} P. 2003.
\newblock \textit{{The Gravitational Million-Body Problem: A Multidisciplinary
  Approach to Star Cluster Dynamics}}.
\newblock The Gravitational Million-Body Problem: A Multidisciplinary Approach
  to Star Cluster Dynamics, by Douglas Heggie and Piet Hut.~ Cambridge
  University Press, 2003, 372 pp.

\bibitem[{{Heggie}(1975)}]{1975MNRAS.173..729H}
{Heggie} DC. 1975.
\newblock \textit{\mnras} 173:729--787

\bibitem[{{Heggie}(1992)}]{1992Natur.359..772H}
{Heggie} DC. 1992.
\newblock \textit{\nat} 359:772--773

\bibitem[{{Heggie}(2001)}]{2001ruag.conf..109H}
{Heggie} DC. 2001.
\newblock In \textit{The Restless Universe}, ed. {B.~A.~Steves \&
  A.~J.~Maciejewski}

\bibitem[{{Heggie} \& {Aarseth}(1992)}]{1992MNRAS.257..513H}
{Heggie} DC, {Aarseth} SJ. 1992.
\newblock \textit{\mnras} 257:513--536

\bibitem[{{Heggie} \& {Giersz}(2008)}]{2008MNRAS.389.1858H}
{Heggie} DC, {Giersz} M. 2008.
\newblock \textit{\mnras} 389:1858--1870

\bibitem[{{Heggie} et~al.(2007){Heggie}, {Hut}, {Mineshige}, {Makino} \&
  {Baumgardt}}]{2007PASJ...59L..11H}
{Heggie} DC, {Hut} P, {Mineshige} S, {Makino} J, {Baumgardt} H. 2007.
\newblock \textit{\pasj} 59:L11--L14

\bibitem[{{Heggie}, {Inagaki} \& {McMillan}(1994)}]{1994MNRAS.271..706H}
{Heggie} DC, {Inagaki} S, {McMillan} SLW. 1994.
\newblock \textit{\mnras} 271:706

\bibitem[{{H{\'e}non}(1965)}]{1965AnAp...28...62H}
{H{\'e}non} M. 1965.
\newblock \textit{Annales d'Astrophysique} 28:62

\bibitem[{{Henon}(1973)}]{1973dses.conf..183H}
{Henon} M. 1973.
\newblock In \textit{Saas-Fee Advanced Course 3: Dynamical Structure and
  Evolution of Stellar Systems}, ed. {G.~Contopoulos, M.~Henon, \&
  D.~Lynden-Bell}

\bibitem[{{Hillenbrand} \& {Hartmann}(1998)}]{1998ApJ...492..540H}
{Hillenbrand} LA, {Hartmann} LW. 1998.
\newblock \textit{\apj} 492:540

\bibitem[{{Hills}(1975)}]{1975AJ.....80..809H}
{Hills} JG. 1975.
\newblock \textit{\aj} 80:809--825

\bibitem[{{Hills}(1980)}]{1980ApJ...235..986H}
{Hills} JG. 1980.
\newblock \textit{\apj} 235:986--991

\bibitem[{{Hills} \& {Day}(1976)}]{1976ApL....17...87H}
{Hills} JG, {Day} CA. 1976.
\newblock \textit{\aplett} 17:87

\bibitem[{{Ho} \& {Filippenko}(1996)}]{1996ApJ...466L..83H}
{Ho} LC, {Filippenko} AV. 1996.
\newblock \textit{\apjl} 466:L83

\bibitem[{{Hodge}(1987)}]{1987PASP...99..724H}
{Hodge} P. 1987.
\newblock \textit{\pasp} 99:724--729

\bibitem[{{Hodge}(1961)}]{1961ApJ...133..413H}
{Hodge} PW. 1961.
\newblock \textit{\apj} 133:413

\bibitem[{{Hodge} et~al.(2009){Hodge}, {Krienke}, {Bellazzini}, {Perina},
  {Barmby} et~al.}]{2009AJ....138..770H}
{Hodge} PW, {Krienke} OK, {Bellazzini} M, {Perina} S, {Barmby} P, et~al. 2009.
\newblock \textit{\aj} 138:770--779

\bibitem[{{Hoekstra} et~al.(2008){Hoekstra}, {Portegies Zwart}, {Bubak} \& {
  Sloot}}]{2008PCAA.book.....H}
{Hoekstra} A. G, {Portegies Zwart} S, {Bubak} M, { Sloot} P. 2008.
\newblock \textit{{Towards Distributed Petascale Computing}}.
\newblock Petascale Computing: Algorithms and Applications, by David A. Bader
  (Ed.). Chapman \& Hall/CRC computational science series 565pp. (ISBN:
  9781584889090, ISBN 10: 1584889098)

\bibitem[{{Holtzman} et~al.(1992){Holtzman}, {Faber}, {Shaya}, {Lauer}, {Groth}
  et~al.}]{1992AJ....103..691H}
{Holtzman} JA, {Faber} SM, {Shaya} EJ, {Lauer} TR, {Groth} J, et~al. 1992.
\newblock \textit{\aj} 103:691--702

\bibitem[{{Huff} \& {Stahler}(2006)}]{2006ApJ...644..355H}
{Huff} EM, {Stahler} SW. 2006.
\newblock \textit{\apj} 644:355--363

\bibitem[{{Hunter} et~al.(2003){Hunter}, {Elmegreen}, {Dupuy} \&
  {Mortonson}}]{2003AJ....126.1836H}
{Hunter} DA, {Elmegreen} BG, {Dupuy} TJ, {Mortonson} M. 2003.
\newblock \textit{\aj} 126:1836--1848

\bibitem[{{Hunter} et~al.(2000){Hunter}, {O'Connell}, {Gallagher} \&
  {Smecker-Hane}}]{2000AJ....120.2383H}
{Hunter} DA, {O'Connell} RW, {Gallagher} JS, {Smecker-Hane} TA. 2000.
\newblock \textit{\aj} 120:2383--2401

\bibitem[{{Hunter} et~al.(1995){Hunter}, {Shaya}, {Holtzman}, {Light}, {O'Neil}
  \& {Lynds}}]{1995ApJ...448..179H}
{Hunter} DA, {Shaya} EJ, {Holtzman} JA, {Light} RM, {O'Neil} Jr. EJ, {Lynds} R.
  1995.
\newblock \textit{\apj} 448:179

\bibitem[{{Hurley}(2007)}]{2007MNRAS.379...93H}
{Hurley} JR. 2007.
\newblock \textit{\mnras} 379:93--99

\bibitem[{{Hurley}, {Aarseth} \& {Shara}(2007)}]{2007ApJ...665..707H}
{Hurley} JR, {Aarseth} SJ, {Shara} MM. 2007.
\newblock \textit{\apj} 665:707--718

\bibitem[{{Hurley}, {Pols} \& {Tout}(2000)}]{2000MNRAS.315..543H}
{Hurley} JR, {Pols} OR, {Tout} CA. 2000.
\newblock \textit{\mnras} 315:543--569

\bibitem[{{Hurley} \& {Shara}(2002)}]{2002ApJ...570..184H}
{Hurley} JR, {Shara} MM. 2002.
\newblock \textit{\apj} 570:184--189

\bibitem[{{Hurley} et~al.(2001){Hurley}, {Tout}, {Aarseth} \&
  {Pols}}]{2001MNRAS.323..630H}
{Hurley} JR, {Tout} CA, {Aarseth} SJ, {Pols} OR. 2001.
\newblock \textit{\mnras} 323:630--650

\bibitem[{{Hurley}, {Tout} \& {Pols}(2002)}]{2002MNRAS.329..897H}
{Hurley} JR, {Tout} CA, {Pols} OR. 2002.
\newblock \textit{\mnras} 329:897--928

\bibitem[{{Hut} \& {Bahcall}(1983)}]{1983ApJ...268..319H}
{Hut} P, {Bahcall} JN. 1983.
\newblock \textit{\apj} 268:319--341

\bibitem[{{Hut} \& {Djorgovski}(1992)}]{1992Natur.359..806H}
{Hut} P, {Djorgovski} S. 1992.
\newblock \textit{\nat} 359:806--808

\bibitem[{{Hut} \& {Inagaki}(1985)}]{1985ApJ...298..502H}
{Hut} P, {Inagaki} S. 1985.
\newblock \textit{\apj} 298:502--520

\bibitem[{{Hut}, {Makino} \& {McMillan}(1988)}]{1988Natur.336...31H}
{Hut} P, {Makino} J, {McMillan} S. 1988.
\newblock \textit{\nat} 336:31--35

\bibitem[{{Hut} et~al.(2003){Hut}, {Shara}, {Aarseth}, {Klessen}, {Lombardi}
  et~al.}]{2003NewA....8..337H}
{Hut} P, {Shara} MM, {Aarseth} SJ, {Klessen} RS, {Lombardi} Jr. JC, et~al.
  2003.
\newblock \textit{New Astronomy} 8:337--370

\bibitem[{{Inagaki} \& {Saslaw}(1985)}]{1985ApJ...292..339I}
{Inagaki} S, {Saslaw} WC. 1985.
\newblock \textit{\apj} 292:339--347

\bibitem[{{Innanen}, {Harris} \& {Webbink}(1983)}]{1983AJ.....88..338I}
{Innanen} KA, {Harris} WE, {Webbink} RF. 1983.
\newblock \textit{\aj} 88:338--360

\bibitem[{{Jord{\'a}n} et~al.(2007){Jord{\'a}n}, {McLaughlin}, {C{\^o}t{\'e}},
  {Ferrarese}, {Peng} et~al.}]{2007ApJS..171..101J}
{Jord{\'a}n} A, {McLaughlin} DE, {C{\^o}t{\'e}} P, {Ferrarese} L, {Peng} EW,
  et~al. 2007.
\newblock \textit{\apjs} 171:101--145

\bibitem[{{Joshi}, {Rasio} \& {Portegies Zwart}(2000)}]{2000ApJ...540..969J}
{Joshi} KJ, {Rasio} FA, {Portegies Zwart} S. 2000.
\newblock \textit{\apj} 540:969--982

\bibitem[{{Kaaret} et~al.(2001){Kaaret}, {Prestwich}, {Zezas}, {Murray}, {Kim}
  et~al.}]{2001MNRAS.321L..29K}
{Kaaret} P, {Prestwich} AH, {Zezas} A, {Murray} SS, {Kim} DW, et~al. 2001.
\newblock \textit{\mnras} 321:L29--L32

\bibitem[{{Karakas} \& {Lattanzio}(2007)}]{2007PASA...24..103K}
{Karakas} A, {Lattanzio} JC. 2007.
\newblock \textit{Publications of the Astronomical Society of Australia}
  24:103--117

\bibitem[{{Kennicutt}(1998)}]{1998ARA&A..36..189K}
{Kennicutt} Jr. RC. 1998.
\newblock \textit{\araa} 36:189--232

\bibitem[{{Kennicutt} \& {Chu}(1988)}]{1988AJ.....95..720K}
{Kennicutt} Jr. RC, {Chu} YH. 1988.
\newblock \textit{\aj} 95:720--730

\bibitem[{{Kharchenko} et~al.(2005){Kharchenko}, {Piskunov}, {R{\"o}ser},
  {Schilbach} \& {Scholz}}]{2005A&A...440..403K}
{Kharchenko} NV, {Piskunov} AE, {R{\"o}ser} S, {Schilbach} E, {Scholz} RD.
  2005.
\newblock \textit{\aap} 440:403--408

\bibitem[{{Kim} et~al.(2002){Kim}, {Einsel}, {Lee}, {Spurzem} \&
  {Lee}}]{2002MNRAS.334..310K}
{Kim} E, {Einsel} C, {Lee} HM, {Spurzem} R, {Lee} MG. 2002.
\newblock \textit{\mnras} 334:310--322

\bibitem[{{Kim}, {Lee} \& {Spurzem}(2004)}]{2004MNRAS.351..220K}
{Kim} E, {Lee} HM, {Spurzem} R. 2004.
\newblock \textit{\mnras} 351:220--236

\bibitem[{{Kim} et~al.(2006){Kim}, {Figer}, {Kudritzki} \&
  {Najarro}}]{2006ApJ...653L.113K}
{Kim} SS, {Figer} DF, {Kudritzki} RP, {Najarro} F. 2006.
\newblock \textit{\apjl} 653:L113--L116

\bibitem[{{King}(2002)}]{2002MNRAS.335L..13K}
{King} AR. 2002.
\newblock \textit{\mnras} 335:L13--L16

\bibitem[{{King} et~al.(2001){King}, {Davies}, {Ward}, {Fabbiano} \&
  {Elvis}}]{2001ApJ...552L.109K}
{King} AR, {Davies} MB, {Ward} MJ, {Fabbiano} G, {Elvis} M. 2001.
\newblock \textit{\apjl} 552:L109--L112

\bibitem[{{King}(1962)}]{1962AJ.....67..471K}
{King} IR. 1962.
\newblock \textit{\aj} 67:471

\bibitem[{{King}(1966)}]{1966AJ.....71...64K}
{King} IR. 1966.
\newblock \textit{\aj} 71:64--75

\bibitem[{{Klessen}(2001)}]{2001ApJ...556..837K}
{Klessen} RS. 2001.
\newblock \textit{\apj} 556:837--846

\bibitem[{{Klose} et~al.(2004){Klose}, {Henden}, {Geppert}, {Greiner},
  {Guetter} et~al.}]{2004ApJ...609L..13K}
{Klose} S, {Henden} AA, {Geppert} U, {Greiner} J, {Guetter} HH, et~al. 2004.
\newblock \textit{\apjl} 609:L13--L16

\bibitem[{{Knigge}, {Leigh} \& {Sills}(2009)}]{2009Natur.457..288K}
{Knigge} C, {Leigh} N, {Sills} A. 2009.
\newblock \textit{\nat} 457:288--290

\bibitem[{{Kouveliotou} et~al.(1999){Kouveliotou}, {Strohmayer}, {Hurley}, {van
  Paradijs}, {Finger} et~al.}]{1999ApJ...510L.115K}
{Kouveliotou} C, {Strohmayer} T, {Hurley} K, {van Paradijs} J, {Finger} MH,
  et~al. 1999.
\newblock \textit{\apjl} 510:L115--L118

\bibitem[{{Kouwenhoven} \& {de Grijs}(2008)}]{2008A&A...480..103K}
{Kouwenhoven} MBN, {de Grijs} R. 2008.
\newblock \textit{\aap} 480:103--114

\bibitem[{{Kroupa}(1995)}]{1995MNRAS.277.1491K}
{Kroupa} P. 1995.
\newblock \textit{\mnras} 277:1491

\bibitem[{{Kroupa}(2001)}]{2001MNRAS.322..231K}
{Kroupa} P. 2001.
\newblock \textit{\mnras} 322:231--246

\bibitem[{{Kroupa}(2008)}]{2008LNP...760..181K}
{Kroupa} P. 2008.
\newblock In \textit{Lecture Notes in Physics, Berlin Springer Verlag}, ed.
  {S.~J.~Aarseth, C.~A.~Tout, \& R.~A.~Mardling}, vol. 760 of \textit{Lecture
  Notes in Physics, Berlin Springer Verlag}

\bibitem[{{Kulkarni}, {Hut} \& {McMillan}(1993)}]{1993Natur.364..421K}
{Kulkarni} SR, {Hut} P, {McMillan} S. 1993.
\newblock \textit{\nat} 364:421--423

\bibitem[{{Kupi}, {Amaro-Seoane} \& {Spurzem}(2006)}]{2006MNRAS.371L..45K}
{Kupi} G, {Amaro-Seoane} P, {Spurzem} R. 2006.
\newblock \textit{\mnras} 371:L45--L49

\bibitem[{{Lada} \& {Lada}(2003)}]{2003ARA&A..41...57L}
{Lada} CJ, {Lada} EA. 2003.
\newblock \textit{\araa} 41:57--115

\bibitem[{{Lada}, {Margulis} \& {Dearborn}(1984)}]{1984ApJ...285..141L}
{Lada} CJ, {Margulis} M, {Dearborn} D. 1984.
\newblock \textit{\apj} 285:141--152

\bibitem[{{Lamers} et~al.(2005){Lamers}, {Gieles}, {Bastian}, {Baumgardt},
  {Kharchenko} \& {Portegies Zwart}}]{2005A&A...441..117L}
{Lamers} HJGLM, {Gieles} M, {Bastian} N, {Baumgardt} H, {Kharchenko} NV,
  {Portegies Zwart} S. 2005.
\newblock \textit{\aap} 441:117--129

\bibitem[{{Lamers}, {Gieles} \& {Portegies Zwart}(2005)}]{2005A&A...429..173L}
{Lamers} HJGLM, {Gieles} M, {Portegies Zwart} SF. 2005.
\newblock \textit{\aap} 429:173--179

\bibitem[{{Lang} et~al.(2005){Lang}, {Johnson}, {Goss} \&
  {Rodr{\'{\i}}guez}}]{2005AJ....130.2185L}
{Lang} CC, {Johnson} KE, {Goss} WM, {Rodr{\'{\i}}guez} LF. 2005.
\newblock \textit{\aj} 130:2185--2196

\bibitem[{{Larsen}(2002)}]{2002AJ....124.1393L}
{Larsen} SS. 2002.
\newblock \textit{\aj} 124:1393--1409

\bibitem[{{Larsen}(2004)}]{2004A&A...416..537L}
{Larsen} SS. 2004.
\newblock \textit{\aap} 416:537--553

\bibitem[{{Larsen}(2006)}]{2006pces.conf...35L}
{Larsen} SS. 2006.
\newblock In \textit{Planets to Cosmology: Essential Science in the Final Years
  of the Hubble Space Telescope}, eds. M~{Livio}, S~{Casertano}

\bibitem[{{Larsen}(2009{\natexlab{a}})}]{2009A&A...494..539L}
{Larsen} SS. 2009{\natexlab{a}}.
\newblock \textit{\aap} 494:539--551

\bibitem[{{Larsen}(2009{\natexlab{b}})}]{2009arXiv0911.0796L}
{Larsen} SS. 2009{\natexlab{b}}.
\newblock \textit{\philtrans, in press (ArXiv:0911.0796)}

\bibitem[{{Larsen} et~al.(2008){Larsen}, {Origlia}, {Brodie} \&
  {Gallagher}}]{2008MNRAS.383..263L}
{Larsen} SS, {Origlia} L, {Brodie} J, {Gallagher} JS. 2008.
\newblock \textit{\mnras} 383:263--276

\bibitem[{{Larsen} \& {Richtler}(2000)}]{2000A&A...354..836L}
{Larsen} SS, {Richtler} T. 2000.
\newblock \textit{\aap} 354:836--846

\bibitem[{{Larsen} \& {Richtler}(2004)}]{2004A&A...427..495L}
{Larsen} SS, {Richtler} T. 2004.
\newblock \textit{\aap} 427:495--504

\bibitem[{{Larsen} \& {Richtler}(2006)}]{2006A&A...459..103L}
{Larsen} SS, {Richtler} T. 2006.
\newblock \textit{\aap} 459:103--111

\bibitem[{{Larson}(1981)}]{1981MNRAS.194..809L}
{Larson} RB. 1981.
\newblock \textit{\mnras} 194:809--826

\bibitem[{{Leonard}(1989)}]{1989AJ.....98..217L}
{Leonard} PJT. 1989.
\newblock \textit{\aj} 98:217--226

\bibitem[{{Leonard}(1996)}]{1996ApJ...470..521L}
{Leonard} PJT. 1996.
\newblock \textit{\apj} 470:521

\bibitem[{{Liu} et~al.(2002){Liu}, {Bregman}, {Irwin} \&
  {Seitzer}}]{2002ApJ...581L..93L}
{Liu} JF, {Bregman} JN, {Irwin} J, {Seitzer} P. 2002.
\newblock \textit{\apjl} 581:L93--L96

\bibitem[{{Liu}, {Bregman} \& {Seitzer}(2004)}]{2004ApJ...602..249L}
{Liu} JF, {Bregman} JN, {Seitzer} P. 2004.
\newblock \textit{\apj} 602:249--256

\bibitem[{{Lombardi} et~al.(2003){Lombardi}, {Thrall}, {Deneva}, {Fleming} \&
  {Grabowski}}]{2003MNRAS.345..762L}
{Lombardi} JC, {Thrall} AP, {Deneva} JS, {Fleming} SW, {Grabowski} PE. 2003.
\newblock \textit{\mnras} 345:762--780

\bibitem[{{Lucas} \& {Roche}(2000)}]{2000MNRAS.314..858L}
{Lucas} PW, {Roche} PF. 2000.
\newblock \textit{\mnras} 314:858--864

\bibitem[{{Lutz}(1991)}]{1991A&A...245...31L}
{Lutz} D. 1991.
\newblock \textit{\aap} 245:31--40

\bibitem[{{Lynden-Bell} \& {Eggleton}(1980)}]{1980MNRAS.191..483L}
{Lynden-Bell} D, {Eggleton} PP. 1980.
\newblock \textit{\mnras} 191:483--498

\bibitem[{{Lynden-Bell} \& {Wood}(1968)}]{1968MNRAS.138..495L}
{Lynden-Bell} D, {Wood} R. 1968.
\newblock \textit{\mnras} 138:495

\bibitem[{{Lyne} \& {Lorimer}(1994)}]{1994Natur.369..127L}
{Lyne} AG, {Lorimer} DR. 1994.
\newblock \textit{\nat} 369:127--129

\bibitem[{{Lynga}(1982)}]{1982A&A...109..213L}
{Lynga} G. 1982.
\newblock \textit{\aap} 109:213--222

\bibitem[{{Ma} et~al.(2009){Ma}, {Fan}, {de Grijs}, {Wu}, {Zhou}
  et~al.}]{2009AJ....137.4884M}
{Ma} J, {Fan} Z, {de Grijs} R, {Wu} Z, {Zhou} X, et~al. 2009.
\newblock \textit{\aj} 137:4884--4896

\bibitem[{{Mackey} \& {Broby Nielsen}(2007)}]{2007MNRAS.379..151M}
{Mackey} AD, {Broby Nielsen} P. 2007.
\newblock \textit{\mnras} 379:151--158

\bibitem[{{Mackey} \& {Gilmore}(2003)}]{2003MNRAS.338...85M}
{Mackey} AD, {Gilmore} GF. 2003.
\newblock \textit{\mnras} 338:85--119

\bibitem[{{Mackey} et~al.(2007){Mackey}, {Wilkinson}, {Davies} \&
  {Gilmore}}]{2007MNRAS.379L..40M}
{Mackey} AD, {Wilkinson} MI, {Davies} MB, {Gilmore} GF. 2007.
\newblock \textit{\mnras} 379:L40--L44

\bibitem[{{Mackey} et~al.(2008){Mackey}, {Wilkinson}, {Davies} \&
  {Gilmore}}]{2008MNRAS.386...65M}
{Mackey} AD, {Wilkinson} MI, {Davies} MB, {Gilmore} GF. 2008.
\newblock \textit{\mnras} 386:65--95

\bibitem[{{Madhusudhan} et~al.(2006){Madhusudhan}, {Justham}, {Nelson},
  {Paxton}, {Pfahl} et~al.}]{2006ApJ...640..918M}
{Madhusudhan} N, {Justham} S, {Nelson} L, {Paxton} B, {Pfahl} E, et~al. 2006.
\newblock \textit{\apj} 640:918--922

\bibitem[{{Ma{\'{\i}}z-Apell{\'a}niz}(2001)}]{2001ApJ...563..151M}
{Ma{\'{\i}}z-Apell{\'a}niz} J. 2001.
\newblock \textit{\apj} 563:151--162

\bibitem[{{Ma{\'{\i}}z-Apell{\'a}niz}(2002)}]{2002IAUS..207..697M}
{Ma{\'{\i}}z-Apell{\'a}niz} J. 2002.
\newblock In \textit{Extragalactic Star Clusters}, eds. DP~{Geisler},
  EK~{Grebel}, D~{Minniti}, vol. 207 of \textit{IAU Symposium}

\bibitem[{{Makino}(1996)}]{1996ApJ...471..796M}
{Makino} J. 1996.
\newblock \textit{\apj} 471:796

\bibitem[{{Makino}(2002)}]{2002NewA....7..373M}
{Makino} J. 2002.
\newblock \textit{New Astronomy} 7:373--384

\bibitem[{{Makino}(2005)}]{2005astro.ph..9278M}
{Makino} J. 2005.
\newblock \textit{ArXiv Astrophysics e-prints}

\bibitem[{{Makino} \& {Aarseth}(1992)}]{1992PASJ...44..141M}
{Makino} J, {Aarseth} SJ. 1992.
\newblock \textit{\pasj} 44:141--151

\bibitem[{{Makino} et~al.(2006){Makino}, {Hut}, {Kaplan} \&
  {Sayg{\i}n}}]{2006NewA...12..124M}
{Makino} J, {Hut} P, {Kaplan} M, {Sayg{\i}n} H. 2006.
\newblock \textit{New Astronomy} 12:124--133

\bibitem[{{Makino} \& {Taiji}(1998)}]{1998sssp.book.....M}
{Makino} J, {Taiji} M. 1998.
\newblock \textit{{Scientific simulations with special-purpose computers : The
  GRAPE systems}}.
\newblock Scientific simulations with special-purpose computers : The GRAPE
  systems /by Junichiro Makino \& Makoto Taiji.~Chichester ; Toronto : John
  Wiley \& Sons, c1998.

\bibitem[{{Makishima} et~al.(2000){Makishima}, {Kubota}, {Mizuno}, {Ohnishi},
  {Tashiro} et~al.}]{2000ApJ...535..632M}
{Makishima} K, {Kubota} A, {Mizuno} T, {Ohnishi} T, {Tashiro} M, et~al. 2000.
\newblock \textit{\apj} 535:632--643

\bibitem[{{Marco}, {Negueruela} \& {Motch}(2007)}]{2007ASPC..367..645M}
{Marco} A, {Negueruela} I, {Motch} C. 2007.
\newblock In \textit{Massive Stars in Interactive Binaries}, eds.
  N~{St.-Louis}, AFJ {Moffat}, vol. 367 of \textit{Astronomical Society of the
  Pacific Conference Series}

\bibitem[{{Mason} et~al.(2009){Mason}, {Hartkopf}, {Gies}, {Henry} \&
  {Helsel}}]{2009AJ....137.3358M}
{Mason} BD, {Hartkopf} WI, {Gies} DR, {Henry} TJ, {Helsel} JW. 2009.
\newblock \textit{\aj} 137:3358--3377

\bibitem[{{Massey} \& {Hunter}(1998)}]{1998ApJ...493..180M}
{Massey} P, {Hunter} DA. 1998.
\newblock \textit{\apj} 493:180

\bibitem[{{Mayor} et~al.(2009){Mayor}, {Udry}, {Lovis}, {Pepe}, {Queloz}
  et~al.}]{2009A&A...493..639M}
{Mayor} M, {Udry} S, {Lovis} C, {Pepe} F, {Queloz} D, et~al. 2009.
\newblock \textit{\aap} 493:639--644

\bibitem[{{McCrady}, {Gilbert} \& {Graham}(2003)}]{2003ApJ...596..240M}
{McCrady} N, {Gilbert} AM, {Graham} JR. 2003.
\newblock \textit{\apj} 596:240--252

\bibitem[{{McCrady} \& {Graham}(2007)}]{2007ApJ...663..844M}
{McCrady} N, {Graham} JR. 2007.
\newblock \textit{\apj} 663:844--856

\bibitem[{{McCrady}, {Graham} \& {Vacca}(2005)}]{2005ApJ...621..278M}
{McCrady} N, {Graham} JR, {Vacca} WD. 2005.
\newblock \textit{\apj} 621:278--284

\bibitem[{{McCrea}(1964)}]{1964MNRAS.128..147M}
{McCrea} WH. 1964.
\newblock \textit{\mnras} 128:147

\bibitem[{{McLaughlin}(2000)}]{2000ApJ...539..618M}
{McLaughlin} DE. 2000.
\newblock \textit{\apj} 539:618--640

\bibitem[{{McLaughlin} et~al.(2006){McLaughlin}, {Anderson}, {Meylan},
  {Gebhardt}, {Pryor} et~al.}]{2006ApJS..166..249M}
{McLaughlin} DE, {Anderson} J, {Meylan} G, {Gebhardt} K, {Pryor} C, et~al.
  2006.
\newblock \textit{\apjs} 166:249--297

\bibitem[{{McLaughlin} \& {van der Marel}(2005)}]{2005ApJS..161..304M}
{McLaughlin} DE, {van der Marel} RP. 2005.
\newblock \textit{\apjs} 161:304--360

\bibitem[{{McMillan}, {Hut} \& {Makino}(1990)}]{1990ApJ...362..522M}
{McMillan} S, {Hut} P, {Makino} J. 1990.
\newblock \textit{\apj} 362:522--537

\bibitem[{{McMillan}, {Hut} \& {Makino}(1991)}]{1991ApJ...372..111M}
{McMillan} S, {Hut} P, {Makino} J. 1991.
\newblock \textit{\apj} 372:111--124

\bibitem[{{McMillan}(1986)}]{1986LNP...267..156M}
{McMillan} SLW. 1986.
\newblock In \textit{The Use of Supercomputers in Stellar Dynamics}, ed.
  {P.~Hut \& S.~L.~W.~McMillan}, vol. 267 of \textit{Lecture Notes in Physics,
  Berlin Springer Verlag}

\bibitem[{{McMillan} \& {Aarseth}(1993)}]{1993ApJ...414..200M}
{McMillan} SLW, {Aarseth} SJ. 1993.
\newblock \textit{\apj} 414:200--212

\bibitem[{{McMillan}, {Vesperini} \& {Portegies
  Zwart}(2007)}]{2007ApJ...655L..45M}
{McMillan} SLW, {Vesperini} E, {Portegies Zwart} SF. 2007.
\newblock \textit{\apjl} 655:L45--L49

\bibitem[{{Mengel} et~al.(2002){Mengel}, {Lehnert}, {Thatte} \&
  {Genzel}}]{2002A&A...383..137M}
{Mengel} S, {Lehnert} MD, {Thatte} N, {Genzel} R. 2002.
\newblock \textit{\aap} 383:137--152

\bibitem[{{Mengel} et~al.(2008){Mengel}, {Lehnert}, {Thatte}, {Vacca},
  {Whitmore} \& {Chandar}}]{2008A&A...489.1091M}
{Mengel} S, {Lehnert} MD, {Thatte} NA, {Vacca} WD, {Whitmore} B, {Chandar} R.
  2008.
\newblock \textit{\aap} 489:1091--1105

\bibitem[{{Mengel} \& {Tacconi-Garman}(2007)}]{2007A&A...466..151M}
{Mengel} S, {Tacconi-Garman} LE. 2007.
\newblock \textit{\aap} 466:151--155

\bibitem[{{Merritt} et~al.(2004){Merritt}, {Piatek}, {Portegies Zwart} \&
  {Hemsendorf}}]{2004ApJ...608L..25M}
{Merritt} D, {Piatek} S, {Portegies Zwart} S, {Hemsendorf} M. 2004.
\newblock \textit{\apjl} 608:L25--L28

\bibitem[{{M{\'e}sz{\'a}ros}(2002)}]{2002ARA&A..40..137M}
{M{\'e}sz{\'a}ros} P. 2002.
\newblock \textit{\araa} 40:137--169

\bibitem[{{Meurer} et~al.(1995){Meurer}, {Heckman}, {Leitherer}, {Kinney},
  {Robert} \& {Garnett}}]{1995AJ....110.2665M}
{Meurer} GR, {Heckman} TM, {Leitherer} C, {Kinney} A, {Robert} C, {Garnett} DR.
  1995.
\newblock \textit{\aj} 110:2665

\bibitem[{{Miller} et~al.(1997){Miller}, {Whitmore}, {Schweizer} \&
  {Fall}}]{1997AJ....114.2381M}
{Miller} BW, {Whitmore} BC, {Schweizer} F, {Fall} SM. 1997.
\newblock \textit{\aj} 114:2381

\bibitem[{{Miller} \& {Scalo}(1979)}]{1979ApJS...41..513M}
{Miller} GE, {Scalo} JM. 1979.
\newblock \textit{\apjs} 41:513--547

\bibitem[{{Miller} et~al.(2003){Miller}, {Fabbiano}, {Miller} \&
  {Fabian}}]{2003ApJ...585L..37M}
{Miller} JM, {Fabbiano} G, {Miller} MC, {Fabian} AC. 2003.
\newblock \textit{\apjl} 585:L37--L40

\bibitem[{{Miller} \& {Hamilton}(2002)}]{2002MNRAS.330..232C}
{Miller} MC, {Hamilton} DP. 2002.
\newblock \textit{\mnras} 330:232--240

\bibitem[{{Milone} et~al.(2009){Milone}, {Bedin}, {Piotto} \&
  {Anderson}}]{2009A&A...497..755M}
{Milone} AP, {Bedin} LR, {Piotto} G, {Anderson} J. 2009.
\newblock \textit{\aap} 497:755--771

\bibitem[{{Milone} et~al.(2008){Milone}, {Piotto}, {Bedin} \&
  {Sarajedini}}]{2008MmSAI..79..623M}
{Milone} AP, {Piotto} G, {Bedin} LR, {Sarajedini} A. 2008.
\newblock \textit{Memorie della Societa Astronomica Italiana} 79:623

\bibitem[{{Mirabel}, {Fuchs} \& {Chaty}(2000)}]{2000AIPC..526..814M}
{Mirabel} IF, {Fuchs} Y, {Chaty} S. 2000.
\newblock In \textit{Gamma-ray Bursts, 5th Huntsville Symposium}, ed.
  {R.~M.~Kippen, R.~S.~Mallozzi, \& G.~J.~Fishman}, vol. 526 of
  \textit{American Institute of Physics Conference Series}

\bibitem[{{Mirabel}, {Rodrigues} \& {Liu}(2004)}]{2004A&A...422L..29M}
{Mirabel} IF, {Rodrigues} I, {Liu} QZ. 2004.
\newblock \textit{\aap} 422:L29--L32

\bibitem[{{Moeckel} \& {Bonnell}(2009)}]{2009MNRAS.400..657M}
{Moeckel} N, {Bonnell} IA. 2009.
\newblock \textit{\mnras} 400:657--664

\bibitem[{{Moll} et~al.(2007){Moll}, {Mengel}, {de Grijs}, {Smith} \&
  {Crowther}}]{2007MNRAS.382.1877M}
{Moll} SL, {Mengel} S, {de Grijs} R, {Smith} LJ, {Crowther} PA. 2007.
\newblock \textit{\mnras} 382:1877--1888

\bibitem[{{Moore} et~al.(1999){Moore}, {Quinn}, {Governato}, {Stadel} \&
  {Lake}}]{1999MNRAS.310.1147M}
{Moore} B, {Quinn} T, {Governato} F, {Stadel} J, {Lake} G. 1999.
\newblock \textit{\mnras} 310:1147--1152

\bibitem[{{Muno} et~al.(2006){Muno}, {Clark}, {Crowther}, {Dougherty}, {de
  Grijs} et~al.}]{2006ApJ...636L..41M}
{Muno} MP, {Clark} JS, {Crowther} PA, {Dougherty} SM, {de Grijs} R, et~al.
  2006.
\newblock \textit{\apjl} 636:L41--L44

\bibitem[{{Muno} et~al.(2007){Muno}, {Gaensler}, {Clark}, {de Grijs}, {Pooley}
  et~al.}]{2007MNRAS.378L..44M}
{Muno} MP, {Gaensler} BM, {Clark} JS, {de Grijs} R, {Pooley} D, et~al. 2007.
\newblock \textit{\mnras} 378:L44--L48

\bibitem[{{Narayan}, {Piran} \& {Shemi}(1991)}]{1991ApJ...379L..17N}
{Narayan} R, {Piran} T, {Shemi} A. 1991.
\newblock \textit{\apjl} 379:L17--L20

\bibitem[{{Narbutis} et~al.(2008){Narbutis}, {Vansevi{\v c}ius}, {Kodaira},
  {Brid{\v z}ius} \& {Stonkut{\.e}}}]{2008ApJS..177..174N}
{Narbutis} D, {Vansevi{\v c}ius} V, {Kodaira} K, {Brid{\v z}ius} A,
  {Stonkut{\.e}} R. 2008.
\newblock \textit{\apjs} 177:174--180

\bibitem[{{Nitadori} \& {Makino}(2008)}]{2008NewA...13..498N}
{Nitadori} K, {Makino} J. 2008.
\newblock \textit{New Astronomy} 13:498--507

\bibitem[{{Nitadori}, {Makino} \& {Hut}(2006)}]{2006NewA...12..169N}
{Nitadori} K, {Makino} J, {Hut} P. 2006.
\newblock \textit{New Astronomy} 12:169--181

\bibitem[{{Normandeau}, {Taylor} \& {Dewdney}(1996)}]{1996Natur.380..687N}
{Normandeau} M, {Taylor} AR, {Dewdney} PE. 1996.
\newblock \textit{\nat} 380:687--689

\bibitem[{{Offner}, {Hansen} \& {Krumholz}(2009)}]{2009ApJ...704L.124O}
{Offner} SSR, {Hansen} CE, {Krumholz} MR. 2009.
\newblock \textit{\apjl} 704:L124--L128

\bibitem[{{Ogilvie} \& {Lin}(2004)}]{2004ApJ...610..477O}
{Ogilvie} GI, {Lin} DNC. 2004.
\newblock \textit{\apj} 610:477--509

\bibitem[{{Oort}, {Kerr} \& {Westerhout}(1958)}]{1958MNRAS.118..379O}
{Oort} JH, {Kerr} FJ, {Westerhout} G. 1958.
\newblock \textit{\mnras} 118:379

\bibitem[{{{\"O}stlin}, {Cumming} \& {Bergvall}(2007)}]{2007A&A...461..471O}
{{\"O}stlin} G, {Cumming} RJ, {Bergvall} N. 2007.
\newblock \textit{\aap} 461:471--483

\bibitem[{{Ostriker}, {Spitzer} \& {Chevalier}(1972)}]{1972ApJ...176L..51O}
{Ostriker} JP, {Spitzer} LJ, {Chevalier} RA. 1972.
\newblock \textit{\apjl} 176:L51

\bibitem[{{Paczynski}(1998)}]{1998ApJ...494L..45P}
{Paczynski} B. 1998.
\newblock \textit{\apjl} 494:L45

\bibitem[{{Park}, {Park} \& {Lee}(2009)}]{2009ApJ...700..103P}
{Park} W, {Park} HS, {Lee} MG. 2009.
\newblock \textit{\apj} 700:103--113

\bibitem[{{Parker} \& {Goodwin}(2007)}]{2007MNRAS.380.1271P}
{Parker} RJ, {Goodwin} SP. 2007.
\newblock \textit{\mnras} 380:1271--1275

\bibitem[{{Parmentier} \& {de Grijs}(2008)}]{2008MNRAS.383.1103P}
{Parmentier} G, {de Grijs} R. 2008.
\newblock \textit{\mnras} 383:1103--1120

\bibitem[{{Parmentier} \& {Gilmore}(2007)}]{2007MNRAS.377..352P}
{Parmentier} G, {Gilmore} G. 2007.
\newblock \textit{\mnras} 377:352--372

\bibitem[{{Patruno} et~al.(2006){Patruno}, {Portegies Zwart}, {Dewi} \&
  {Hopman}}]{2006MNRAS.370L...6P}
{Patruno} A, {Portegies Zwart} S, {Dewi} J, {Hopman} C. 2006.
\newblock \textit{\mnras} 370:L6--L9

\bibitem[{{Peacock} et~al.(2010){Peacock}, {Maccarone}, {Knigge}, {Kundu},
  {Waters} et~al.}]{2010MNRAS.tmp...28P}
{Peacock} MB, {Maccarone} TJ, {Knigge} C, {Kundu} A, {Waters} CZ, et~al. 2010.
\newblock \textit{\mnras} :28

\bibitem[{{Perina} et~al.(2009){Perina}, {Barmby}, {Beasley}, {Bellazzini},
  {Brodie} et~al.}]{2009A&A...494..933P}
{Perina} S, {Barmby} P, {Beasley} MA, {Bellazzini} M, {Brodie} JP, et~al. 2009.
\newblock \textit{\aap} 494:933--948

\bibitem[{{Pfahl}, {Rappaport} \& {Podsiadlowski}(2002)}]{2002ApJ...573..283P}
{Pfahl} E, {Rappaport} S, {Podsiadlowski} P. 2002.
\newblock \textit{\apj} 573:283--305

\bibitem[{{Pfahl}, {Scannapieco} \& {Bildsten}(2009)}]{2009arXiv0903.1104P}
{Pfahl} E, {Scannapieco} E, {Bildsten} L. 2009.
\newblock \textit{ArXiv e-prints}

\bibitem[{{Pfalzner}(2009)}]{2009A&A...498L..37P}
{Pfalzner} S. 2009.
\newblock \textit{\aap} 498:L37--L40

\bibitem[{{Piotto}(2008)}]{2008MmSAI..79..334P}
{Piotto} G. 2008.
\newblock \textit{Memorie della Societa Astronomica Italiana} 79:334

\bibitem[{{Piotto} et~al.(2005){Piotto}, {Villanova}, {Bedin}, {Gratton},
  {Cassisi} et~al.}]{2005ApJ...621..777P}
{Piotto} G, {Villanova} S, {Bedin} LR, {Gratton} R, {Cassisi} S, et~al. 2005.
\newblock \textit{\apj} 621:777--784

\bibitem[{{Piskunov} et~al.(2006){Piskunov}, {Kharchenko}, {R{\"o}ser},
  {Schilbach} \& {Scholz}}]{2006A&A...445..545P}
{Piskunov} AE, {Kharchenko} NV, {R{\"o}ser} S, {Schilbach} E, {Scholz} RD.
  2006.
\newblock \textit{\aap} 445:545--565

\bibitem[{{Piskunov} et~al.(2008){Piskunov}, {Schilbach}, {Kharchenko},
  {R{\"o}ser} \& {Scholz}}]{2008A&A...477..165P}
{Piskunov} AE, {Schilbach} E, {Kharchenko} NV, {R{\"o}ser} S, {Scholz} RD.
  2008.
\newblock \textit{\aap} 477:165--172

\bibitem[{{Plummer}(1911)}]{1911MNRAS..71..460P}
{Plummer} HC. 1911.
\newblock \textit{\mnras} 71:460--470

\bibitem[{{Pooley} et~al.(2003){Pooley}, {Lewin}, {Anderson}, {Baumgardt},
  {Filippenko} et~al.}]{2003ApJ...591L.131P}
{Pooley} D, {Lewin} WHG, {Anderson} SF, {Baumgardt} H, {Filippenko} AV, et~al.
  2003.
\newblock \textit{\apjl} 591:L131--L134

\bibitem[{{Portegies Zwart} et~al.(2004){Portegies Zwart}, {Baumgardt}, {Hut},
  {Makino} \& {McMillan}}]{2004Natur.428..724P}
{Portegies Zwart} SF, {Baumgardt} H, {Hut} P, {Makino} J, {McMillan} SLW. 2004.
\newblock \textit{\nat} 428:724--726

\bibitem[{{Portegies Zwart}, {Belleman} \&
  {Geldof}(2007)}]{2007NewA...12..641P}
{Portegies Zwart} SF, {Belleman} RG, {Geldof} PM. 2007.
\newblock \textit{New Astronomy} 12:641--650

\bibitem[{{Portegies Zwart}, {Hut} \& {Verbunt}(1997)}]{1997A&A...328..130P}
{Portegies Zwart} SF, {Hut} P, {Verbunt} F. 1997.
\newblock \textit{\aap} 328:130--142

\bibitem[{{Portegies Zwart} et~al.(1999){Portegies Zwart}, {Makino}, {McMillan}
  \& {Hut}}]{1999A&A...348..117P}
{Portegies Zwart} SF, {Makino} J, {McMillan} SLW, {Hut} P. 1999.
\newblock \textit{\aap} 348:117--126

\bibitem[{{Portegies Zwart} et~al.(2008){Portegies Zwart}, {McMillan}, {Groen},
  {Gualandris}, {Sipior} \& {Vermin}}]{2008NewA...13..285P}
{Portegies Zwart} SF, {McMillan} S, {Groen} D, {Gualandris} A, {Sipior} M,
  {Vermin} W. 2008.
\newblock \textit{New Astronomy} 13:285--295

\bibitem[{{Portegies Zwart} et~al.(2009){Portegies Zwart}, {McMillan},
  {Harfst}, {Groen}, {Fujii} et~al.}]{2009NewA...14..369P}
{Portegies Zwart} SF, {McMillan} S, {Harfst} S, {Groen} D, {Fujii} M, et~al.
  2009.
\newblock \textit{New Astronomy} 14:369--378

\bibitem[{{Portegies Zwart} \& {McMillan}(2002)}]{2002ApJ...576..899P}
{Portegies Zwart} SF, {McMillan} SLW. 2002.
\newblock \textit{\apj} 576:899--907

\bibitem[{{Portegies Zwart} et~al.(2001){Portegies Zwart}, {McMillan}, {Hut} \&
  {Makino}}]{2001MNRAS.321..199P}
{Portegies Zwart} SF, {McMillan} SLW, {Hut} P, {Makino} J. 2001.
\newblock \textit{\mnras} 321:199--226

\bibitem[{{Portegies Zwart}, {McMillan} \&
  {Makino}(2007)}]{2007MNRAS.374...95P}
{Portegies Zwart} SF, {McMillan} SLW, {Makino} J. 2007.
\newblock \textit{\mnras} 374:95--106

\bibitem[{{Portegies Zwart}, {Pooley} \& {Lewin}(2002)}]{2002ApJ...574..762P}
{Portegies Zwart} SF, {Pooley} D, {Lewin} WHG. 2002.
\newblock \textit{\apj} 574:762--770

\bibitem[{{Portegies Zwart} \& {Takahashi}(1999)}]{1999CeMDA..73..179P}
{Portegies Zwart} SF, {Takahashi} K. 1999.
\newblock \textit{Celestial Mechanics and Dynamical Astronomy} 73:179--186

\bibitem[{{Portegies Zwart} \& {van den Heuvel}(2007)}]{2007Natur.450..388P}
{Portegies Zwart} SF, {van den Heuvel} EPJ. 2007.
\newblock \textit{\nat} 450:388--389

\bibitem[{{Portegies Zwart} \& {Verbunt}(1996)}]{1996A&A...309..179P}
{Portegies Zwart} SF, {Verbunt} F. 1996.
\newblock \textit{\aap} 309:179--196

\bibitem[{{Prantzos} \& {Charbonnel}(2006)}]{2006A&A...458..135P}
{Prantzos} N, {Charbonnel} C. 2006.
\newblock \textit{\aap} 458:135--149

\bibitem[{{Price} \& {Bate}(2009)}]{2009MNRAS.398...33P}
{Price} DJ, {Bate} MR. 2009.
\newblock \textit{\mnras} 398:33--46

\bibitem[{{Proszkow} et~al.(2009){Proszkow}, {Adams}, {Hartmann} \&
  {Tobin}}]{2009ApJ...697.1020P}
{Proszkow} EM, {Adams} FC, {Hartmann} LW, {Tobin} JJ. 2009.
\newblock \textit{\apj} 697:1020--1031

\bibitem[{{Quinlan}(1996)}]{1996NewA....1..255Q}
{Quinlan} GD. 1996.
\newblock \textit{New Astronomy} 1:255--270

\bibitem[{{Rauw} et~al.(2005){Rauw}, {Crowther}, {De Becker}, {Gosset},
  {Naz{\'e}} et~al.}]{2005A&A...432..985R}
{Rauw} G, {Crowther} PA, {De Becker} M, {Gosset} E, {Naz{\'e}} Y, et~al. 2005.
\newblock \textit{\aap} 432:985--998

\bibitem[{{Read} et~al.(2006){Read}, {Wilkinson}, {Evans}, {Gilmore} \&
  {Kleyna}}]{2006MNRAS.366..429R}
{Read} JI, {Wilkinson} MI, {Evans} NW, {Gilmore} G, {Kleyna} JT. 2006.
\newblock \textit{\mnras} 366:429--437

\bibitem[{{Roberts} \& {Yusef-Zadeh}(2005)}]{2005AJ....129..805R}
{Roberts} DA, {Yusef-Zadeh} F. 2005.
\newblock \textit{\aj} 129:805--808

\bibitem[{{Sabbi} et~al.(2008){Sabbi}, {Sirianni}, {Nota}, {Tosi}, {Gallagher}
  et~al.}]{2008AJ....135..173S}
{Sabbi} E, {Sirianni} M, {Nota} A, {Tosi} M, {Gallagher} J, et~al. 2008.
\newblock \textit{\aj} 135:173--181

\bibitem[{{Safronov}(1969)}]{1969QB981.S26......}
{Safronov} VS. 1969.
\newblock \textit{{Evoliutsiia doplanetnogo oblaka.}}

\bibitem[{{Salpeter}(1955)}]{1955ApJ...121..161S}
{Salpeter} EE. 1955.
\newblock \textit{\apj} 121:161

\bibitem[{{San Roman} et~al.(2009){San Roman}, {Sarajedini}, {Garnett} \&
  {Holtzman}}]{2009ApJ...699..839S}
{San Roman} I, {Sarajedini} A, {Garnett} DR, {Holtzman} JA. 2009.
\newblock \textit{\apj} 699:839--849

\bibitem[{{Sana} et~al.(2008){Sana}, {Gosset}, {Naz{\'e}}, {Rauw} \&
  {Linder}}]{2008MNRAS.386..447S}
{Sana} H, {Gosset} E, {Naz{\'e}} Y, {Rauw} G, {Linder} N. 2008.
\newblock \textit{\mnras} 386:447--460

\bibitem[{{Sandage}(1953)}]{1953AJ.....58...61S}
{Sandage} AR. 1953.
\newblock \textit{\aj} 58:61--75

\bibitem[{{Sarajedini} \& {Mancone}(2007)}]{2007AJ....134..447S}
{Sarajedini} A, {Mancone} CL. 2007.
\newblock \textit{\aj} 134:447--456

\bibitem[{{Scally}, {Clarke} \& {McCaughrean}(2005)}]{2005MNRAS.358..742S}
{Scally} A, {Clarke} C, {McCaughrean} MJ. 2005.
\newblock \textit{\mnras} 358:742--754

\bibitem[{{Schechter}(1976)}]{1976ApJ...203..297S}
{Schechter} P. 1976.
\newblock \textit{\apj} 203:297--306

\bibitem[{{Scheepmaker} et~al.(2007){Scheepmaker}, {Haas}, {Gieles}, {Bastian},
  {Larsen} \& {Lamers}}]{2007A&A...469..925S}
{Scheepmaker} RA, {Haas} MR, {Gieles} M, {Bastian} N, {Larsen} SS, {Lamers}
  HJGLM. 2007.
\newblock \textit{\aap} 469:925--940

\bibitem[{{Schilbach} \& {R{\"o}ser}(2008)}]{2008A&A...489..105S}
{Schilbach} E, {R{\"o}ser} S. 2008.
\newblock \textit{\aap} 489:105--114

\bibitem[{{Schweizer}(1982)}]{1982ApJ...252..455S}
{Schweizer} F. 1982.
\newblock \textit{\apj} 252:455--460

\bibitem[{{Shapiro}(1985)}]{1985IAUS..113..373S}
{Shapiro} SL. 1985.
\newblock In \textit{Dynamics of Star Clusters}, ed. {J.~Goodman \& P.~Hut},
  vol. 113 of \textit{IAU Symposium}

\bibitem[{{Shara} \& {Hurley}(2002)}]{2002ApJ...571..830S}
{Shara} MM, {Hurley} JR. 2002.
\newblock \textit{\apj} 571:830--842

\bibitem[{{Silich}, {Tenorio-Tagle} \&
  {Rodr{\'{\i}}guez-Gonz{\'a}lez}(2004)}]{2004ApJ...610..226S}
{Silich} S, {Tenorio-Tagle} G, {Rodr{\'{\i}}guez-Gonz{\'a}lez} A. 2004.
\newblock \textit{\apj} 610:226--232

\bibitem[{{Sills} et~al.(2003){Sills}, {Deiters}, {Eggleton}, {Freitag},
  {Giersz} et~al.}]{2003NewA....8..605S}
{Sills} A, {Deiters} S, {Eggleton} P, {Freitag} M, {Giersz} M, et~al. 2003.
\newblock \textit{New Astronomy} 8:605--628

\bibitem[{{Smith} \& {Gallagher}(2001)}]{2001MNRAS.326.1027S}
{Smith} LJ, {Gallagher} JS. 2001.
\newblock \textit{\mnras} 326:1027--1040

\bibitem[{{Smith} et~al.(2006){Smith}, {Westmoquette}, {Gallagher},
  {O'Connell}, {Rosario} \& {de Grijs}}]{2006MNRAS.370..513S}
{Smith} LJ, {Westmoquette} MS, {Gallagher} JS, {O'Connell} RW, {Rosario} DJ,
  {de Grijs} R. 2006.
\newblock \textit{\mnras} 370:513--527

\bibitem[{{Sollima} et~al.(2007){Sollima}, {Beccari}, {Ferraro}, {Fusi Pecci}
  \& {Sarajedini}}]{2007MNRAS.380..781S}
{Sollima} A, {Beccari} G, {Ferraro} FR, {Fusi Pecci} F, {Sarajedini} A. 2007.
\newblock \textit{\mnras} 380:781--791

\bibitem[{{Sollima} et~al.(2008){Sollima}, {Lanzoni}, {Beccari}, {Ferraro} \&
  {Fusi Pecci}}]{2008A&A...481..701S}
{Sollima} A, {Lanzoni} B, {Beccari} G, {Ferraro} FR, {Fusi Pecci} F. 2008.
\newblock \textit{\aap} 481:701--704

\bibitem[{{Sommariva} et~al.(2009){Sommariva}, {Piotto}, {Rejkuba}, {Bedin},
  {Heggie} et~al.}]{2009A&A...493..947S}
{Sommariva} V, {Piotto} G, {Rejkuba} M, {Bedin} LR, {Heggie} DC, et~al. 2009.
\newblock \textit{\aap} 493:947--958

\bibitem[{{Spitzer}(1987)}]{1987degc.book.....S}
{Spitzer} L. 1987.
\newblock \textit{{Dynamical evolution of globular clusters}}.
\newblock Princeton, NJ, Princeton University Press, 1987, 191 p.

\bibitem[{{Spitzer}(1975)}]{1975IAUS...69....3S}
{Spitzer} Jr. L. 1975.
\newblock In \textit{Dynamics of the Solar Systems}, ed. {A.~Hayli}, vol.~69 of
  \textit{IAU Symposium}

\bibitem[{{Spitzer}(1940)}]{1940MNRAS.100..396S}
{Spitzer} LJ. 1940.
\newblock \textit{\mnras} 100:396

\bibitem[{{Spitzer}(1958)}]{1958ApJ...127...17S}
{Spitzer} LJ. 1958.
\newblock \textit{\apj} 127:17

\bibitem[{{Spitzer}(1969)}]{1969ApJ...158L.139S}
{Spitzer} LJ. 1969.
\newblock \textit{\apjl} 158:L139

\bibitem[{{Spitzer} \& {Chevalier}(1973)}]{1973ApJ...183..565S}
{Spitzer} LJ, {Chevalier} RA. 1973.
\newblock \textit{\apj} 183:565--582

\bibitem[{{Spitzer} \& {Hart}(1971)}]{1971ApJ...164..399S}
{Spitzer} LJ, {Hart} MH. 1971.
\newblock \textit{\apj} 164:399

\bibitem[{{Spurzem}(1999)}]{1999JCoAM.109..407S}
{Spurzem} R. 1999.
\newblock \textit{Journal of Computational and Applied Mathematics}
  109:407--432

\bibitem[{{Spurzem} et~al.(2009){Spurzem}, {Giersz}, {Heggie} \&
  {Lin}}]{2009ApJ...697..458S}
{Spurzem} R, {Giersz} M, {Heggie} DC, {Lin} DNC. 2009.
\newblock \textit{\apj} 697:458--482

\bibitem[{{Stodolkiewicz}(1982)}]{1982AcA....32...63S}
{Stodolkiewicz} JS. 1982.
\newblock \textit{Acta Astronomica} 32:63--91

\bibitem[{{Stodolkiewicz}(1986)}]{1986AcA....36...19S}
{Stodolkiewicz} JS. 1986.
\newblock \textit{Acta Astronomica} 36:19--41

\bibitem[{{Stolte} et~al.(2005){Stolte}, {Brandner}, {Grebel}, {Lenzen} \&
  {Lagrange}}]{2005ApJ...628L.113S}
{Stolte} A, {Brandner} W, {Grebel} EK, {Lenzen} R, {Lagrange} AM. 2005.
\newblock \textit{\apjl} 628:L113--L117

\bibitem[{{Stolte} et~al.(2002){Stolte}, {Grebel}, {Brandner} \&
  {Figer}}]{2002A&A...394..459S}
{Stolte} A, {Grebel} EK, {Brandner} W, {Figer} DF. 2002.
\newblock \textit{\aap} 394:459--478

\bibitem[{{Strohmayer} \& {Mushotzky}(2003)}]{2003ApJ...586L..61S}
{Strohmayer} TE, {Mushotzky} RF. 2003.
\newblock \textit{\apjl} 586:L61--L64

\bibitem[{{Swartz}, {Tennant} \& {Soria}(2009)}]{2009ApJ...703..159S}
{Swartz} DA, {Tennant} AF, {Soria} R. 2009.
\newblock \textit{\apj} 703:159--168

\bibitem[{{Takahashi}(1996)}]{1996PASJ...48..691T}
{Takahashi} K. 1996.
\newblock \textit{\pasj} 48:691--700

\bibitem[{{Takahashi}(1997)}]{1997PASJ...49..547T}
{Takahashi} K. 1997.
\newblock \textit{\pasj} 49:547--560

\bibitem[{{Takahashi} \& {Portegies Zwart}(1998)}]{1998ApJ...503L..49T}
{Takahashi} K, {Portegies Zwart} SF. 1998.
\newblock \textit{\apjl} 503:L49

\bibitem[{{Takahashi} \& {Portegies Zwart}(2000)}]{2000ApJ...535..759T}
{Takahashi} K, {Portegies Zwart} SF. 2000.
\newblock \textit{\apj} 535:759--775

\bibitem[{{Theuns}(1991)}]{1991MmSAI..62..909T}
{Theuns} T. 1991.
\newblock \textit{Memorie della Societa Astronomica Italiana} 62:909--914

\bibitem[{{Thorsett}, {Arzoumanian} \& {Taylor}(1993)}]{1993ApJ...412L..33T}
{Thorsett} SE, {Arzoumanian} Z, {Taylor} JH. 1993.
\newblock \textit{\apjl} 412:L33--L36

\bibitem[{{Trenti} et~al.(2007){Trenti}, {Ardi}, {Mineshige} \&
  {Hut}}]{2007MNRAS.374..857T}
{Trenti} M, {Ardi} E, {Mineshige} S, {Hut} P. 2007.
\newblock \textit{\mnras} 374:857--866

\bibitem[{{Udry} \& {Santos}(2007)}]{2007ARA&A..45..397U}
{Udry} S, {Santos} NC. 2007.
\newblock \textit{\araa} 45:397--439

\bibitem[{{{\v S}ubr}, {Kroupa} \& {Baumgardt}(2008)}]{2008MNRAS.385.1673S}
{{\v S}ubr} L, {Kroupa} P, {Baumgardt} H. 2008.
\newblock \textit{\mnras} 385:1673--1680

\bibitem[{{Vall{\'e}e}(2008)}]{2008AJ....135.1301V}
{Vall{\'e}e} JP. 2008.
\newblock \textit{\aj} 135:1301--1310

\bibitem[{{van den Bergh}(1957)}]{1957ApJ...125..445V}
{van den Bergh} S. 1957.
\newblock \textit{\apj} 125:445

\bibitem[{{van den Bergh}(1971)}]{1971A&A....12..474V}
{van den Bergh} S. 1971.
\newblock \textit{\aap} 12:474

\bibitem[{{van den Bergh}(1991)}]{1991ApJ...369....1V}
{van den Bergh} S. 1991.
\newblock \textit{\apj} 369:1--12

\bibitem[{{van den Bergh} \& {McClure}(1980)}]{1980A&A....88..360V}
{van den Bergh} S, {McClure} RD. 1980.
\newblock \textit{\aap} 88:360--362

\bibitem[{{van den Berk}, {Portegies Zwart} \&
  {McMillan}(2007)}]{2007MNRAS.379..111V}
{van den Berk} J, {Portegies Zwart} SF, {McMillan} SLW. 2007.
\newblock \textit{\mnras} 379:111--122

\bibitem[{{Vanbeveren} et~al.(2009){Vanbeveren}, {Belkus}, {van Bever} \&
  {Mennekens}}]{2009Ap&SS.tmp..114V}
{Vanbeveren} D, {Belkus} H, {van Bever} J, {Mennekens} N. 2009.
\newblock \textit{\apss} :114

\bibitem[{{Vansevi{\v c}ius} et~al.(2009){Vansevi{\v c}ius}, {Kodaira},
  {Narbutis}, {Stonkut{\.e}}, {Brid{\v z}ius} et~al.}]{2009ApJ...703.1872V}
{Vansevi{\v c}ius} V, {Kodaira} K, {Narbutis} D, {Stonkut{\.e}} R, {Brid{\v
  z}ius} A, et~al. 2009.
\newblock \textit{\apj} 703:1872--1883

\bibitem[{{Ventura} et~al.(2001){Ventura}, {D'Antona}, {Mazzitelli} \&
  {Gratton}}]{2001ApJ...550L..65V}
{Ventura} P, {D'Antona} F, {Mazzitelli} I, {Gratton} R. 2001.
\newblock \textit{\apjl} 550:L65--L69

\bibitem[{{Vesperini}(1997)}]{1997MNRAS.287..915V}
{Vesperini} E. 1997.
\newblock \textit{\mnras} 287:915--928

\bibitem[{{Vesperini} et~al.(2009){Vesperini}, {D'Ercole}, {D'Antona},
  {McMillan} \& {Recchi}}]{2009AAS...21333105V}
{Vesperini} E, {D'Ercole} A, {D'Antona} F, {McMillan} SLW, {Recchi} S. 2009.
\newblock In \textit{Bulletin of the American Astronomical Society}, vol.~41 of
  \textit{Bulletin of the American Astronomical Society}

\bibitem[{{Vesperini} \& {Heggie}(1997)}]{1997MNRAS.289..898V}
{Vesperini} E, {Heggie} DC. 1997.
\newblock \textit{\mnras} 289:898--920

\bibitem[{{Vesperini}, {McMillan} \& {Portegies
  Zwart}(2009)}]{2009ApJ...698..615V}
{Vesperini} E, {McMillan} SLW, {Portegies Zwart} S. 2009.
\newblock \textit{\apj} 698:615--622

\bibitem[{{Vesperini} \& {Zepf}(2003)}]{2003ApJ...587L..97V}
{Vesperini} E, {Zepf} SE. 2003.
\newblock \textit{\apjl} 587:L97--L100

\bibitem[{{Vesperini} et~al.(2003){Vesperini}, {Zepf}, {Kundu} \&
  {Ashman}}]{2003ApJ...593..760V}
{Vesperini} E, {Zepf} SE, {Kundu} A, {Ashman} KM. 2003.
\newblock \textit{\apj} 593:760--771

\bibitem[{{Vink{\'o}} et~al.(2009){Vink{\'o}}, {S{\'a}rneczky}, {Balog},
  {Immler}, {Sugerman} et~al.}]{2009ApJ...695..619V}
{Vink{\'o}} J, {S{\'a}rneczky} K, {Balog} Z, {Immler} S, {Sugerman} BEK, et~al.
  2009.
\newblock \textit{\apj} 695:619--635

\bibitem[{{von Hoerner}(1957)}]{1957ApJ...125..451V}
{von Hoerner} S. 1957.
\newblock \textit{\apj} 125:451

\bibitem[{{von Hoerner}(1958)}]{1958ZA.....44..221V}
{von Hoerner} S. 1958.
\newblock \textit{Zeitschrift fur Astrophysik} 44:221--242

\bibitem[{{von Hoerner}(1963)}]{1963ZA.....57...47V}
{von Hoerner} S. 1963.
\newblock \textit{Zeitschrift fur Astrophysik} 57:47--82

\bibitem[{{Vrba} et~al.(2000){Vrba}, {Henden}, {Luginbuhl}, {Guetter},
  {Hartmann} \& {Klose}}]{2000ApJ...533L..17V}
{Vrba} FJ, {Henden} AA, {Luginbuhl} CB, {Guetter} HH, {Hartmann} DH, {Klose} S.
  2000.
\newblock \textit{\apjl} 533:L17--L20

\bibitem[{{Wang} et~al.(2005){Wang}, {Yang}, {Zhang}, {Ma}, {Zhou}
  et~al.}]{2005ApJ...626L..89W}
{Wang} X, {Yang} Y, {Zhang} T, {Ma} J, {Zhou} X, et~al. 2005.
\newblock \textit{\apjl} 626:L89--L92

\bibitem[{{Weidner} \& {Kroupa}(2004)}]{2004MNRAS.348..187W}
{Weidner} C, {Kroupa} P. 2004.
\newblock \textit{\mnras} 348:187--191

\bibitem[{{Weidner} \& {Kroupa}(2006)}]{2006MNRAS.365.1333W}
{Weidner} C, {Kroupa} P. 2006.
\newblock \textit{\mnras} 365:1333--1347

\bibitem[{{Weidner}, {Kroupa} \& {Larsen}(2004)}]{2004MNRAS.350.1503W}
{Weidner} C, {Kroupa} P, {Larsen} SS. 2004.
\newblock \textit{\mnras} 350:1503--1510

\bibitem[{{Weigelt} \& {Baier}(1985)}]{1985A&A...150L..18W}
{Weigelt} G, {Baier} G. 1985.
\newblock \textit{\aap} 150:L18--L20

\bibitem[{{Weinberg}(1994)}]{1994AJ....108.1403W}
{Weinberg} MD. 1994.
\newblock \textit{\aj} 108:1403--1413

\bibitem[{{Weldrake} et~al.(2005){Weldrake}, {Sackett}, {Bridges} \&
  {Freeman}}]{2005ApJ...620.1043W}
{Weldrake} DTF, {Sackett} PD, {Bridges} TJ, {Freeman} KC. 2005.
\newblock \textit{\apj} 620:1043--1051

\bibitem[{{White} \& {van Paradijs}(1996)}]{1996ApJ...473L..25W}
{White} NE, {van Paradijs} J. 1996.
\newblock \textit{\apjl} 473:L25

\bibitem[{{Whitmore}(2003)}]{2003dhst.symp..153W}
{Whitmore} BC. 2003.
\newblock In \textit{A Decade of Hubble Space Telescope Science}, eds.
  M~{Livio}, K~{Noll}, M~{Stiavelli}

\bibitem[{{Whitmore}, {Chandar} \& {Fall}(2007)}]{2007AJ....133.1067W}
{Whitmore} BC, {Chandar} R, {Fall} SM. 2007.
\newblock \textit{\aj} 133:1067--1084

\bibitem[{{Whitmore} \& {Schweizer}(1995)}]{1995AJ....109..960W}
{Whitmore} BC, {Schweizer} F. 1995.
\newblock \textit{\aj} 109:960--980

\bibitem[{{Whitmore} et~al.(1999){Whitmore}, {Zhang}, {Leitherer}, {Fall},
  {Schweizer} \& {Miller}}]{1999AJ....118.1551W}
{Whitmore} BC, {Zhang} Q, {Leitherer} C, {Fall} SM, {Schweizer} F, {Miller} BW.
  1999.
\newblock \textit{\aj} 118:1551--1576

\bibitem[{{Wielen}(1971)}]{1971A&A....13..309W}
{Wielen} R. 1971.
\newblock \textit{\aap} 13:309--322

\bibitem[{{Wielen}(1988)}]{1988IAUS..126..393W}
{Wielen} R. 1988.
\newblock In \textit{IAU Symp. 126: The Harlow-Shapley Symposium on Globular
  Cluster Systems in Galaxies}, eds. JE~{Grindlay}, AGD {Philip}

\bibitem[{{Wilkinson} et~al.(2003){Wilkinson}, {Hurley}, {Mackey}, {Gilmore} \&
  {Tout}}]{2003MNRAS.343.1025W}
{Wilkinson} MI, {Hurley} JR, {Mackey} AD, {Gilmore} GF, {Tout} CA. 2003.
\newblock \textit{\mnras} 343:1025--1037

\bibitem[{{Wu} et~al.(2002){Wu}, {Xue}, {Xia}, {Deng} \&
  {Mao}}]{2002ApJ...576..738W}
{Wu} H, {Xue} SJ, {Xia} XY, {Deng} ZG, {Mao} S. 2002.
\newblock \textit{\apj} 576:738--744

\bibitem[{{Wyder}, {Hodge} \& {Zucker}(2000)}]{2000PASP..112.1162W}
{Wyder} TK, {Hodge} PW, {Zucker} DB. 2000.
\newblock \textit{\pasp} 112:1162--1176

\bibitem[{{Yungelson} et~al.(2008){Yungelson}, {van den Heuvel}, {Vink},
  {Portegies Zwart} \& {de Koter}}]{2008A&A...477..223Y}
{Yungelson} LR, {van den Heuvel} EPJ, {Vink} JS, {Portegies Zwart} SF, {de
  Koter} A. 2008.
\newblock \textit{\aap} 477:223--237

\bibitem[{{Zepf} et~al.(1999){Zepf}, {Ashman}, {English}, {Freeman} \&
  {Sharples}}]{1999AJ....118..752Z}
{Zepf} SE, {Ashman} KM, {English} J, {Freeman} KC, {Sharples} RM. 1999.
\newblock \textit{\aj} 118:752--764

\bibitem[{{Zezas} et~al.(2002){Zezas}, {Fabbiano}, {Rots} \&
  {Murray}}]{2002ApJ...577..710Z}
{Zezas} A, {Fabbiano} G, {Rots} AH, {Murray} SS. 2002.
\newblock \textit{\apj} 577:710--725

\bibitem[{{Zhang} \& {Fall}(1999)}]{1999ApJ...527L..81Z}
{Zhang} Q, {Fall} SM. 1999.
\newblock \textit{\apjl} 527:L81--L84

\end{thebibliography}
\end{document}